\documentclass[10pt,journal,compsoc]{IEEEtran}

\usepackage{graphicx}
\usepackage{subfigure}

\usepackage{amsmath}

\usepackage{cite}
\usepackage{float}
\usepackage{amssymb}
\usepackage{multirow}
\usepackage{mathrsfs}
\usepackage{algorithmic}
\usepackage[ruled]{algorithm2e}
\usepackage{rotating}
\usepackage{balance}
\usepackage{threeparttable}
\usepackage{color}
\usepackage{colortbl}
\definecolor{mygray}{gray}{.9}
\usepackage{makecell}

\usepackage[pagewise]{lineno}
\usepackage{stfloats}
\usepackage{verbatim}
\usepackage{setspace}
\usepackage{threeparttable}
\usepackage{supertabular,booktabs}
\usepackage{rotating}

\usepackage{supertabular}
\usepackage{url}

\begin{document}
\title{A Survey on Adaptive Random Testing}
\author{Rubing Huang, ~\IEEEmembership{Senior Member,~IEEE,}
        Weifeng Sun,
        Yinyin Xu,
       Haibo Chen,
       Dave Towey,
       Xin Xia

\IEEEcompsocitemizethanks{


\IEEEcompsocthanksitem Rubing Huang is with the School of Computer Science and Communication Engineering, and also with Jiangsu Key Laboratory of Security Technology for Industrial Cyberspace, Jiangsu University, Zhenjiang, Jiangsu 212013, China.\protect\\
E-mail: rbhuang@ujs.edu.cn.

\IEEEcompsocthanksitem Weifeng Sun, Yinyin Xu, and Haibo Chen are with the School of Computer Science and Communication Engineering, Jiangsu University, Zhenjiang, Jiangsu 212013, China.\protect\\
E-mail: \{2211808031, 2221808040, 2221808034\}@stmail.ujs.edu.cn.

\IEEEcompsocthanksitem D. Towey is with the School of Computer Science, University of Nottingham Ningbo China, Ningbo, Zhejiang 315100, China.\protect\\
E-mail: dave.towey@nottingham.edu.cn.

\IEEEcompsocthanksitem Xin Xia is with the Faculty of Information Technology, Monash University, Melbourne, VIC 3168, Australia. \protect\\
E-mail: xin.xia@monash.edu.
}
}

\markboth{A Survey on Adaptive Random Testing}%
{Shell \MakeLowercase{\textit{Huang et al.}}: A Survey on Adaptive Random Testing}

\IEEEtitleabstractindextext{%
\begin{abstract}
Random testing (RT) is a well-studied testing method that has been widely applied to the testing of many applications, including embedded software systems, SQL database systems, and Android applications.
Adaptive random testing (ART) aims to enhance RT's failure-detection ability by more evenly spreading the test cases over the input domain.
Since its introduction in 2001, there have been many contributions to the development of ART, including various approaches, implementations, assessment and evaluation methods, and applications.
This paper provides a comprehensive survey on ART, classifying techniques, summarizing application areas, and analyzing experimental evaluations.
This paper also addresses some misconceptions about ART, and identifies open research challenges to be further investigated in the future work.
\end{abstract}


\begin{IEEEkeywords}
Adaptive random testing, random testing, survey.
\end{IEEEkeywords}}

\maketitle

\IEEEdisplaynontitleabstractindextext

%
\IEEEpeerreviewmaketitle

\IEEEraisesectionheading{\section{Introduction}\label{SEC:Introduction}}


\IEEEPARstart{S}{oftware} testing is a popular technique used to assess and assure the quality of the (software) system under test (SUT).
One fundamental testing approach involves simply constructing test cases in a random manner from the \textit{input domain} (the set of all possible program inputs):
This approach is called \textit{random testing} (RT)~\cite{Orso2014}.
RT may be the only testing approach used not only for operational testing, where the software reliability is estimated, but also for debug testing, where software failures are targeted with the purpose of removing the underlying bugs\footnote{According to the IEEE~\cite{STD2010}, the relationship amongst \emph{mistake}, \emph{fault}, \emph{bug}, \emph{defect}, \emph{failure} and \emph{error} can be briefly explained as follows: A software developer makes a \emph{mistake}, which may introduce a \emph{fault} (\emph{defect} or \emph{bug}) in the software. When a fault is encountered, a \emph{failure} may be produced, i.e., the software behaves unexpectedly. ``An \emph{error} is the difference between a computed, observed, or measured value or condition and the true, specified, or theoretically correct value or condition.'' \cite[p.128]{STD2010}.}~\cite{Frankl1998}.
Although conceptually very simple, RT has been used in the testing of many different environments and systems, including:
Windows NT applications~\cite{Forrester2000};
embedded software systems~\cite{Regehr2005};
SQL database systems~\cite{Bati2007}; and
Android applications~\cite{Muangsiri2017}.


RT has generated a lot of discussion and controversy, notably in the context of its effectiveness as a debug testing method~\cite{Myers2004}.
Many approaches have been proposed to enhance RT's testing effectiveness, especially for failure detection.
\textit{Adaptive random testing} (ART)~\cite{Chen2001} is one such proposed improvement over RT.
ART was motivated by observations reported independently by many researchers from multiple different areas
regarding the behavior and patterns of software failures:
Program inputs that trigger failures (\textit{failure-causing inputs}) tend to cluster into contiguous regions (\textit{failure regions})~\cite{White1980, Ammann1988, Finelli1991, Bishop1993, Schneckenburger2007}.
Furthermore, if the failure regions are contiguous, then it follows that non-failure regions should also be adjacent throughout the input domain.
Specifically:
if a test case ${tc}$ is a failure-causing input, then its neighbors have a high probability of also being failure-causing;
similarly, if ${tc}$ is not failure-causing, then its neighbors have a high probability of also not being failure-causing.
In other words, a program input that is far away from non-failure-causing inputs may have a higher probability of causing failure than the  neighboring test inputs.
Based on this, ART aims to achieve an even spread of (random) test cases over the input domain.
ART generally involves two processes:
one for the random generation of test inputs; and
another to ensure an even-spreading of the inputs throughout the input domain~\cite{Chen2010}.

\begin{figure*}[b]
\centering
\graphicspath{{Graphs/failurePattern/}}
    \subfigure[Strip pattern]
    {
        \includegraphics[width=0.25\textwidth]{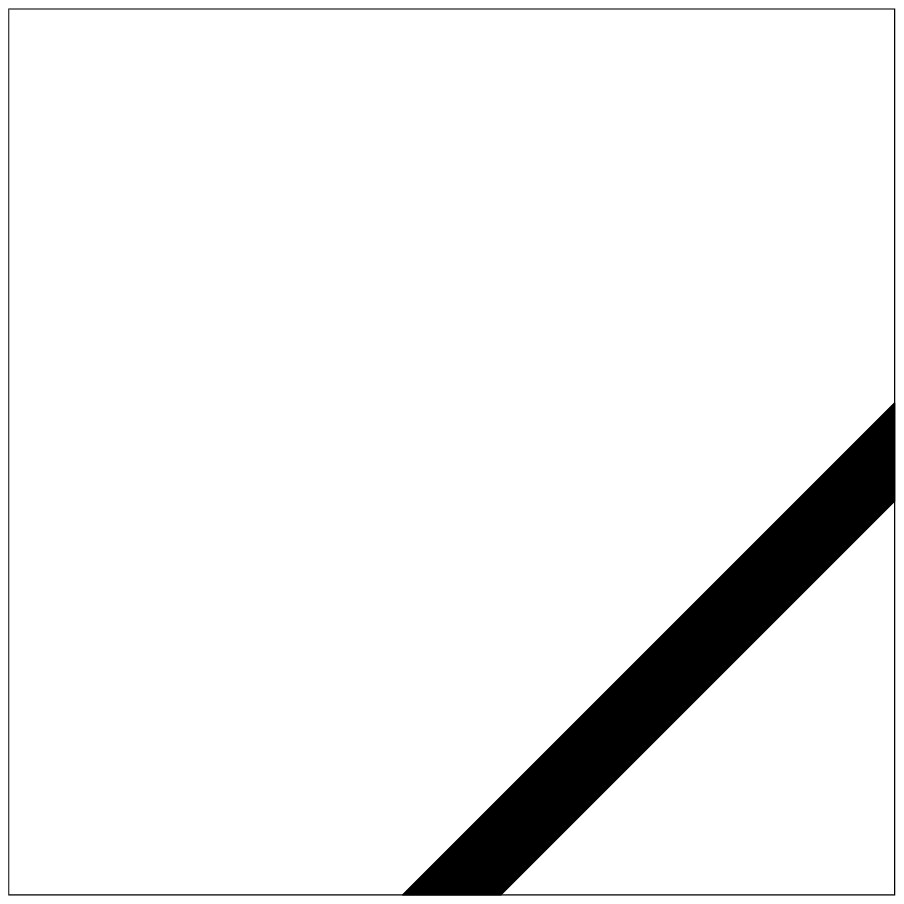}
        \label{figure:1.1}
    }
    \hspace*{25pt}
    \subfigure[Block pattern]
    {
        \includegraphics[width=0.25\textwidth]{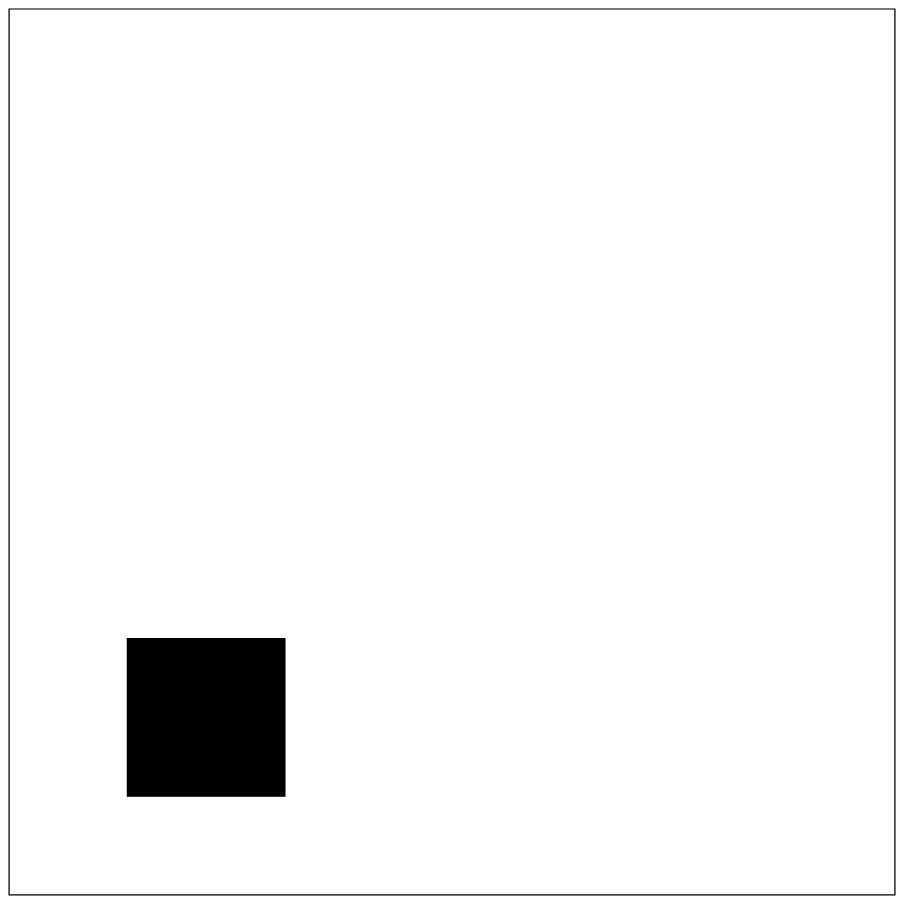}
        \label{figure:1.2}
    }
    \hspace*{25pt}
    \subfigure[Point pattern]
    {
        \includegraphics[width=0.25\textwidth]{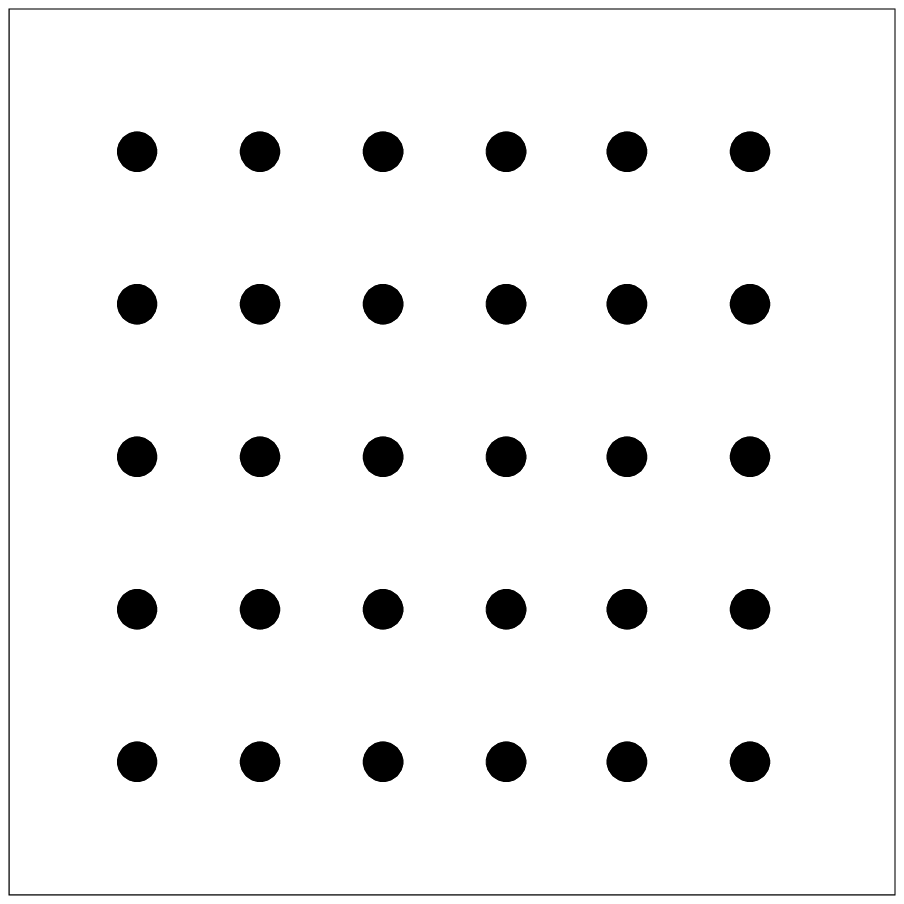}
        \label{figure:1.3}
    }
    \caption{Three types of failure patterns in the two-dimensional input domain.}
    \label{FIG:Failurepattern}
\end{figure*}

ART's invention and appearance in the literature can be traced back to a journal paper by Chen et al., published in 2001~\cite{Chen2001}.
Some papers present overviews of ART, but are either preliminary, or do not make ART the main focus~\cite{Chan2003,Towey2007,Chen2010,Anand2013,Roslina2015,Chen2015}. 
For example, to draw attention to the fundamental role of diversity in test case selection strategies, Chen et al.~\cite{Chen2010} presented a synthesis of some of the most important ART research results before 2010.
Similarly, Anand et al.~\cite{Anand2013} presented an orchestrated survey of the most popular techniques for automatic test case generation that included a brief report on the then state-of-the-art of ART.
Roslina et al.~\cite{Roslina2015} also conducted a study of ART techniques based on 61 papers.
Consequently, there is currently no up-to-date, exhaustive survey analyzing both the state-of-the-art and new potential research directions of ART.
This paper fills this gap in the literature.

In this paper, we present a comprehensive survey on ART covering 140 papers published between 2001 and 2017.
The paper includes the following:
(1) a summary and analysis of the selected 140 papers;
(2) a description, classification, and summary of the techniques used to derive the main ART strategies; 
(3) a summary of the application and testing domains in which ART has been applied;
(4) an analysis of the empirical studies conducted into ART;
(5) a discussion of some misconceptions surrounding ART; and
(6) a summary of some open research challenges that should be addressed in future ART work.
To the best of our knowledge, this is the first large-scale and comprehensive survey on ART.

The rest of this paper is organized as follows.
Section~\ref{SEC:ARTbrief} briefly introduces the preliminaries and gives an overview of ART.
Section~\ref{SEC:Literture} discusses this paper's literature review methodology.
Section~\ref{SEC:overview} examines the evolution and distribution of ART studies. 
Section~\ref{SEC:Techniques} analyzes the-state-of-the-art of ART techniques.
Section~\ref{SEC:Applications} presents the situations and problems to which ART has been applied.
Section~\ref{SEC:Evaluations} gives a detailed analysis of the various empirical evaluations of ART.
Section~\ref{SEC:Misconceptions} discusses some misconceptions.
Section~\ref{SEC:Challenges} provides some potential challenges to be addressed in future work.
Finally, Section~\ref{SEC:Conclusions} concludes the paper.

\section{Background\label{SEC:ARTbrief}}

This section presents some preliminary concepts, and provides an introduction to ART.

\subsection{Preliminaries}

For a given SUT, many software testing methods have been implemented according to the following four steps:
(1) define the testing objectives;
(2) choose inputs for the SUT (\textit{test cases});
(3) run the SUT with these test cases; and
(4) analyze the results.
Each test case is selected from the set of all possible inputs that form the \textit{input domain}.
When the SUT's output or behavior  when executing a test case $tc$ is not as expected (determined by the \textit{test oracle}~\cite{Weyuker1982}), then the test is considered to \textit{fail}, otherwise it is \textit{passes}.
When a test fails, $tc$ is called a \textit{failure-causing input}. 

Given some faulty software, two fundamental features can be used to describe the properties of the fault(s):
the \textit{failure rate} (the number of failure-causing inputs as a proportion of all possible inputs); and
the \textit{failure pattern} (the distributions of failure-causing inputs across the input domain, including their geometric shapes and locations).
Before testing, these two features are fixed, but unknown.

Chan et al.~\cite{Chan1996} identified three broad categories of failure patterns:
\textit{strip}, \textit{block} and \textit{point}.
Fig.~\ref{FIG:Failurepattern} illustrates these three failure patterns in a two-dimensional input domain (the bounding box represents  the input domain boundaries; and the black strip, block, or dots represent the failure-causing inputs).
Previous studies have indicated that strip and block patterns are more commonly encountered than point patterns~\cite{White1980, Ammann1988, Finelli1991, Bishop1993, Schneckenburger2007}.

Generally speaking, failure regions are identified or constructed in empirical studies (experiments or simulations).
Experiments involve real faults or mutants (seeded using mutation testing~\cite{Jia2011}) in real-life subject programs.
Simulations, in contrast, create artificial failure regions using pre-defined values for the dimensionality $d$ and failure rate $\theta$, and a predetermined failure pattern type:
A $d$-dimensional unit hypercube is often used to simulate the input domain $\mathcal{D}$
($\mathcal{D}=\{(x_1,x_2,\cdots,x_d)|0 \leq x_1,x_2,\cdots,x_d < 1.0\}$), with the failure regions randomly placed inside $\mathcal{D}$, and their sizes and shapes determined by $\theta$ and the selected failure pattern, respectively.
During testing, when a generated test case is inside a failure region, a failure is said to be detected.

\subsection{Adaptive Random Testing (ART)\label{SEC:2.2}}

ART is a family of testing methods, with many different implementations based on various intuitions and criteria.
In this section, we present an ART implementation to illustrate the fundamental principles.

\begin{figure*}[!t]
\centering
\graphicspath{{Graphs/FSCSprocess/}}
    \subfigure[]
    {
        \includegraphics[width=0.25\textwidth]{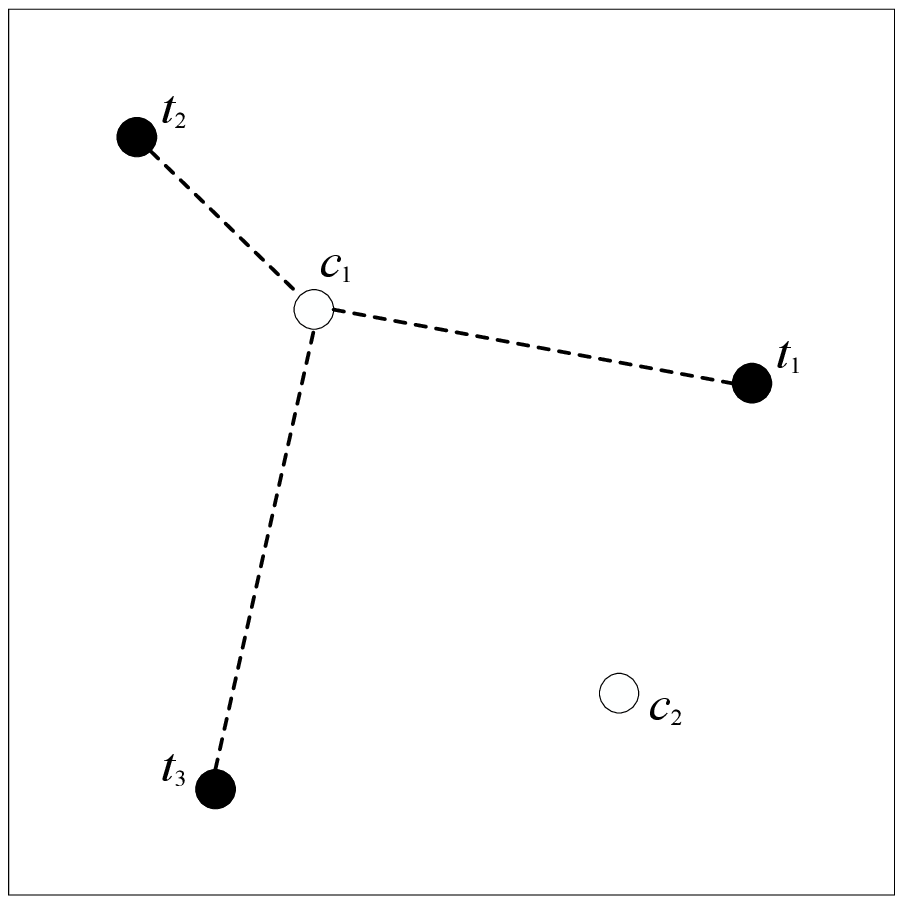}
        \label{FIG:FSCS-1}
    }
    \hspace*{25pt}
    \subfigure[]
    {
        \includegraphics[width=0.25\textwidth]{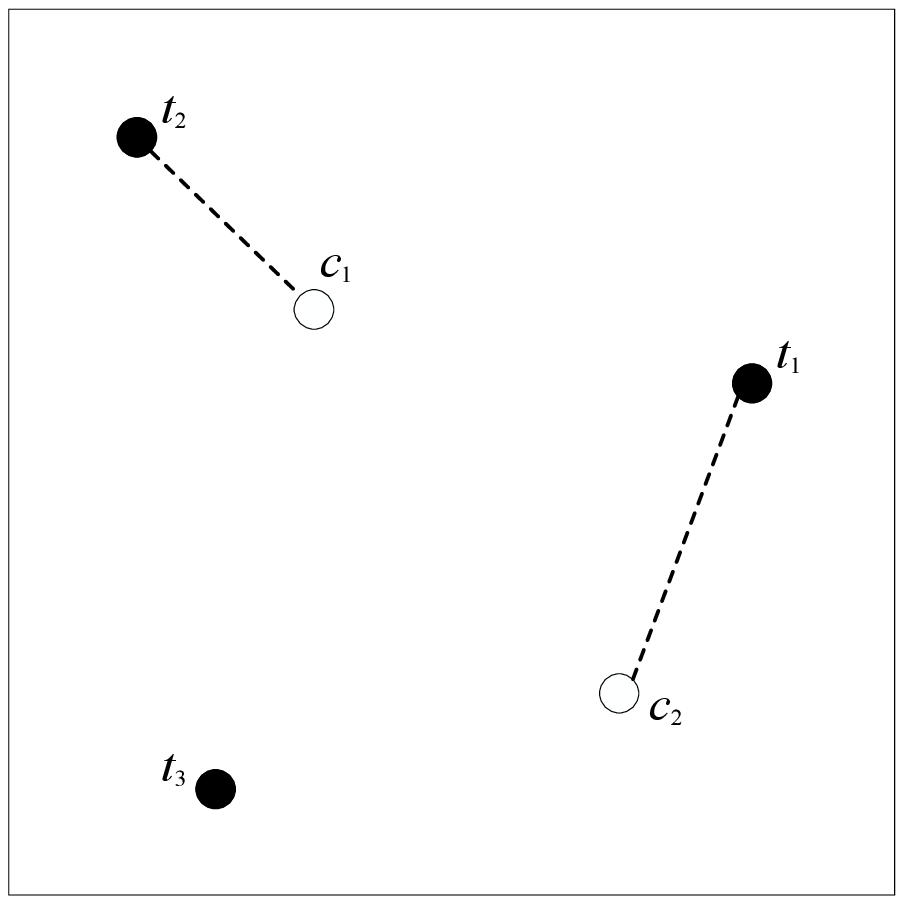}
        \label{FIG:FSCS-2}
    }
    \hspace*{25pt}
    \subfigure[]
    {
        \includegraphics[width=0.25\textwidth]{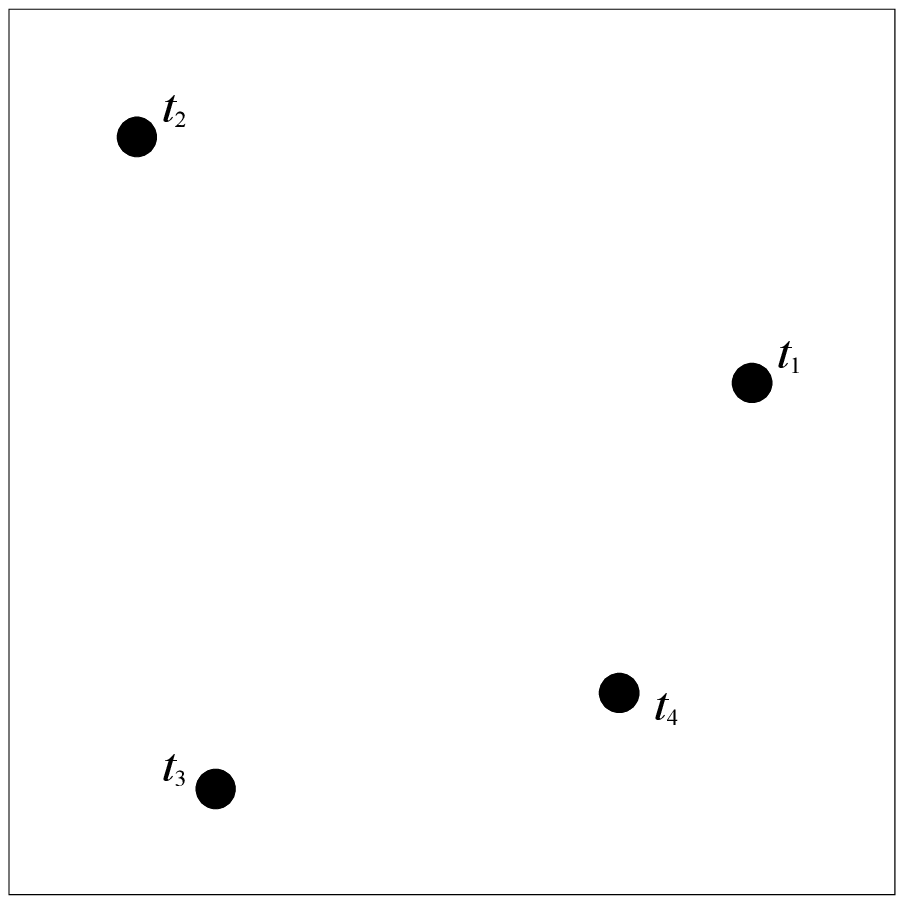}
        \label{FIG:FSCS-3}
    }
    \caption{Illustration of FSCS in a two-dimensional input domain.}
    \label{FIG:FSCS}
\end{figure*}

The first implementation of ART was the \textit{Fixed-Size-Candidate-Set} (FSCS)~\cite{Chen2001} version, which makes use of the concept of distance between test inputs.
FSCS uses two sets of test cases:
the candidate set $C$; and
the executed set, $E$.
$C$ is a set of $k$ tests randomly generated from the input domain (according to the specific distribution); and
$E$ is the set of those inputs that have already been executed, but without causing any failure.
$E$ is initially empty.
The first test input is generated randomly.
In each iteration of FSCS ART, an element from $C$ is selected as the next test case such that it is farthest away from all previously executed tests (those in $E$).
Formally, the element $c'$ from $C$ is chosen as the next test case such that it satisfies the following constraint:
\begin{equation}\tag{2.1}
    \forall c \in C,~\min_{e \in E}dist(c',e) \geq \min_{e \in E}dist(c,e),
    \label{EQ:1}
\end{equation}
where $dist(x, y)$ is a function to measure the distance between two test inputs $x$ and $y$.
The \textit{Euclidean distance} is typically used in $dist(x, y)$ for numerical input domains.

Fig.~\ref{FIG:FSCS} illustrates the FSCS process in a two-dimensional input domain:
Suppose that there are three previously executed test cases, $t_1,t_2$, and $t_3$ (i.e., $E=\{t_1,t_2,t_3\}$), and two randomly generated test candidates, $c_1$ and $c_2$ (i.e., $C=\{c_1,c_2\}$) (Fig.~\ref{FIG:FSCS-1}).
To select the next test case from $C$, the distance between each candidate and each previously executed each test case in $E$ is calculated, and the minimum value for each candidate is recorded as the \textit{fitness value} (Fig.~\ref{FIG:FSCS-2}).
Finally, the candidate with the maximum fitness value is selected to be the next test case (Fig.~\ref{FIG:FSCS-3}):
in this example, $c_2$ is used for testing ($t_4=c_2$).

As Chen et al. have explained~\cite{Chen2010}, ART aims to more evenly spread randomly generated test cases than RT across the input domain. In other words, ART attempts to generate more diverse test cases than RT.

\section{Methodology\label{SEC:Literture}}

Guided by Kitchenham and Charters~\cite{Kitchenham2007} and Petersen et al.~\cite{Petersen2015}, in this paper, we followed a structured and systematic method to perform the ART survey.
We also referred to recent survey papers on other software engineering topics, including:
mutation analysis~\cite{Jia2011};
metamorphic testing~\cite{Segura2016,Chen2018};
constrained interaction testing~\cite{Ahmed2017}; and
test case prioritization for regression testing~\cite{Khatibsyarbini2018}.
The detailed methodology used is described in this section.

\subsection{Research Questions}

The goal of this survey paper is to structure and categorize the available ART details and evidence.
To achieve this, we used the following research questions (RQs):
\begin{itemize}
    \item \textit{RQ1: What has been the evolution and distribution of ART topics in the published studies?}
    \item \textit{RQ2: What different ART strategies and approaches exist?}
    \item \textit{RQ3: In what domains and applications has ART been applied?}
    \item \textit{RQ4: How have empirical evaluations in ART studies been performed?}
    \item \textit{RQ5: What misconceptions surrounding ART exist?}
    \item \textit{RQ6: What are the remaining challenges and other future ART work?}
\end{itemize}

The answer to RQ1 will provide an overview of the published ART papers.
RQ2 will identify the state-of-the-art in ART strategies and techniques, giving a description, summary, and classification.
RQ3 will identify where and how ART has been applied.
RQ4 will explore how the various ART studies involving simulations and experiments with real programs were conducted and evaluated.
RQ5 will examine common ART misconceptions, and, finally, RQ6 will identify some remaining challenges and potential research opportunities.

\subsection{Literature Search and Selection}

Following previous survey studies~\cite{Segura2016,Ahmed2017,Khatibsyarbini2018}, we also selected the following five online literature repositories belonging to publishers of technical research:
\begin{itemize}
    \item ACM Digital Library
    \item Elsevier Science Direct
    \item IEEE Xplore Digital Library
    \item Springer Online Library
    \item Wiley Online Library
\end{itemize}
The choice of these repositories was influenced by the fact that a number of important journal articles about ART are available through Elsevier Science Direct, Springer Online Library, and Wiley Online Library.
Also, ACM Digital Library and IEEE Xplore not only offer articles from conferences, symposia, and workshops, but also provide access to some important relevant journals.
%

\begin{table}[!b]
    \centering
    \caption{Initial Number of Papers for Each Search Engine\label{TAB:InitialNumber}}
    \begin{tabular*}{0.48\textwidth}{lcc} \Xhline{1.0pt}
        \textbf{Search Engine} &~~~~~~~~~~~~~~~~~~~~~~~~~ &\textbf{Studies}\\\Xhline{0.6pt}
        ACM Digital Library	&&29   \\
        Elsevier Science Direct	&&57   \\
        IEEE Xplore Digital Library	&&60   \\
        Springer Online Library	&&89   \\
        Wiley Online Library	&&31   \\
        Google Scholar  &&1010\\
        \textbf{Total}  &&\textbf{1276}\\
        \Xhline{1.0pt}
    \end{tabular*}
\end{table}

\begin{figure*}[!b]
\centering
\graphicspath{{Graphs/summary/}}
    \subfigure[Number of publications per year.]
    {
        \includegraphics[width=0.485\textwidth]{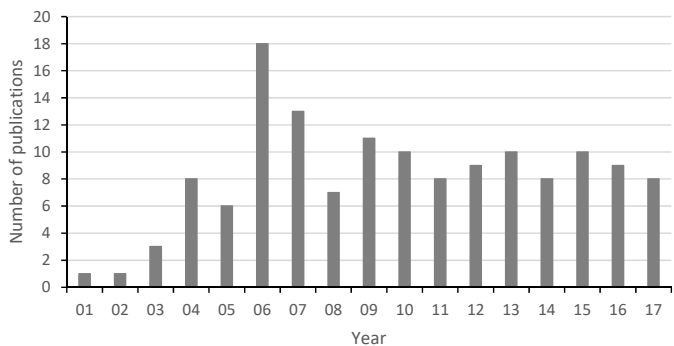}
        \label{FIG:paperNumber}
    }
    \subfigure[Cumulative number of publications per year.]
    {
        \includegraphics[width=0.485\textwidth]{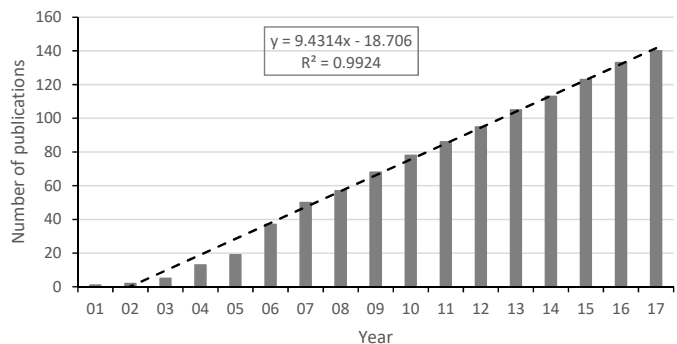}
        \label{FIG:paperNumberC}
    }
    \caption{ART papers published between January 1st 2001 and December 31st 2017.}
    \label{FIG:paperGraph}
\end{figure*}

After determining the literature repositories, each repository was searched using the exact phrase ``adaptive random testing" and for titles or keywords containing ``random test".
To avoid missing papers that were not included in these five repositories, we augmented the set of papers using a search in the Google Scholar database with the phrase ``adaptive random testing" as the search string\footnote{The search was performed on May 1st, 2018.}.
The results were combined to form the candidate set of published studies shown in Table~\ref{TAB:InitialNumber}.
Duplicates were removed, and then, to further reduce the candidate set size, we then applied the following exclusion criteria:
\begin{itemize}
    \item[1)] Studies not written in English.
    \item[2)] Studies not related to the field of computer science.
    \item[3)] Studies not related to ART.
    \item[4)] Books, book chapters, or technical reports (most of which have been published as articles).
    \item[5)] Master or Ph.D. theses.
    \item[6)] Keynote records (because, generally, these are only extremely brief summaries or overviews --- e.g., Chen's keynote at the 8th International Conference on Quality Software~\cite{Chen2008c}).
    \item[7)] Studies without a full-text. 
\end{itemize}

Removal of duplicates and application of the exclusion criteria reduced the initial 1,276 candidate studies to 138 published papers.
Finally, a snowballing process~\cite{Petersen2015} was conducted by checking the references of the selected 138 papers, resulting in the addition of two more papers.
In total, 140 publications (\textit{primary studies}) were selected for inclusion in the survey.

We acknowledge the apparent infeasibility of finding all ART papers through our search.
However, we are confident that we have included the majority of relevant published papers, and that our survey provides the overall trends and the-state-of-the-art of ART.

\subsection{Data Extraction and Collection}

\begin{table}[!t]
    \centering
    \renewcommand\arraystretch{1.1}
    \footnotesize
    \caption{Data Collection for Research Questions\label{TAB:DataCollection}}
    \begin{tabular*}{0.48\textwidth}{ll} \Xhline{1.0pt}
        \textbf{RQs} &\textbf{Type of Data Extracted}\\\Xhline{0.6pt}
        \multirow{2}*
        {RQ1}	&Fundamental information for each paper (publication year,\\
        		&~~~~type of paper, author name, and affiliation).\\
        RQ2 	&Motivation, description, and analysis of each technique.\\
        RQ3 	&Language, function, and environment for each application.\\
        \multirow{2}*
        {RQ4} 	&Simulation details, subject programs, evaluation metrics,\\
            		&~~~~fault types, and analysis.  \\
        RQ5 	&Current misconception details.\\
        RQ6 	&Details of remaining challenges.\\
        \Xhline{1pt}
    \end{tabular*}
\end{table}

All 140 primary studies were carefully read and inspected, with data extracted according to our research questions.
As summarized in Table~\ref{TAB:DataCollection}, we identified the following information from each study:
motivation, contribution, empirical evaluation details, misconceptions, and remaining challenges.
To avoid missing information and reduce error as much as possible, this process was performed by two different co-authors, and subsequently verified by the other co-authors at least twice.

\section{Answer to RQ1: What has been the evolution and distribution of ART topics in the published studies?\label{SEC:overview}}

In this section, we address RQ1 by summarizing the primary studies according to publication trends, authors, venues, and types of contributions to ART.


\begin{figure*}[!b]
\centering
\graphicspath{{Graphs/summary/}}
    \subfigure[Overall venue distribution.]
    {
        \includegraphics[width=0.485\textwidth]{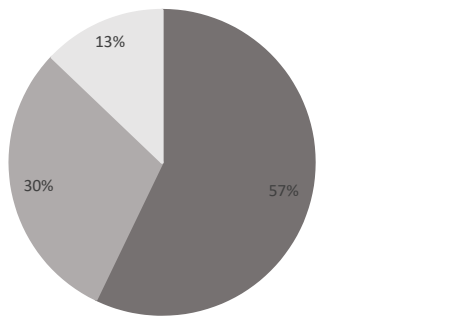}
        \label{FIG:venue1}
    }
    \subfigure[Venue distribution per year.]
    {
        \includegraphics[width=0.485\textwidth]{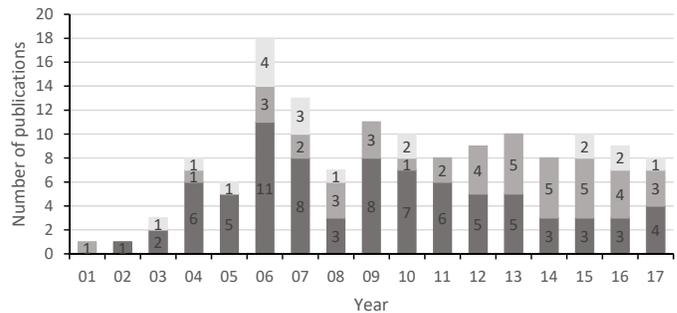}
        \label{FIG:venue2}
    }
    \caption{Venue distribution for ART papers.}
    \label{FIG:CJWdistribution}
\end{figure*}

\subsection{Publication Trends}

Fig.~\ref{FIG:paperGraph} presents the ART publication trends between January 1st, 2001 and December 31st, 2017, with
Fig.~\ref{FIG:paperNumber} showing the number of publications each year, and Fig.~\ref{FIG:paperNumberC} showing the cumulative number.
It can be observed that, after the first three years there are at least six publications per year, with the number reaching a peak in 2006. 
Furthermore, since 2009, the number of publications each year has remained relatively fixed, ranging from seven to 10.
An analysis of the cumulative publications (Fig.~\ref{FIG:paperNumberC}) shows that a line function with high determination coefficient ($R^2=0.9924$) can be identified.
This indicates that the topic of ART has been experiencing a strong linear growth, attracting continued interest and showing healthy development.
Following this trend, it is anticipated that there will be about 180 ART papers by 2021, two decades after its appearance in Chen et al.~\cite{Chen2001}.

\begin{table*}[!t]
    \centering
    \caption{Top Ten ART Authors\label{TAB:TopCoauthor}}
    \begin{tabular*}{0.8\textwidth}{rlllc} \Xhline{1.0pt}
        \textbf{Rank}    &\textbf{Name} &\textbf{Current Affiliation} &\textbf{Country or Region} &\textbf{Papers}\\\Xhline{0.6pt}
        1.  &T. Y. Chen	&Swinburne University of Technology	&Australia	&62   \\
        2.  &F.-C. Kuo	&Swinburne University of Technology	&Australia	&35   \\
        3.  &H. Liu	&Victoria University	&Australia	&19   \\
        4.  &R. G. Merkel	&Monash University	&Australia	&15   \\
        5.  &D. Towey	&University of Nottingham Ningbo China	&PRC	&15   \\
        6.  &J. Mayer	&Ulm University	&Germany	&13   \\
        7.  &K. P. Chan	&The University of Hong Kong	&Hong Kong	&10   \\
        8.  &Z. Q. Zhou	&University of Wollongong	&Australia	&9   \\
        9.  &R. Huang	&Jiangsu University	&PRC	&8   \\
        10.  &L. C. Briand	&University of Luxembourg	&Luxembourg	&7   \\
        \Xhline{1.0pt}
    \end{tabular*}
\end{table*}

\subsection{Researchers and Organizations}

Based on the 140 primary studies, 167 ART authors were identified, representing 82 different affiliations.
Table~\ref{TAB:TopCoauthor} lists the top 10 ART authors and their most recent affiliation (with country or region).
It is clear that T. Y. Chen, from Swinburne University of Technology in Australia, is the most prolific ART author, with 62 papers.

\subsection{Geographical Distribution of Publications}

We examined the geographical distribution of the 140 publications according to the affiliation country of the first author, as shown in Table~\ref{TAB:Geographic}.
We found that all primary studies could be associated with a total of 18 different countries or regions, with Australia ranking first, followed by the People's Republic of China (PRC). 
Overall, about 41\% of ART papers came from Asia;
32\% from Oceania;
19\% from Europe; and about
8\% from America.
This distribution of papers suggests that the ART community may only be represented by a modest number of countries spread throughout the world.

\begin{figure*}[!b]
\centering
\graphicspath{{Graphs/summary/}}
    \subfigure[Type of main contribution.]
    {
        \includegraphics[width=0.485\textwidth]{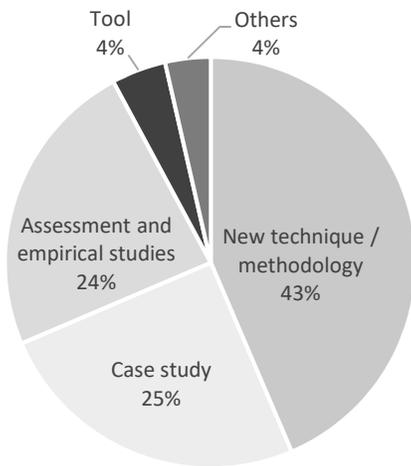}
        \label{FIG:contribution}
    }
    \subfigure[Research topic.]
    {
        \includegraphics[width=0.485\textwidth]{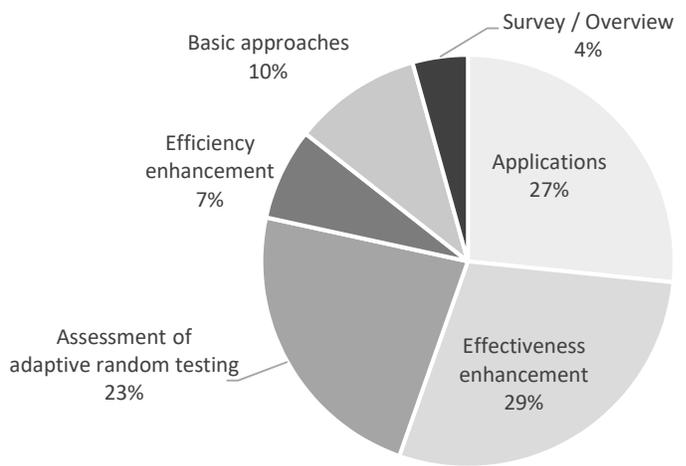}
        \label{FIG:topic}
    }
    \caption{Distribution of primary studies by main contribution and research topic.}
    \label{FIG:contributionTopic}
\end{figure*}

\begin{table}[!t]
    \centering
    \caption{Geographical Distribution of Publications\label{TAB:Geographic}}
    \begin{tabular*}{0.48\textwidth}{rclcc} \Xhline{1.0pt}
        \textbf{Rank}    &~~~~~~~~~~    &\textbf{Country or Region} &~~~~~~~~~~ &\textbf{Papers}\\\Xhline{0.6pt}
        1.  &&Australia	&&45   \\
        2.  &&PRC	&&30   \\
        3.  &&Hong Kong	&&13   \\
        4.  &&Germany	&&13   \\
        5.  &&Norway	&&6   \\
        6.  &&United States	&&5   \\
        7.  &&Canada	&&4   \\
        8.  &&United Kingdom	&&4   \\
        9.  &&Malaysia	&&3   \\
        10.  &&Iran	&&3   \\
        11.  &&Indonesia	&&3   \\
        12.  &&Japan	&&2   \\
        13.  &&Switzerland	&&2   \\
        14.  &&Brazil	&&2   \\
        15.  &&Korea	&&2   \\
        16.  &&India	&&1   \\
        17.  &&Italy	&&1   \\
        18.  &&Luxembourg	&&1   \\
        \Xhline{1.0pt}
    \end{tabular*}
\end{table}

\subsection{Distribution of Publication Venues}

The 140 primary studies under consideration were published in 72 different venues (41 conferences or symposia, 20 journals, and 11 workshops).
Fig.~\ref{FIG:CJWdistribution} illustrates the distribution of publication venues, with Fig.~\ref{FIG:venue1} showing the overall venue distribution, and Fig.~\ref{FIG:venue2} giving the venue distribution per year.
As Fig.~\ref{FIG:venue1} shows, most papers have been published in conferences or symposia proceedings (57\%), followed by journals (30\%), and then workshops (13\%).
Fig.~\ref{FIG:venue2} shows that, between 2002 and 2012, most ART publications each year were conference and symposium papers, followed by journals and workshops.
Fig.~\ref{FIG:venue2} also shows that, since 2012, this trend has changed, with the number of journal papers per year increasing, usually outnumbering the conference papers.
The workshop papers generally form the least number of publications each year.

\begin{table*}[!t]
    \centering
    \caption{Top Venues with a Minimum of Three ART Papers\label{TAB:TopVenues}}
    \begin{tabular*}{0.94\textwidth}{rllc} \Xhline{1.0pt}
        \textbf{Rank}    &\textbf{Acronym} &\textbf{Full name} &\textbf{Papers}\\\Xhline{0.6pt}
        1.  &QSIC	&International Conference on Quality Software	&11   \\
        2.  &SEKE	&International Conference on Software Engineering and Knowledge Engineering	&8   \\
        3.  &JSS	&Journal of Systems and Software	&7   \\
        4.  &COMPSAC	&Annual Computer Software and Applications Conference	&6   \\
        5.  &IEEE-TR	&IEEE Transactions on Reliability	&5   \\
        6.  &SAC	&ACM Symposium on Applied Computing	&5   \\
        7.  &RT	&International Workshop on Random Testing	&5   \\
        8.  &ASE	&International Conference on Automated Software Engineering	&4   \\
        9.  &IST	&Information and Software Technology	&4   \\
        10.  &TSE	&IEEE Transactions on Software Engineering	&3   \\
        11.  &TOSEM	&ACM Transactions on Software Engineering and Methodology	&3   \\
        12.  &TOC	&IEEE Transactions on Computers	&3   \\
        13.  &Ada-Europe	&Ada-Europe International Conference on Reliable Software Technologies	&3   \\
        14.  &APSEC	&Asia-Pacific Software Engineering Conference	&3   \\
        15.  &ICST	&International Conference on Software Testing, Verification and Validation	&3   \\
        \Xhline{1.0pt}
    \end{tabular*}
\end{table*}

Table~\ref{TAB:TopVenues} lists the ranking of publication venues where at least three ART papers have appeared.
Most of these venues are well-known and highly regarded in the field of software engineering or software testing.

\subsection{Types of Contributions}

Fig.~\ref{FIG:contributionTopic} categorizes the primary studies according to main contribution (Fig.~\ref{FIG:contribution}) and research topic (Fig.~\ref{FIG:topic})\footnote{If a paper has multiple types of contributions or research topics, then only the main contribution or topic is identified.}.

As Fig.~\ref{FIG:contribution} shows, the main contribution of 43\% of the studies was to present new ART techniques or methodologies, 25\% were case studies, and 24\% were assessments and empirical studies.
4\% of studies were surveys or overviews of ART, and the main contribution of five primary studies (4\%) was to present a tool.

Fig.~\ref{FIG:topic} shows that the primary research topic of 10\% of the studies was about basic ART approaches.
The effectiveness and efficiency enhancement of ART approaches were the focus of 29\% and 7\%, respectively; and
application and assessment of ART were 27\% and 23\%, respectively.
Finally, 4\% of the papers report on techniques, achievements, and research directions.

\vspace{0.2cm}\noindent \fbox{\parbox[c]{0.97\linewidth} {
\textit{\textbf{Summary of answers to RQ1:}}
\begin{itemize}
    \item[1)] \textit{ART has attracted sustained interest, with the topic showing healthy development.}
    \item[2)] \textit{Over 160 ART authors have been identified, representing more than 80 different affiliations, with T. Y. Chen being the most prolific.}
    \item[3)] \textit{Primary studies have come from 18 countries or regions, with Australia ranking first, followed by the PRC (People's Republic of China).}
    \item[4)] \textit{Most studies were published at conferences and symposia, followed by journals, and workshops.}
    \item[5)] \textit{The main contribution of most primary studies was to propose a new technique.
    The most popular research topics have been ART effectiveness enhancement, application, and assessment.}
\end{itemize}
}}

\section{Answer to RQ2: What different ART strategies and approaches exist?\label{SEC:Techniques}}

In this section, we present the state-of-the-art of ART approaches, including a description, classification, and summary of their strengths and weaknesses. 

ART attempts  to spread test cases evenly over the entire input domain~\cite{Chen2010}, which should result in a better failure detection ability~\cite{Chen2007}.
There are two basic rationales to achieving this even spread of test cases:

\textbf{\emph{Rationale 1: New test cases should be far away from previously executed, non-failure-causing test cases.}}
As discussed in the literature \cite{White1980, Ammann1988, Finelli1991, Bishop1993,Schneckenburger2007}, failure regions tend to be contiguous, which means that new test cases farther away from those already executed (but without causing failure) should have a higher probability of being failure-causing.

\textbf{\emph{Rationale 2: New test cases should contribute to a good distribution of all test cases over the entire input domain.}}
Previous studies~\cite{Chen2007} have empirically determined that better test case distributions result in better failure detection ability. 

{\em Discrepancy} is a metric commonly used to measure the equidistribution of sample inputs~\cite{Chen2007}, with lower values indicating better distributions.
The discrepancy of a test set $T$ is calculated as:
    \begin{equation}\tag{5.1}\label{Eq:5.1}
     	Discrepancy(T) = \max_{1 \leq i \leq m}{\bigg| \frac{|T_i|}{|T|} - \frac{|\mathcal{D}_i|}{|\mathcal{D}|}\bigg|},
    \end{equation}
    where
$\mathcal{D}_1, \mathcal{D}_2, \cdots, \mathcal{D}_m$ are $m$ randomly defined subdomains of the input domain $\mathcal{D}$; and
$T_1,T_2,\cdots T_m$ are the corresponding subsets of $T$, such that each $T_i~(i=1,2,\cdots,m)$ is in $\mathcal{D}_i$.
Discrepancy checks whether or not the number of test cases in a subdomain is proportionate to the relative size of the subdomain area ---
larger subdomains should have more test cases; and smaller ones should have relatively fewer.


Both \textbf{Rationale 1} and \textbf{Rationale 2} achieve a degree of \textit{diversity} of test cases over the input domain~\cite{Chen2010}.
Based on these rationales, many strategies have been proposed:
\textit{Select-Test-From-Candidates Strategy} (STFCS), \textit{Partitioning-Based Strategy} (PBS), \textit{Test-Profile-Based Strategy} (TPBS), \textit{Quasi-Random Strategy} (QRS), \textit{Search-Based Strategy} (SBS), and \textit{Hybrid Strategies} (HSs).


%


\subsection{Select-Test-From-Candidates Strategy}
The \textit{Select-Test-From-Candidates Strategy} (STFCS) chooses the next test case from a set of candidates based on some criteria or evaluation involving the previously executed test cases.


\subsubsection{Framework}

Fig.~\ref{FIG:STFCS} presents a  pseudocode framework for STFCS, showing two main components:
the \textit{random-candidate-set-construction}, and \textit{test-case-selection}.
The STFCS framework maintains two sets of test cases:
the candidate set ($C$) of randomly generated candidate test cases; and
the executed set ($E$) of those test cases already executed (without causing failure).
The first test case is selected randomly from the input domain $\mathcal{D}$ according to a uniform distribution ---
all inputs in $\mathcal{D}$ have equal probability of selection.
The test case is then applied to the SUT, and the output and behavior are examined to confirm whether or not a failure has been caused.
Until a stopping condition is satisfied (e.g., a failure has been caused), the framework repeatedly uses the random-candidate-set-construction component to prepare the candidates, and then uses the test-case-selection component to choose one of these candidates as the next test case to be applied to the SUT.

Two basic (and popular) approaches to implementing the STFCS framework are \textit{Fixed-Size-Candidate-Set (FSCS) ART}~\cite{Chen2001,Chen2004a}, and
\textit{Restricted Random Testing (RRT) }~\cite{Chan2002,Chan2003a,Chan2004,Chan2006b}\footnote{Previous ART studies have generally considered RRT to represent an \textit{ART by exclusion} category~\cite{Anand2013,Roslina2015}.
However, both RRT and FSCS belong to the STFCS category.}.
Clearly, there are different ways to realize the random-candidate-set-construction and test-case-selection components, leading to different STFCS implementations.
A number of enhanced versions of both FSCS and RRT have also been developed.

\begin{figure}[!t]
\centering
\fbox{
\centering
\parbox{0.95\linewidth}{
\begin{algorithmic}[1]
    \renewcommand{\algorithmicrequire}{\textbf{Input:}}
    \renewcommand{\algorithmicensure}{\textbf{Output:}}
    \renewcommand{\algorithmicelsif}{\algorithmicelse}
    \renewcommand{\algorithmicthen}{}
\STATE Set $C\leftarrow\{\}$, and $E\leftarrow\{\}$
\STATE Randomly generate a test case ${tc}$ from $\mathcal{D}$, according to uniform distribution
\STATE Add ${tc}$ into $E$, i.e., $E\leftarrow E\bigcup\{{tc}\}$
\WHILE {The stopping condition is not satisfied}
\STATE Randomly choose a specific number of elements from $\mathcal{D}$ to form $C$ according to the specific criterion\\\boxed{\textit{Random-candidate-set-construction component}}
\STATE Find a ${tc}\in C$ as the next test case satisfying the specific criterion
\boxed{\textit{Test-case-selection component}}
\STATE $E\leftarrow E\bigcup\{{tc}\}$
\ENDWHILE
\STATE Report the result and exit
\end{algorithmic}
}
}
\caption{Framework pseudocode of the STFCS category.}
\label{FIG:STFCS}
\end{figure}

\subsubsection{Random-Candidate-Set-Construction Component}

Several different implementations of the random-candidate-construction component have been developed.

1) \textit{Uniform distribution}~\cite{Chen2001}:
This involves construction of the candidate set by randomly selecting test cases according to a uniform distribution.

2) \textit{Non-uniform distribution}~\cite{Chen2009b}:
When not using a uniform distribution to generate the candidates, the non-uniform distribution is usually dynamically updated throughout the test case generation process.
Chen et al.~\cite{Chen2009b}, for example, used a dynamic, non-uniform distribution to have candidates be more likely to come from the center of $\mathcal{D}$ than from the boundary region.

3) \textit{Filtering by eligibility}~\cite{Kuo2007a,Kuo2008}:
Using an eligibility criterion (specified using a tester-defined parameter), this filtering ensures that candidates (and therefore the eventually generated test cases) are drawn only from the eligible regions of $\mathcal{D}$.
The criterion used is that the selected candidate's coordinates are as different as possible to those of all previously executed test cases.
Given a test case in a $d$-dimensional input domain, $(x_1,x_2,\cdots,x_d)$, filtering by eligibility selects candidates such that each $i$-th  coordinate, $x_i~(1\leq i \leq d)$, is different to the $i$-th coordinate of every previously selected test case. 
This ensures that all test cases have different values for all coordinates.

4) \textit{Construction using data pools}~\cite{Ciupa2008,Hou2013}:
``Data pools'' are first constructed by identifying and adding both specific special values for the particular data type (such as -1, 0, 1, etc. for integers), and the boundary values (such as the minimum and maximum possible values). 
This method then selects candidates randomly from either just the data pools, with a probability of $p$, or from the entire input domain, with probability of $1-p$. 
Selected candidates are removed from the data pool.
Once the data pool is exhausted, or falls below a threshold size, it is then updated by adding new elements.

5) \textit{Achieving a specific degree of coverage}~\cite{Jiang2009}:
This involves selecting candidates randomly (with uniform distribution) from the input domain until some specific coverage criteria (such as branch, statement, or function coverage) are met.

\subsubsection{Candidate Set Size}

The size of the candidate set, $k$, may either be a fixed number (e.g., determined in advance, perhaps by the testers), or a flexible one. 
Although, intuitively, increasing the size of $k$ should improve the testing effectiveness, as reported by Chen et al.~\cite{Chen2004a}, the improvement in FSCS ART performance is not significant when $k > 10$:
In most studies, therefore, $k$ has been assigned a value of $10$.
However, when the value of $k$ is flexible, there are different methods to design and determine its value, based on the execution conditions or environment. 

\subsubsection{Test-Case-Identification Component}
The test-case-identification component chooses one of the candidates as the next test case, according to the specific criterion.
There are generally two different implementations:
\textit{Implementation 1}: After measuring all candidates, identifying the \textit{best} one  (as implemented in FSCS); and
\textit{Implementation 2}: Checking candidates until the first \textit{suitable} (or \textit{valid}) one is identified (as implemented in RRT).
The goal of the test-case-identification component is to achieve the even spreading of test cases over the input domain, which it does based on the \textit{fitness} value.
In other words, the \textit{fitness function} measures each candidate $c$ from the candidate set $C$ against the executed set $E$.
We next list the seven different fitness functions, $fitness(c,E)$, used in STFCS, with the first six following \textit{Implementation 1}; and the last one following \textit{Implementation 2}.

1) \textit{Minimum-Distance}~\cite{Chen2001}:
This involves calculating the distance between $c$ and each element $e$ from $E$ ($e\in E$), and then choosing the minimum distance as the fitness value for $c$.
In other words, the fitness function of $c$ against $E$ is the distance between $c$ and its nearest neighbor in $E$:
    \begin{equation}\tag{5.2}
        \label{EQ:4.1}
         fitness(c,E)=\min_{e \in E}{dist}(c,e).
    \end{equation}

2) \textit{Average-Distance}~\cite{Ciupa2008}:
Similar to \textit{Minimum-Distance}, this also computes the distance between $c$ and each element $e$ in $E$, but instead of the minimum, the average of these distances is used as the fitness value for $c$: 
    \begin{equation}\tag{5.3}
        fitness(c,E)=\frac{1}{|E|}\sum\limits_{e \in E}{dist}(c,e).
    \end{equation}

3) \textit{Maximum-Distance}~\cite{Jiang2009}:
This assigns the {\em maximum} distance as the fitness value for $c$.
In other words, this fitness function chooses the distance between $c$ and its neighbor in $E$ that is farthest away.
    \begin{equation}\tag{5.4}
        fitness(c,E)=\max\limits_{e \in E}{dist}(c,e).
    \end{equation}

4) \textit{Centroid-Distance}~\cite{Chan2004a,Putra2013}:
This uses the distance between $c$ and the centroid (center of the gravity) of $E$ as the fitness value for $c$:
    \begin{equation}\tag{5.5}
        fitness(c,E)= {dist}\Bigg(c, \frac{1}{|E|}\sum_{e\in E}{e}\Bigg),
    \end{equation}
where $\frac{1}{|E|}\sum\limits_{e\in E}{e}$ returns the centroid of $E$.


5) \textit{Discrepancy}~\cite{Chen2007f,Chen2009}:
This involves choosing the next test case such that it achieves the lowest discrepancy when added to $E$.
Therefore, this fitness function of $c$ can be defined as:
    \begin{equation}\tag{5.6}
         fitness(c,E) = 1 - Discrepancy\Big(E \bigcup \{c\}\Big).
    \end{equation}



6) \textit{Membership-Grade}~\cite{Chan2004b}:
\textit{Fuzzy Set Theory}~\cite{Zadeh1965} can be used to define some \textit{fuzzy features} to construct a \textit{membership grade function}, allowing a candidate with the highest (or threshold) score to be selected as the next test case.
Chan et al.~\cite{Chan2004b} defined some fuzzy features based on distance, combining them to calculate the membership grade function for candidate test cases.
Two of the features they used are:
the \textit{Dynamic Minimum Separating Distance} (DMSD), which is a minimum distance between executed test cases, decreasing in magnitude as the number of executed test cases increases; and
the \textit{Absolute Minimum Separating Distance} (AMSD), which is an absolute minimum distance between test cases, regardless of how many test cases have been executed.
During the evaluation of a candidate $c$ against $E$, if $\forall e \in E$, $dist(c,e) \geq \textit{\textrm{DMSD}}$, then selection of $c$ will be strongly favored;
however, if $\exists e \in E$ such that $dist(c,e) \leq \textit{\textrm{AMSD}}$, then $c$ will be strongly disfavored.
The candidate most highly favored (with the highest membership grade) is then selected as the next test case.

7) \textit{Restriction}~\cite{Chan2002,Zhou2010,Chan2004b}:
This involves checking whether or not a candidate $c$ violates the pre-defined restriction criteria related to $E$, denoted $restriction(c,E)$.
The fitness function of $c$ against $E$ can be defined as:
    \begin{equation}\tag{5.7}
        fitness(c,E) = \left\{%
         \begin{array}{ll}
        0,&{\textrm{if }} {restriction}(c,E) {\textrm{is true}}, \\
        1,&{\textrm{otherwise}}.  \\
         \end{array}
         \right.
    \end{equation}
The random-candidate-construction criterion successively selects candidates from the input domain (according to uniform or non-uniform distribution) until one that is not restricted is identified.
Three approaches to using \textit{Restriction} in ART are:
\begin{itemize}
\item
Previous studies~\cite{Chan2002,Chan2003a,Chan2004,Chan2006} have implemented restriction by checking whether or not $c$ is located outside of all (equally-sized) \textit{exclusion regions} defined around all test cases in $E$.
In a 2-dimensional numeric input domain $\mathcal{D}$, for example, Chan et al.~\cite{Chan2002} used circles around each already selected test case as exclusion regions, thereby defining the $restriction(c,E)$ as:
    \begin{equation}\tag{5.8}\label{EQ:5.8}
      \forall e \in E,~{dist}(c,e) < \sqrt{\frac{R \cdot |\mathcal{D}|}{\pi \cdot |E|}},
    \end{equation}
     where $R$ is the \textit{target exclusion ratio} (set by the tester~\cite{Chen2007b}), and $dist(c,e)$ is the \textit{Euclidean distance} between $c$ and $e$.
     In fact, Eq. (\ref{EQ:5.8}) relates to the DMSD fuzzy feature~\cite{Chan2004b}.

\item
Zhou et al.~\cite{Zhou2010,Zhou2010a,Zhou2013} designed an acceptance probability $P_\beta$ based on Markov Chain Monte Carlo (MCMC) methods~\cite{Stephen1998} to control the identification of random candidates.
Given a candidate $c$, the method generates a new candidate $c'$ according to the applied distribution (uniform or non-uniform), resulting in $restriction(c,E)$ being defined as:
    \begin{equation}
        \tag{5.9}
        \mathcal{U} > P_\beta=\min\bigg\{ \frac{P(X(c')=1|E)}{P(X(c)=1|E)},1 \bigg\},
    \end{equation}
    where $\mathcal{U}$ is a uniform random number in the interval $[0, 1.0)$,
    $X(c)$ is the execution output of $c$ ($X(c)=1$ means that $c$ is a failure-causing input, and $X(c)=0$ means that it is not), and
    $P(X(c)=1|E)$ represents the probability that $c$ is failure-causing, given the set $E$ of already executed test cases.
    According to Bayes' rule, we have:
    \begin{equation}
        \tag{5.10}
        \label{EQ:5.10}
        P(X(c)=1|E) = P(E|X(c)=1)P(X(c)=1)/Z,
    \end{equation}
    where $Z$ is a normalizing constant.
    Assuming all elements in $E$ are conditionally independent for a test output of $c$, we then have:
    \begin{equation}
        \tag{5.11}
        P(E|X(c)=1) = \prod_{e \in E}P(X(e)|X(c)=1).
    \end{equation}
    As illustrated by Zhou et al.~\cite{Zhou2010,Zhou2010a,Zhou2013}, $P(X(e)|X(c)=1)$ is defined as:
    \begin{equation}\tag{5.12}
         P(X(e)=1|X(c)=1) = \exp(-dist(e,c)/\beta_1),
    \end{equation}
    and
    \begin{equation}\tag{5.13}
         P(X(e)=0|X(c)=1) = 1-\exp(-dist(e,c)/\beta_1),
    \end{equation}
    where $\beta_1$ is a constant.
    If one candidate is a greater distance from the non-failure-causing test cases than another candidate, then it has a higher probability of being selected as the next test case.

\item
Using \textit{Fuzzy Set Theory}~\cite{Zadeh1965}, Chan et al.~\cite{Chan2004b} applied a dynamic threshold $\lambda$ to determine whether or not a candidate was acceptable, accepting the candidate if its membership grade was greater than $\lambda$.
If a predetermined number of candidates are rejected for being below $\lambda$, they adjusted the threshold according to the specified principles.
It should be noted that any \textit{Implementation 1} approach to choosing candidates can be transformed into \textit{Implementation 2} by applying a threshold mechanism.
\end{itemize}

Six of the seven fitness functions (\textit{Minimum-Distance}, \textit{Average-Distance}, \textit{Maximum-Distance}, \textit{Centroid-Distance}, \textit{Membership-Grade}, and \textit{Restriction}) satisfy \textbf{Rationale 1};
one (\textit{Discrepancy}) satisfies \textbf{Rationale 2}.



\subsection{Partitioning-Based Strategy}

The \textit{Partitioning-Based Strategy} (PBS) divides the input domain into a number of subdomains, choosing one as the location within which to generate the next test case.
Core elements of PBS, therefore, are to partition the input domain and to select the subdomain.

\subsubsection{Framework}

Fig.~\ref{FIG:PBS} presents a pseudocode framework for PBS, showing two main components:
the \textit{partitioning-schema}, and \textit{subdomain-selection}.
The partitioning-schema component defines how to partition the input domain into  subdomains, and the subdomain-selection component
defines how to choose the target subdomain where the next test case will be generated.

\begin{figure}[!t]
    \centering
    \fbox{
    \centering
    \parbox{0.95\linewidth}{
    \begin{algorithmic}[1]
            \renewcommand{\algorithmicrequire}{\textbf{Input:}}
            \renewcommand{\algorithmicensure}{\textbf{Output:}}
            \renewcommand{\algorithmicelsif}{\algorithmicelse}
        \STATE Set $E\leftarrow\{\}$
        \STATE Randomly generate a test case ${tc}$ from $\mathcal{D}$, according to uniform distribution
        \STATE Add ${tc}$ into $E$, i.e., $E\leftarrow E\bigcup\{{tc}\}$
        \WHILE {The stopping condition is not satisfied}
        \IF {The partitioning condition is triggered}
        \STATE Partition the input domain $\mathcal{D}$ into $m$ disjoint subdomains $D_1,D_2,\cdots,D_m$, according to the specific criterion \boxed{\textit{Partitioning-schema component}}
        \ENDIF
        \STATE Choose a subdomain $D_i$ according the specific criterion \boxed{\textit{Subdomain-selection component}}
        \STATE Randomly generate the next test case ${tc}$ from $D_i$, according to uniform distribution
        \STATE $E\leftarrow E\bigcup\{{tc}\}$
        \ENDWHILE
        \STATE Report the result and exit
    \end{algorithmic}
    }
    }
    \caption{Framework pseudocode of the PBS category.}
    \label{FIG:PBS}
\end{figure}


After partitioning, the input domain $\mathcal{D}$ will be divided into $m$ disjoint subdomains $\mathcal{D}_1,\mathcal{D}_2,\cdots, \mathcal{D}_m~(m > 1)$, according to the partitioning-schema criteria:
$\mathcal{D}_i \bigcap \mathcal{D}_j = \emptyset~(1\leq i \neq j \leq m)$, and $\mathcal{D}=\mathcal{D}_1\bigcup \mathcal{D}_2 \bigcup \cdots \bigcup \mathcal{D}_m$.
Next, based on the subdomain-selection criteria, PBS chooses a \textit{suitable} subdomain  within which to generate the next test case.

\subsubsection{Partitioning-Schema Component}

Many different criteria can be used to partition the input domain, which can be achieved using either \textit{static} or \textit{dynamic} partitioning.
Static partitioning~\cite{Chen2003,Chen2004b,Kuo2009,Sabor2015} means that the input domain is divided before test case generation, with no further partitioning required once testing begins.
Dynamic partitioning involves dividing the input domain dynamically, often at the same time that each new test case is generated.
There have been many dynamic partitioning schemas proposed, including \textit{random partitioning}~\cite{Chen2004d}, \textit{bisection partitioning}~\cite{Chen2004d,Chow2013}, and \textit{iterative partitioning}~\cite{Chen2006a,Mayer2006d}.


1) \textit{Static partitioning}~\cite{Sabor2015}:
Static partitioning divides the input domain into a fixed number of equally-sized subdomains, with these subdomains then remaining unchanged throughout the entire testing process. 
This is simple, but influenced by the tester:
testers need to divide the input domain before testing, and different numbers of subdomains may result in different ART performance. 

2) \textit{Random partitioning}~\cite{Chen2004d}:
Random partitioning uses the generated test case ${tc}$ as the breakpoint to divide the input (sub-)domain $\mathcal{D}_i$ into smaller subdomains.
This partitioning usually results in the input domain $\mathcal{D}$ being divided into subdomains of unequal size.

3) \textit{Bisection partitioning}~\cite{Chen2004d,Chow2013}:
Similar to static partitioning, bisection partitioning divides the input domain into equally-sized subdomains.
However, unlike static partitioning, bisection partitioning {\em dynamically} bisects the input domain whenever the partitioning condition is triggered.
There are a number of bisection partitioning implementations.
Chen et al.~\cite{Chen2004d}, for example, successively bisected dimensions of the input domain;
whenever the $i$-th bisection of a dimension resulted in $2^i$ parts, the $(i+1)$-th bisection was then applied to another dimension.
Chow et al.~\cite{Chow2013} bisected all dimensions at the same time, with the input domain $\mathcal{D}$ being divided into $2^{i*d}$ subdomains (where $d$ is the dimensionality of $\mathcal{D}$) after the $i$-th bisection.
Bisection partitioning does not change existing partitions during bisection, only bisecting the subdomains in the next round. 

4) \textit{Iterative partitioning}~\cite{Chen2006a,Mayer2006d}:
In contrast to bisection partitioning, iterative partitioning modifies existing partitions, resulting in the input domain being divided into equally-sized subdomains.
Each round of iterative partitioning divides the entire input domain $\mathcal{D}$ using a new schema.
After the $i$-th round of partitioning, for example, Chen et al.~\cite{Chen2006a} divided the input domain into $i^d$ subdomains, with each dimension divided into $i$ equally-sized parts.
Mayer et al.~\cite{Mayer2006d}, however, divided only the largest dimension into equally-sized parts, leaving other dimensions unchanged, resulting in a dimension with $j$ parts being divided into $j+1$ parts. 

Although random partitioning may divide the input domain into subdomains with different sizes, the other three partitioning approaches result in equally-sized subdomains.


\subsubsection{Subdomain-Selection Component}

After partitioning the input domain $\mathcal{D}$ into $m$ subdomains $\mathcal{D}_1, \mathcal{D}_2,\cdots,\mathcal{D}_m$, the next step is to choose the subdomain where the next test case will be generated.
The following criteria can be used to support this subdomain selection process:

1) \textit{Maximum size}~\cite{Chen2004d}: Given the set $T$ of previously generated test cases, among those subdomains $\mathcal{D}_i~(1 \leq i \leq m)$ without any test cases, the largest one is selected for generation of the next test case:
$\forall j\in\{1,2,\cdots,m\}$ satisfying $\mathcal{D}_i\bigcap T = \mathcal{D}_j \bigcap T = \emptyset$, and $|\mathcal{D}_i| \geq|\mathcal{D}_j|$.

2) \textit{Fewest previously generated test cases}~\cite{Chen2004d,Chow2013}:
Given the set $T$ of previously generated test cases, this criterion chooses a subdomain $\mathcal{D}_i$ containing the fewest test cases:
$\forall j\in\{1,2,\cdots,m\}$, $|\mathcal{D}_i \bigcap T| \leq |\mathcal{D}_j \bigcap T|$.

3) \textit{No test cases in target or neighbor subdomains}~\cite{Sabor2015,Chen2006a,Mayer2006d}:
This ensures that the selected subdomain not only contains no test cases, but also does not neighbor other subdomains containing test cases.

4) \textit{Proportional selection}~\cite{Kuo2007}:
Proportional selection uses two dynamic probability values, $p_1$ and $p_2$, to represent the likelihood that some (or all) elements of the failure region are located in the edge or center regions, respectively. 
Kuo et al.~\cite{Kuo2007}, for example, used two equally-sized subdomains in their proportional selection implementation, with each test case selected from either the edge or center region based on the value of $p_1/p_2$.

Three criteria (\textit{fewest previously generated test cases}, \textit{no test cases in target or neighbor subdomains}, and \textit{proportional selection}) require that all subdomains be the same size; only one (\textit{maximum size}) has no such requirement.
Furthermore, two criteria (\textit{maximum size} and \textit{no test cases in target or neighbor subdomains})
generally select one test case per subdomain;
the other two (\textit{fewest previously generated test cases} and \textit{proportional selection}) may select multiple test cases from each subdomain.

Intuitively speaking, three criteria (\textit{maximum size}, \textit{fewest previously generated test cases}, and \textit{proportional selection}) follow \textbf{Rationale 2}.
The \textit{maximum size} criterion chooses the largest subdomain without any test cases as the target location for the next test case ---
test selection from a larger subdomain may have a better chance of achieving a good distribution of test cases.
Similarly, the \textit{fewest previously generated test cases}, and \textit{proportional selection} criteria ensure that subdomains with fewer test cases have a higher probability of being selected.
The third criterion (\textit{no test cases in target or neighbor subdomains}) follows both \textbf{Rationale 1} and \textbf{Rationale 2}, choosing a target subdomain without (and away from) any test cases, thereby achieving a good test case distribution.
Furthermore, because this criterion also avoids subdomains neighboring those containing test cases, the subsequently generated test cases generally have a minimum distance from all others.

\subsection{Test-Profile-Based Strategy}

The \textit{Test-Profile-Based Strategy} (TPBS)~\cite{Liu2010} generates test cases based on a well-designed test profile (different from the uniform distribution), dynamically updating the profile after each test case selection.



\begin{figure}[!b]
    \centering
    \fbox{
    \centering
    \parbox{0.95\linewidth}{
    \begin{algorithmic}[1]
            \renewcommand{\algorithmicrequire}{\textbf{Input:}}
            \renewcommand{\algorithmicensure}{\textbf{Output:}}
            \renewcommand{\algorithmicelsif}{\algorithmicelse}
        \STATE Set $E\leftarrow\{\}$
        \STATE Randomly generate a test case ${tc}$ from $\mathcal{D}$, according to uniform distribution
        \STATE Add ${tc}$ into $E$, i.e., $E\leftarrow E\bigcup\{{tc}\}$
        \WHILE {The stopping condition is not satisfied}
        \STATE Adjust the test profile based on already selected test cases from $E$ \boxed{\textit{Test-profile-adjustment component}}
        \STATE Randomly generate the next test case ${tc}$ based on adjusted test profile
        \STATE $E\leftarrow E\bigcup\{{tc}\}$
        \ENDWHILE
        \STATE Report the result and exit
    \end{algorithmic}
    }
    }
    \caption{Framework pseudocode of the TPBS category.}
    \label{FIG:TPBS}
\end{figure}

\subsubsection{Framework}

Fig.~\ref{FIG:TPBS} presents a pseudocode framework for TPBS.
Because TPBS generates test cases based on the test profile, the core part of TBPS focuses on how to design the dynamic test profile.
A test profile can be considered as the selection probability distribution for all test inputs in the input domain $\mathcal{D}$, with test cases in different locations having different probabilities.
When a test case is executed without causing failure, its selection probability is then assigned a value of $0$.



\subsubsection{Test-Profile-Adjustment Component}

Based on the intuitions underlying ART~\cite{Chen2001}, a test profile should be adjusted to satisfy the following~\cite{Liu2010}:
\begin{itemize}
    \item The closer a test input is to the previously executed test cases, the {\em lower} the selection probability that it is assigned should be.
    \item The farther away a test input is from previously executed test cases, the {\em higher} the selection probability that it is assigned should be.
    \item The probability distribution should be dynamically adjusted to maintain these two features.
\end{itemize}

A number of test profiles exist to describe the probability distribution of test cases, including the \textit{triangle profile}~\cite{Liu2009}, \textit{cosine profile}~\cite{Liu2009}, \textit{semicircle profile}~\cite{Liu2009}, and \textit{power-law profile}~\cite{Merkel2011}.
Furthermore, the \textit{probabilistic ART} implementation~\cite{Chan2006} uses a similar mechanism to TPBS.

Because the test profiles use the location of non-failure-causing test cases when assigning the selection probability of each test input from the input domain, TPBS obviously follows \textbf{Rationale 1}:
If a test input is farther away from non-failure-causing test cases than other candidates, it has a higher probability of being chosen as the next test case.

\begin{figure}[!b]
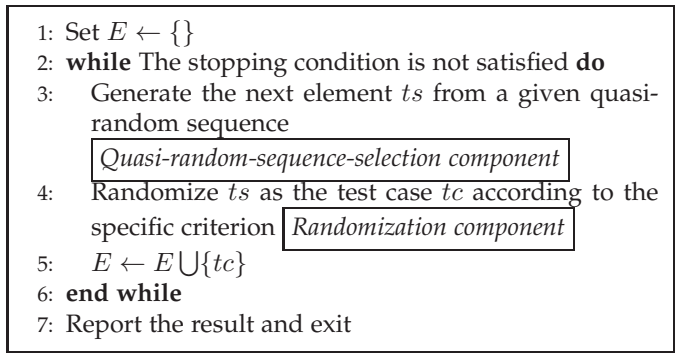

    \centering
    \fbox{
    \centering
    \parbox{0.95\linewidth}{
    \begin{algorithmic}[1]
            \renewcommand{\algorithmicrequire}{\textbf{Input:}}
            \renewcommand{\algorithmicensure}{\textbf{Output:}}
            \renewcommand{\algorithmicelsif}{\algorithmicelse}
        \STATE Set $E\leftarrow\{\}$
        \WHILE {The stopping condition is not satisfied}
        \STATE Generate the next element $ts$ from a given quasi-random sequence\\\boxed{\textit{Quasi-random-sequence-selection component}}
        \STATE Randomize $ts$ as the test case ${tc}$ according to the specific criterion \boxed{\textit{Randomization component}}
        \STATE $E\leftarrow E\bigcup\{{tc}\}$
        \ENDWHILE
        \STATE Report the result and exit
    \end{algorithmic}
    }
    }
    \caption{Framework pseudocode of the QRS category.}
    \label{FIG:QRS}
\end{figure}

\subsection{Quasi-Random Strategy}

The \textit{Quasi-Random Strategy} (QRS)~\cite{Chen2005a,Chen2007g} applies \textit{quasi-random sequences} to the implementation of ART.
Quasi-random sequences are point sequences with low discrepancy and low dispersion:
As discussed by Chen et al.~\cite{Chen2007}, a set of points with lower discrepancy and dispersion generally has a more even distribution.
Furthermore, the computational overheads incurred when generating $n$ quasi-random test cases is only $O(n)$, which is similar to that of pure RT.
In other words,  QRS can achieve an even-spread of test cases with a low computational cost.

\subsubsection{Framework}

Fig.~\ref{FIG:QRS} presents a pseudocode framework for QRS, showing the two main components:
\textit{quasi-random-sequence-selection} and \textit{randomization}.
QRS first takes a quasi-random sequence to construct each point, then randomizes it to create the next test case according to the specific criterion.
The main motivation for involving randomization in the process is that quasi-random sequences are usually generated by deterministic algorithms,
which means that the sequences violate a core principle of ART:
the randomness of the test cases.

\subsubsection{Quasi-Random-Sequence-Selection Component}

A number of quasi-random sequences have been examined, including Halton~\cite{Halton1964}, Sobol~\cite{Sobol1976}, and Niederreiter~\cite{Niederreiter1988}. 
In this section, we only describe some representative sequences for quasi-random testing.

1) \textit{Halton sequence}~\cite{Halton1964}:
The Halton sequence can be considered the $d$-dimensional extension of the Van der Corput sequence, a one-dimensional quasi-random sequence~\cite{Chi2006} defined as:
\begin{equation}\tag{5.14}
    \phi_b(i)=\sum_{j=0}^\omega{i_jb^{-j-1}},
\end{equation}
where $b$ is a prime number, $\phi_b(i)$ denotes the $i$-th element of the Van der Corput sequence, $i_j$ is the $j$-th digit of $i$ (in base $b$), and $\omega$ denotes the lowest integer for which $\forall j > \omega,~i_j = 0$ is true.
For a $d$-dimensional input domain, therefore, the $i$-th element of the Halton sequence can be defined as $(\phi_{b_1}(i),\phi_{b_2}(i), \cdots, \phi_{b_d}(i))$, where the bases, $b_1, b_2, \cdots,b_d$, are pairwise coprime.
Previous studies~\cite{Chi2006,Shahbazi2013} have used the Halton sequence to generate test cases.

2) \textit{Sobol sequence}~\cite{Sobol1976}:
The Sobol sequence can be considered a permutation of the binary Van der Corput sequence, $\phi_2(i)$, in each dimension~\cite{Chi2006}, and is defined as:
\begin{equation}
    \tag{5.15}
    {Sobol}(i)=\mathop{\mathrm{XOR}}\limits_{j=1,2,\cdots,\omega}(i_j\delta_j),
\end{equation}
\begin{equation}
    \tag{5.16}
    \delta_j=\mathop{\mathrm{XOR}}\limits_{k=1,2,\cdots,r}(\frac{\beta_k\delta_{j-k}}{2^j})\oplus\frac{\delta_{j-r}}{2^{j+r}},
\end{equation}
where ${{Sobol}}(i)$ represents the $i$-th element of the Sobol sequence, $i_j$ is the $j$-th digit of $i$ in binary, $\omega$ denotes the number of digits of $i$ in binary, and $\beta_1,\beta_2,\cdots,\beta_r$ come from the coefficients of a degree $r$ primitive polynomial over the finite field.
Previous studies~\cite{Chi2006,Shahbazi2013,Liu2016} have used this sequence for test case generation.

3) \textit{Niederreiter sequence}~\cite{Niederreiter1988}:
The Niederreiter sequence may be considered to provide a good reference for other quasi-random sequences:
because all the other approaches can be described in terms of what Niederreiter calls $(t,d)$-sequences~\cite{Niederreiter1988}. 
As discussed by Chen and Merkel~\cite{Chen2005a}, the Niederreiter sequence has lower discrepancy than other sequences.
Previous investigations~\cite{Chen2005a,Shahbazi2013} have used Niederreiter sequences to conduct software testing.

\subsubsection{Randomization Component}

The randomization step involves randomizing the quasi-random sequences into actual test cases.
The following three representative methods illustrate this.

1) \textit{Cranley-Patterson Rotation}~\cite{Cranley1976,Kollig2002}:
This generates a random $d$-dimensional vector $V=(v_1,v_2,\cdots,v_d)$ to permute each coordinate of the $i$-th point $T_i=(t_i^1,t_i^2,\cdots,t_i^d)$ to a new $i$-th point $P_i=(p_i^1,p_i^2,\cdots,p_i^d)$, where
    \begin{equation}\tag{5.17}
        p_i^j = \left\{%
    \begin{array}{ll}
        t_i^j+v_j,&{\textrm{if }} t_i^j+v_j<1, \\
        t_i^j+v_j - 1,&{\textrm{if }} t_i^j+v_j \geq 1.  \\
     \end{array}
     \right.
    \end{equation}

2) \textit{Owen Scrambling}~\cite{Owen1995}:
Owen Scrambling applies the randomization process to the Niederreiter sequence (a $(t,d)$-sequence in base $b$).
The $i$-th point in the sequence can be written as $T_i=(t_i^1,t_i^2,\cdots,t_i^d)$, where $t_i^j=\sum\limits_{k=1}^{\infty}{a_{ijk}b^{-k}}$.
The permutation process is applied to the parameter $a_{ijk}$ for each point according to some criteria.
Compared with Cranley-Patterson rotation, Owen Scrambling more precisely maintains the low discrepancy and low dispersion of quasi-random sequences~\cite{Chen2007g}.

3) \textit{Random Shaking and Rotation}~\cite{Liu2009a,Liu2016}:
This first uses a non-uniform distribution (such as the \textit{cosine distribution}) to shake the coordinates of each item in the quasi-random sequence into a random number within a specific value range.
Then, a random vector based on the non-uniform distribution is used to permute the coordinates of all points in the sequence.

QRS generates a list of test cases (a quasi-random sequence) with a good distribution (including discrepancy~\cite{Chen2007}), indicating that it follows \textbf{Rationale 2}.


\begin{figure}[!b]
    \centering
    \fbox{
    \centering
    \parbox{0.95\linewidth}{
    \begin{algorithmic}[1]
            \renewcommand{\algorithmicrequire}{\textbf{Input:}}
            \renewcommand{\algorithmicensure}{\textbf{Output:}}
            \renewcommand{\algorithmicelsif}{\algorithmicelse}
        \STATE Set the number of test cases $N$
        \STATE $E \leftarrow\{\}$
        \STATE Generate an initial population of test sets $PT=\{T_1,T_2,\cdots,T_{ps}\}$, each of which is randomly generated with size $N$ according to uniform distribution, where $ps$ is the population size
        \WHILE {The stopping condition is not satisfied}
        \STATE Evolve $PT$ to construct a new population of test sets $PT'$ by using a given search-based algorithm\\\boxed{\textit{Evolution component}}
        \STATE $PT \leftarrow PT'$
        \ENDWHILE
        \STATE $E \leftarrow$ the best solution of $PT$
        \STATE Report the result and exit
    \end{algorithmic}
    }
    }
    \caption{Framework pseudocode of the SBS category.}
    \label{FIG:SBS}
\end{figure}

\subsection{Search-Based Strategy}

The \textit{Search-Based Strategy} (SBS), which comes from \textit{Search Based Software Testing} (SBST)~\cite{McMinn2004,Harman2010}, uses search-based algorithms to achieve the even-spreading of test cases over the input domain.
In contrast to other ART strategies, SBS aims to address the question:
Given a test set $E$, of size $N$ ($|E|=N$), due to limited testing resources, how can $E$ achieve an even spread of test cases over the input domain, thereby enhancing its fault detection ability?
SBS needs to assign a parameter (the number of test cases $N$) before test case generation begins.


\subsubsection{Framework}

Fig.~\ref{FIG:SBS} shows a pseudocode framework for SBS.
Because ART requires that test cases that have some randomness, SBS generates an initial test set population $PT$ (of size $ps$) where each test set (of size $N$) is randomly generated.
A search-based algorithm is then used to iteratively evolve $PT$ into its next generation.
Once a stopping condition is satisfied, the best solution from $PT$ is selected as the final test set.
Two core elements of SBS, therefore, are the choice of search-based algorithm for evolving $PT$, and the evaluation (\textit{fitness}) function for each solution.
Because the fitness function is also involved in the evolution process (to evaluate the $PT$ updates), we do not consider it a separate SBS component.

\subsubsection{Evolution Component}

A number of search-based algorithms have been used to evolve ART test sets, including the following:

1) \textit{Hill Climbing} (HC)~\cite{Schneckenburger2008}:
HC makes use of a single initial test set $T$, rather than a population of test sets $PT$ (i.e., $ps=1$). 
The basic idea behind HC is to calculate the fitness of $T$, and to shake it for as long as the fitness value increases.
One HC fitness function is the minimum distance between any two test cases in $T$, where the distance is a specific Euclidean distance~\cite{Mayer2006a}:
\begin{equation}
    \tag{5.18}
    {fitness}(T)=\min_{{tc}_i\neq {tc}_j \in T}{dist}({tc}_i,{tc}_j).
\end{equation}

2) \textit{Simulated Annealing} (SA)~\cite{Bueno2014}:
Similar to HC, SA also only uses a single test set $T$ ($ps=1$).
During each iteration, SA constructs a new test set $T'$ by randomly selecting input variables from $T$ with a mutation probability and modifying their values.
The fitness values of both $T$ and $T'$ are then calculated.
If the fitness of $T'$ is greater than that of $T$, then $T'$ is accepted as the current solution for the next iteration.
If the fitness of $T'$ is {\em not} greater than that of $T$, then the acceptance of $T'$ is determined by a controlled probability function using random numbers and the temperature parameter adopted in SA. 
Bueno et al.~\cite{Bueno2014} defined the fitness function of $T$ as the sum of distances between each test case and its nearest neighbor:
\begin{equation}\tag{5.19}
    fitness(T)=\sum_{{tc}_i \in T}\min_{{tc}_j\neq {tc}_i \in T}{dist}({tc}_i,{tc}_j).
    \label{EQ:diversity}
\end{equation}

3) \textit{Genetic Algorithm} (GA)~\cite{Bueno2014}:
GA uses a population of test sets rather than just a single one, and three main operations:
\textit{selection}, \textit{crossover}, and \textit{mutation}.
GA first chooses the test sets for the next generation by assigning a selection probability ---
Bueno et al.~\cite{Bueno2014} used a selection probability proportional to the fitness of $T$ (calculated with Eq.~(\ref{EQ:diversity})).
The crossover operation then generates \textit{offspring} by exchanging partial values of test cases between pairs of test sets, and then through mutation by randomly changing some partial values.

4) \textit{Simulated Repulsion} (SR)~\cite{Bueno2007}:
Similar to GA, SR makes use of a population of test sets, $PT$, with each solution $T_i \in PT~(1\leq i \leq ps)$ evolving independently from, and concurrently to, the other test sets.
In each SR iteration, each test case from each solution $T_i$ updates its value based on Newton mechanics with electrostatic force from Coulomb's Law. 
The principle of moving a test case ${tc} \in T_i$ is as follows:
\begin{equation}
    \tag{5.20}
    {tc}_{{{new}}} = {tc} + (\overrightarrow{{{RF}}}({tc})/m),
\end{equation}
where $m$ is a constant (the mass of all test cases), and $\overrightarrow{{{RF}}}({tc})$ is the resultant force of ${tc}$, defined as:
\begin{equation}
    \tag{5.21}
    \overrightarrow{{{RF}}}({tc}) = \sum_{{tc}' \neq {tc} \in T_i}{\frac{Q^2}{{dist}({tc},{tc}')^2}},
\end{equation}
where $Q$ is the current value of electric charge for the test cases.

5) \textit{Local Spreading} (LS)~\cite{Huang2017}:
Similar to HC and SA, LS also only uses a single initial test set $T$.
LS successively moves each point $tc \in T$ that is allowed to move according to the following:
$tc$'s first and second nearest neighbors in $T$, $tc_f$ and $tc_s$, are identified, and the corresponding distances, $d_f$ and $d_s$, are calculated.
A direction of movement is identified related to $tc_f$. 
Then, $tc$ is moved a small distance (related to $d_s-d_f$) in the identified direction, slightly increasing the minimum distance from $tc$ to its nearest neighbor (the distance between $tc$ and $tc_f$).
These steps are repeated until there are no points remaining that can still move.
LS effectively attempts to increase the minimum distance among all test cases in $T$, thereby producing a more evenly-spread test set.

6) \textit{Random Border Centroidal Voronoi Tessellations} (RBCVT)~\cite{Shahbazi2013}:
RBCVT uses an initial test set $T$ of size $N$, 
and makes use of Centroidal Voronoi Tessellations (CVT)~\cite{Du1999} to achieve an even spread of the $N$ test cases over the input domain, $\mathcal{D}$.
It constructs $N$ disjoint cells around the initial $N$ test cases using a Voronoi tessellation with random border point set,
$V_1,V_2,\cdots,V_N$, satisfying  $i\neq j$, $V_i \bigcap V_j =\emptyset$; and
$\bigcup_{i=1}^N V_i=\mathcal{D}$, where $1\leq i,j\leq N$.
Each cell $V_i$ corresponds to a point ${tc} \in T$ such that
\begin{equation}
    \tag{5.22}
    V_i =\{ x \in \mathcal{D}|\forall tc'\neq tc \in T: dist(x, tc) < {dist}(x,tc')\}.
\end{equation}
RBCVT then calculates the centroid of each Voronoi region to obtain $N$ new points for the next generation and evolution.


Since SBS achieves an even spread of the $N$ test cases over the input domain, many studies have used test suites generated by other ART approaches to replace the random test suites, to speed up the evolution process. 
Shahbazi et al.~\cite{Shahbazi2013}, for example, used RBCVT to improve the quality of test suites obtained from STFCS and QRS.
Huang et al.~\cite{Huang2017} have also argued that it would be better to use adaptive random test suites than random test suites as the input for LS.


\subsection{Hybrid Strategies}

\textit{Hybrid Strategies} (HSs) aim at improving the testing effectiveness (such as fault detection capability) or efficiency (such as test generation cost) by combining multiple ART approaches.

\subsubsection{STFCS + PBS}

The STFCS + PBS hybrids aim to enhance the effectiveness of either STFCS or PBS.

From the perspective of STFCS enhancement, Chen et al.~\cite{Chen2007e,Chen2008a},
when generating the $m$-th test case, divided the input domain $\mathcal{D}$ into $m$ disjoint, equally-sized subdomains, $\mathcal{D}_1,\mathcal{D}_2,\cdots,\mathcal{D}_m$,
from the edge to the center of $\mathcal{D}$,
such that:
$\mathcal{D}=\bigcup_{i=1}^{m}\mathcal{D}_i$;
$D_i \bigcap D_j = \emptyset$ for $1\leq i \neq j \leq m$; and
$|\mathcal{D}_1|=|\mathcal{D}_2|=\cdots=|\mathcal{D}_m|$.
Next, for the STFCS random-candidate-construction component, they generated random test cases in those subdomains not already containing test cases.
Mayer~\cite{Mayer2006g} used bisection partitioning to control the STFCS test-case-identification component, only checking the distance from a candidate $c$ to points in its neighboring regions, instead of to all points.
These methods could significantly reduce the STFCS computational overheads, for both FSCS and RRT.
Mao et al.~\cite{Mao2017} proposed a similar method, \textit{distance-aware forgetting}, to reduce the FSCS computational overheads, but they used static, rather than bisection, partitioning.
Chow et al.~\cite{Chow2013} proposed a new efficient and effective method called \textit{ART with divide-and-conquer} that independently applies STFCS to each subdomain (using bisection partitioning).
Previous studies~\cite{Chen2003,Chen2004b,Kuo2009} have combined STFCS with static partitioning, using the concept of \textit{mirroring} to reduce the computational costs.
Enhancements to mirroring have included a revised distance metric~\cite{Nie2009}, and dynamic partitioning with new mirroring functions~\cite{Huang2015}.
Chan et al.~\cite{Chan2004b} applied bisection partitioning to each dimension of the input domain, then checking the amount of executed test cases in each subdomain:
candidates in subdomains with fewer executed test cases were then more likely to be selected.

Regarding the enhancement of PBS, Chen and Huang~\cite{Chen2004c} applied the test-case-identification component to improving the effectiveness of PBS with random partitioning, 
selecting test cases based on the principle of Minimum-Distance and Restriction.
Mayer~\cite{Mayer2005} used a similar mechanism to improve the effectiveness of PBS with bisection partitioning.
Mao and Zhan~\cite{Mao2017a} also used this mechanism to enhance PBS by bisection partitioning, but instead of Euclidean distance, they used the coordinate distance to boundaries (\textit{boundary distance}).
Similarly, Mayer~\cite{Mayer2005a,Mayer2006}, for PBS with random and bisection partitioning, used exclusion regions in a possible subdomain to generate a new test case. 
Mao~\cite{Mao2012}, to overcome the drawbacks of random partitioning, proposed a new partitioning schema, {\em two-point partitioning}, based on the STFCS test-case-identification component: 
When generating a new test case $tc$, it randomly chooses two candidates from the subdomain that needs to be partitioned, and then uses the midpoint of $tc$ and the farthest candidate as the break point to partition the subdomain.

In addition to the hybrid methods listed above, other, new ART techniques have been proposed based on other combinations of different concepts.
Chen et al.~\cite{Chen2007d}, for example, introduced a new test-case-identification criterion (identifying the test case that is more adjacent to the subdomain centroid), and combined it with PBS with bisection partitioning to form a new technique:
\textit{ART by balancing}. 
Mayer~\cite{Mayer2005b} proposed a new approach, \textit{lattice-based ART}, that uses bisection partitioning to divide the input domain for lattice generation.
It then generates test cases by permuting the lattices within a restricted region. 
Chen et al.~\cite{Chen2009a} enhanced Mayer's lattice-based ART~\cite{Mayer2005b} by refining the restricted regions for each lattice.
Sabor and Mohsenzadeh~\cite{Sabor2012,Sabor2012a} proposed an enhanced version of Chen and Huang's method~\cite{Chen2004c} by including an \textit{enlarged input domain}~\cite{Mayer2006e}.

\subsubsection{STFCS + SBS}

The STFCS + SBS hybrids either enhance STFCS, or represent new methods.
Tappenden and Miller~\cite{Tappenden2009} proposed \textit{Evolutionary ART}, a new method that aims to construct an evolved test set for the STFCS test-case-identification.
The method initially generates a fixed-size random test set, according to a uniform distribution.
Until a stopping condition is satisfied (for example, 100 generations have been completed~\cite{Tappenden2009}), each iteration uses an evolutionary algorithm, a \textit{Genetic Algorithm} (GA), to evolve the test set.
The fitness function used during the evolution stage is the same as Eq.~(\ref{EQ:4.1}), and the candidate with the highest fitness value is then selected as the next test case.

Iqbal et al.~\cite{Iqbal2012} combined STFCS with SBS to produce a hybrid that initially uses GA to generate test cases, but if no fitter test cases are found after running a number of iterations, then the algorithm switches to FSCS to generate the following test cases.

\subsubsection{TPBS + PBS or STFCS}

The TPBS + PBS or STFCS hybrids aim to augment the TPBS test profiles using PBS or STFCS principles.
Liu et al.~\cite{Liu2011} proposed three methods to design test profiles,
based on STFCS restriction (\textit{exclusion}),
and on PBS subdomain-selection criteria (\textit{maximum size}~\cite{Chen2004d} and \textit{least number of previously generated test cases}~\cite{Chen2004d,Chow2013}).
Using \textit{exclusion}, all points inside the exclusion regions should have no chance of being selected as test cases:
their probability of selection is $0$.
Using the \textit{maximum size}~\cite{Chen2004d}, all points inside the largest subdomain have a probability to be chosen as test cases, and all other points (those in  other subdomains) have no chance.
Similarly, when using the \textit{least number of previously generated test cases}~\cite{Chen2004d,Chow2013}, all points within the subdomains with the least number of previously generated test cases have a chance to be selected, and all others have no chance.

\subsection{Strengths and Weaknesses}

Previous studies~\cite{Anand2013} have confirmed that ART is more effective than RT in general, according to several different evaluations.
As discussed by Chen et al.~\cite{Chen2007c}, however, both favorable and unfavorable conditions exist for ART.
In this section, therefore, we summarize the strengths and weaknesses of ART.

\subsubsection{Strengths}

ART outperforms RT from the following perspectives:

1) \textit{Test case distribution}:
ART generally delivers a more even distribution of test cases across the input domain than RT.
All ART approaches deliver better test case dispersion~\cite{Chen2007} than RT~\cite{Chen2009,Liu2010,Liu2011,Liu2009a,Chen2007f,Kuo2008,Chen2007,Liu2016}.
When the input domain dimensionality $d$ is low ($d$ is equal to $1$ or $2$),
all ART approaches generate test cases with a more even spread over the input domain than RT, in terms of test case discrepancy~\cite{Chen2007}.
However, as the dimensionality increases, some ART approaches have worse performance than RT, including FSCS~\cite{Chen2009,Chen2007f,Kuo2008,Chen2007}, RRT~\cite{Chen2007,Liu2011}, and TPBS~\cite{Liu2010}.
Nevertheless, even when the dimensionality is $3$ or $4$, a number of ART approaches do still have better discrepancy than RT, including PBS~\cite{Chen2007,Liu2011} and QRS~\cite{Liu2016}.

2) \textit{Fault detection capability}:
It is natural that ART should have better fault detection ability than RT when the failure region is clustered ---
ART was  specifically designed to make use of this clustering information.
Chen et al.~\cite{Chen2007c} investigated the factors impacting on ART fault detection ability, identifying a number of favorable conditions for ART, including:
(a) when the failure rate is small;
(b) when the failure region is compact;
(c) when the number of failure regions is low; and
(d) when a predominant region exists among all the failure regions.
When any of these conditions are satisfied, ART generally has better fault detection performance than RT. 
Even when none of the conditions are satisfied, ART can achieve comparable fault detection to RT.


3) \textit{Code coverage}:
Studies have shown that ART achieves greater \textit{structure-based code coverage} than RT for the same number of test cases~\cite{Bueno2014,Bueno2007,Chen2008,Chen2013}.
Bueno et al.~\cite{Bueno2014,Bueno2007} have observed that SBS outperforms RT for \textit{data-flow coverage}~\cite{Laski1983} (including \textit{all-uses coverage} and \textit{all-du-paths coverage}).
Chen et al.~\cite{Chen2008,Chen2013} have reported that FSCS is more effective than RT for both \textit{control-flow coverage}~\cite{Huang1975} (including \textit{block coverage} and \textit{decision coverage}), and data-flow coverage (\textit{c-uses coverage} and \textit{p-uses coverage}~\cite{Laski1983}).

4) \textit{Reliability estimation and assessment}:
\label{Subsection:ReliabilityEstimation}
For the same number of test cases, ART has greater code coverage than RT~\cite{Bueno2014,Bueno2007,Chen2008,Chen2013}.
It has also been observed that coverage can be used to improve the effectiveness of software reliability estimation~\cite{Chen2001a}.
Compared with RT, therefore, ART should enable better software reliability estimation, and higher confidence in the reliability of the SUT, even when no failure is detected.
Unfortunately, this characteristic (strength) of ART was obtained from the perspective of theoretical results rather than empirical studies, which means that no ART studies have yet investigated the reliability estimation and assessment.

5) \textit{Cost-effectiveness}:
The cost-effectiveness of testing considers both effectiveness (e.g., fault detection) and efficiency (including test case generation and execution time).
ART cost-effectiveness has often been examined using the \textit{F-time}~\cite{Chen2006a}, which is the amount of computer execution time required to detect the first failure (including the time for both generation and execution of test cases).
Because ART involves additional computation to evenly spread the test cases over the input domain~\cite{Towey2007,Anand2013},  ART should naturally take more time than RT to generate the same number of test cases, suggesting that it may have worse cost-effectiveness than RT.
However, studies~\cite{Huang2015,Lin2009,Selay2014,Liu2010a,Anand2013} have shown that compared with RT, ART typically requires less time to identify the first failure (\textit{F-time}) ---
therefore, ART can be more cost-effective than RT.
In general, three main conditions can result in ART achieving a better cost-effectiveness than RT:
(a) ART using fewer test cases than RT to detect the first failure (\textit{F-measure});
(b) the computational overhead of the ART approach being acceptable (comparable or slightly higher than that of RT); or
(c) the combined program execution and test setup time being more than the time required by ART to generate a test case.

\subsubsection{Weaknesses}

There are three main challenges associated with some ART approaches:
\textit{boundary effect}, \textit{computational overheads}, and \textit{high dimension problem}.

1) \textit{Boundary effect}~\cite{Chen2007a}:
Some ART approaches tend to generate more test cases near the boundary than near the input domain center, a situation known as the \textit{boundary effect}.
One reason for the boundary effect, as explained in the context of RRT \cite{Chan2002,Chan2003a,Chan2004,Chan2006b}, is that  test cases cannot be generated outside the boundary, thus reducing the number of sources of restriction from close to boundary regions.
Both FSCS and RRT have been shown to suffer from the boundary effect, especially when the failure rate and dimensionality are high~\cite{Chen2007a}.

A number of attempts have been made to address the boundary effect.
Some studies increased the selection probability of candidates from the input domain center over those from the boundary~\cite{Chen2009b,Kuo2007,Chen2007e,Chen2008a,Kuo2007a,Kuo2008}.
Other studies have removed the boundaries themselves, either by joining boundaries~\cite{Geng2010,Mayer2006a,Chen2011}, or by extending the input domain beyond the original boundaries~\cite{Mayer2006e,Chen2007a,Mayer2006b}.
Chen et al.~\cite{Chen2007d} preferred to choose test cases close to the input domain centroid.
Mayer~\cite{Mayer2006c} initially selected test cases within a small region around the center of the input domain, extending the region if no failures were identified.

2) \textit{Computational overheads}~\cite{Towey2007,Anand2013}:
ART approaches typically incur heavier computational costs than RT to generate test cases.
This is particularly the case for some ART approaches, such as FSCS, and RRT.
When testing, the actual test execution time can be an important factor that may mitigate the computational overheads:
When the test execution time is very long, for example, the ART computational costs may be more acceptable.
However, because the test execution time depends mainly on  characteristics of the SUT, and not on the test generation, we do not discuss it here.

The time complexity of FSCS and RRT are of the order of $O(n^2)$ and $n^2\log n$, respectively (where $n$ is the number of generated test cases)~\cite{Chen2004a,Mayer2006f}.
The PBS and QRS techniques of ART are significantly more efficient than others (such as STFCS and SBS).
Many hybrid approaches have been developed to reduce the overheads incurred by FSCS and RRT.
Other techniques for ART overhead reduction have also been explored.
Chan et al.~\cite{Chan2005,Chan2006b}, for example, used a square exclusion region version of RRT to reduce the distance calculation overheads (though not the algorithm's complexity, $O(n^2)$).
Chen and Merkel~\cite{Chen2006d} applied Voronoi diagrams~\cite{Aurenhammer1991} to reduce the FSCS distance calculations, lowering the complexity from $O(n^2)$ to $O(n^{\frac{4}{3}})$.
{\em Mirroring}~\cite{Chen2003,Chen2004b,Kuo2009,Huang2015} has been used to directly generate \textit{mirror test cases} of a \textit{source test case} based on a principle of symmetry of subdomains.
Although the time complexity for FSCS mirroring can still be $O(\frac{n^2}{m^2})$~\cite{Chen2003,Chen2004b,Kuo2009} (where $m$ is the number of subdomains),
an enhanced mirroring technique can reduce this to $O(n)$~\cite{Huang2015}. 
Shahbazi et al.~\cite{Shahbazi2013} have also proposed a linear-order ($O(n)$) ART approach:
a fast search algorithm for RBCVT (RBCVT-Fast).

While the overhead-reduction approaches listed above can only be applied to numeric input domains, \textit{forgetting}~\cite{Chan2006a}, which reduces overheads by omitting some previous test cases from calculations, has no such limitation.
Based on the \textit{Category-Partition Method}~\cite{Ostrand1988}, Barus et al.~\cite{Barus2016} have also recently introduced a linear-order FSCS algorithm using the test-case-identification with \textit{Average-Distance} for nonnumeric inputs.

3) \textit{High dimension problem}~\cite{Chen2005,Chen2007}:
It has been observed that the effectiveness of some ART approaches may decrease when the number of dimensions of the input domain increases, due to the \textit{curse of dimensionality}~\cite{Bellman1957}.
Because the center of a high dimensional input domain has a higher probability of being a failure region than the boundary~\cite{Mayer2006c,Chen2007d}, it has been suggested that the boundary effect may impact (or even cause) the high dimension problem. 
Approaches for addressing the boundary effect~\cite{Huang2013,Chen2009b,Kuo2007,Chen2007e,Chen2008a,Kuo2007a,Kuo2008,Geng2010,Mayer2006a,Chen2011,Mayer2006e,Chen2007a,Mayer2006b,Chen2007d,Mayer2006c}, therefore, may also help to alleviate the high dimension problem.
However, because their effectiveness is not constant across dimensions, although they may alleviate, current approaches do not {\em solve} the dimensionality problem~\cite{Schneckenburger2008}.
Furthermore, finding a solution with consistent effectiveness across all dimensions seems unlikely~\cite{Schneckenburger2008}, so some decrease in effectiveness in higher dimensions may need to be tolerated.
Nevertheless, Schneckenburger and Schweiggert~\cite{Schneckenburger2008} have combined \textit{Hill Climbing} with \textit{continuous distance}~\cite{Mayer2006f} to produce a search-based ART approach that (slightly) reduces the dependency on dimensionality.

\vspace{0.2cm}\noindent \fbox{\parbox[c]{0.97\linewidth} {
\textit{\textbf{Summary of answers to RQ2:}}
\begin{itemize}
    \item[1)] \textit{Based on different concepts for the even-spreading of test cases, the various ART approaches can be classified into the following six categories:
        \textit{Select-Test-From-Candidates Strategy} (STFCS);
        \textit{Partitioning-Based Strategy} (PBS);
        \textit{Test-Profile-Based Strategy} (TPBS);
         \textit{Quasi-Random Strategy} (QRS);
          \textit{Search-Based Strategy} (SBS); and \textit{Hybrid Strategies} (HSs).}
    \item[2)] \textit{For each of the first five categories, a framework has been presented showing the basic steps involved.}
    \item[3)] \textit{Compared with RT, ART generally performs better when certain conditions hold (identified as ``favorable conditions'' for ART), according to test case distribution, fault detection capability, code coverage, reliability estimation and assessment, and cost-effectiveness.}
    \item[4)] \textit{ART suffers from three main weaknesses: boundary effect, computational overheads, and the high dimension problem.}
\end{itemize}
}}

\section{Answer to RQ3: In what domains and applications has ART been applied?\label{SEC:Applications}}

\begin{figure*}[!t]
    \centering
    \graphicspath{{Graphs/summary/}}
    \includegraphics[width=0.95\textwidth]{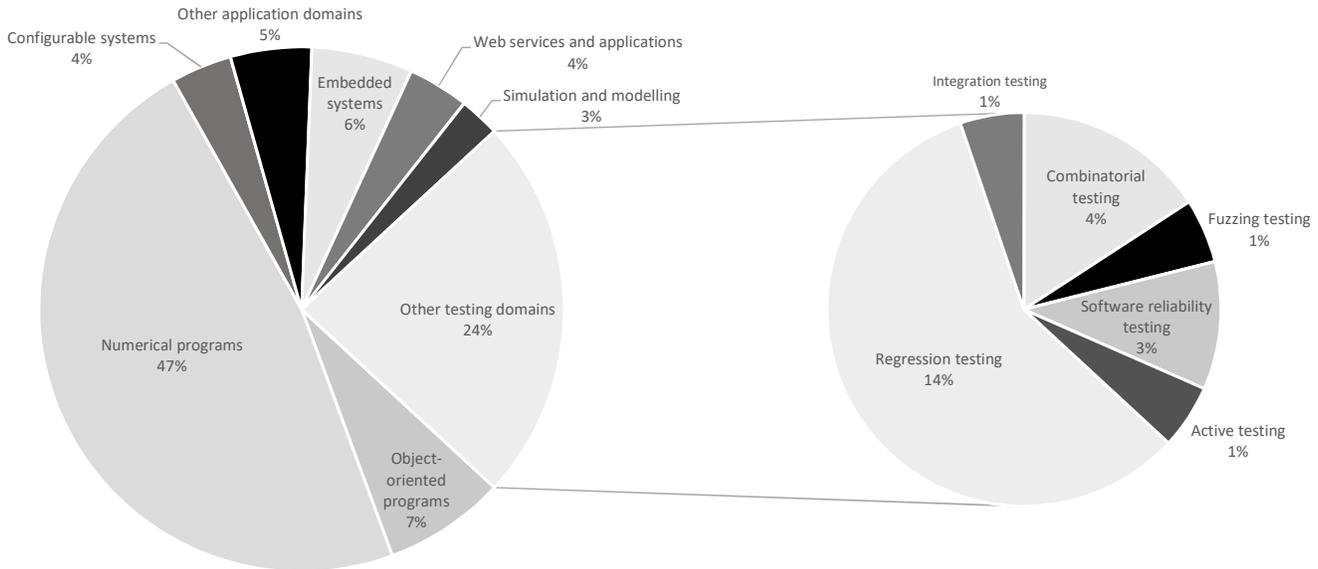}
    \caption{Application distribution of ART.}
    \label{FIG:applicationDistribution}
\end{figure*}

Of the 140 papers in our survey, 80 involved application of ART to specific testing problems.
Fig.~\ref{FIG:applicationDistribution} presents the distribution of the ART applications, showing that 76\% of studies focused on various testing environments (or programs), and 24\% involved application of ART to other testing techniques.

\subsection{Application Domains}

Based on the classification  shown in Fig.~\ref{FIG:applicationDistribution}, numeric programs (47\%) have been the most popular domains for ART application, followed by object-oriented programs (7\%), embedded systems (6\%), web services and applications (4\%), configurable systems (4\%), and simulations and modelling (3\%).
5\% of the papers  revealed other application domains, including mobile applications and aspect-oriented programs. 

\subsubsection{Numeric Programs}

Chen et al.~\cite{Chen2004a} applied ART to the testing of twelve open-source numeric analysis programs from Numerical Recipes~\cite{Press1986} and ACM Collected Algorithms~\cite{ACM1980}, written in C or C++. 
These programs have also been widely used in other ART studies, and, in addition to C and C++, have also been implemented in Java. 
Zhou et al.~\cite{Zhou2010a} applied ART to three other numeric programs from the Numerical Recipes~\cite{Press1986}.
Arcuri and Briand~\cite{Arcuri2011} conducted experiments on a further nine numeric programs for basic algorithms~\cite{Cormen2001} and for mathematical routines from the Numerical Recipes~\cite{Press1986}. 
Chen et al.~\cite{Chen2008} used ART with two numeric programs from the GNU Scientific Library~\cite{GNU}, and one from the Software-artifact Infrastructure Repository (SIR)~\cite{Do2005}.
Walkinshaw and Fraser~\cite{Walkinshaw2017} investigated six units within the Apache Commons Math framework~\cite{Apacheframework} and two units within JodaTime~\cite{JodaTime}. 

\subsubsection{Object-Oriented Programs}

Ciupa et al.~\cite{Ciupa2006,Ciupa2008} defined a new similarity metric for object-oriented (OO) inputs, and integrated it into ART (ARTOO).
They compared ARTOO with RT using eight real-life, faulty OO programs from the EiffelBase Library~\cite{EiffelBase}.
Lin et al.~\cite{Lin2009} used their ART approach on six OO programs containing manually-seeded faults ---
five of these programs were from the Apache Common library~\cite{ApacheCommons}, and one was a wide-area event notification system, Siena. 
Chen et al.~\cite{Chen2017} also proposed a new similarity metric for OO inputs, using it to apply ART to 17 OO programs (five C++ libraries and 12 C\# programs) from the following open-source repositories:
Codeforge~\cite{Codeforge}, Sourceforge~\cite{Sourceforge}, Codeplex~\cite{Codeplex}, Codeproject~\cite{Codeproject}, and Github~\cite{Github}.
Putra and Mursanto~\cite{Putra2013} compared two ART techniques applied to eight OO programs, written in Java, from the Apache Common library~\cite{ApacheCommons}.
Jaygarl et al.~\cite{Jaygarl2009} evaluated different RT and ART techniques applied to four open-source OO programs (written in Java). 

\subsubsection{Configurable Systems}

Chen et al.~\cite{Chen2013} compared the code coverage achieved by ART and RT using ten UNIX utility programs that can be considered configurable systems ---
they are influenced by different configurations or factors, obtained using the \textit{Category-Partition Method} (CPM)~\cite{Ostrand1988}.
Huang et al.~\cite{Huang2014} applied ART to \textit{combinatorial input domains} (\textit{configurable input domains}), testing five small C programs \cite{Chris}, and four versions of another configurable system, Flex (a fast lexical analyzer), from the SIR~\cite{Do2005}.
They also identified the configurable input domains using CPM.
Barus et al.~\cite{Barus2016} proposed an efficient ART approach, and  applied it to 14 configurable systems and programs ---
seven Siemens programs from the SIR~\cite{Do2005};
six UNIX utilities; and
one regular expression processor from the GNU Scientific Library~\cite{GNU}.

\subsubsection{Web Services and Applications}

Tappenden and Miller~\cite{Tappenden2014} applied an evolutionary ART algorithm to cookie collection testing, applying it to six open-source web applications written in C\# and PHP.
Selay et al.~\cite{Selay2014} used ART in image comparisons when testing a set of real-world, industrial web applications.
Chen et al.~\cite{Chen2014} developed a system to test web service vulnerabilities that generates ART test cases based on \textit{Simple Object Access Protocol} (SOAP) messages:
twenty web services (both open-source and specifically written services) were examined in their study.

\subsubsection{Embedded Systems}

Hemmati et al.~\cite{Hemmati2010,Hemmati2011,Hemmati2013} applied ART to two industrial embedded systems:
a core subsystem of a video-conference system, implemented in C; and
a safety monitoring component of a safety-critical control system written in C++.
Arcuri et al.~\cite{Arcuri2010,Iqbal2012} compared RT, ART, and search-based testing using a real-life real-time embedded system.
This very large and complex seismic acquisition system, implemented in Java, interacts with many sensors and actuators.

\subsubsection{Simulations and Models}

Matinnejad et al.~\cite{Matinnejad2015} applied an ART approach to test three Stateflow models of mixed discrete-continuous controllers:
two industrial models from Delphi~\cite{Delphi}, Supercharger Clutch Controller (SCC) and Auto Start-Stop Control (ASS);
and one public domain model, Guidance Control System (GCS), from Mathworks~\cite{Stateflow}.
Sun et al.~\cite{Sun2012} proposed an enhanced ART approach for testing \textit{Architecture Analyze and Design Language} (AADL) models~\cite{AADL}, reporting on a case study applying it to an Unmanned Aerial Vehicle (UAV) cruise control system, which includes three sensor devices (radar, GPS, and speed devices), and two subsystems (read data calculation, and flight control systems).

\subsubsection{Other Domains}

Parizi and Ghani~\cite{Parizi2011} conducted a preliminary study of ART for aspect-oriented programs, identifying some potential research directions.
Shahbazi and Miller~\cite{Shahbazi2016} investigated the application of ART to programs with string inputs, comparing the performance of different ART approaches on 19 open-source programs.
Liu et al.~\cite{Liu2010a} used ART to test mobile applications, providing a new distance metric for event sequences.
Their study examined six real-life mobile applications implemented in Java.
Koo and Park~\cite{Koo2016} investigated ART for SDN (\textit{Software-Defined Networking}) OpenFlow switches, aiming to generate test packages for switches. 

\begin{table*}[!t]
\scriptsize
\centering
\caption{Summary of Main ART Results for each Application Domain\label{TAB:applicationResults}}
\begin{tabular}{l|l|l|l|r|r} \Xhline{1.0pt}
\multirow{2}*{\textbf{Application Domain}} &\multirow{2}*{\textbf{Original study}} &\multirow{2}*{\textbf{Method}} &\multirow{2}*{\textbf{Program Name}} &\textbf{\%Effectiveness} &\textbf{\%Efficiency} \\
&&&&\textbf{Improvement}&\textbf{Improvement}\\\Xhline{0.6pt}

\multirow{26}*{Numeric Programs}
&\multirow{12}*{Chen et al.~\cite{Chen2004a}} &\multirow{12}*{FSCS}	&Airy  &42.14\% &\multirow{26}*{NR$^\dag$}\\\cline{4-5}
&	&	&Bessj	&41.83\%	 &\\\cline{4-5}
&	&	&Bessj0	&42.24\%	&\\\cline{4-5}
&	&	&Cel	&47.55\%	&\\\cline{4-5}
&	&	&El2	&52.02\%	&\\\cline{4-5}
&	&	&Erfcc	&44.30\%	 &\\\cline{4-5}
&	&	&Gammq	&11.38\%	 &\\\cline{4-5}
&	&	&Golden	&1.67\%	 &\\\cline{4-5}
&	&	&Plgndr	&34.09\%	&\\\cline{4-5}
&	&	&Probks	&45.25\%	&\\\cline{4-5}
&	&	&Sncndn	&1.18\%	 &\\\cline{4-5}
&	&	&Tanh	&45.00\%	&\\

\cline{2-5}

&\multirow{3}*{Zhou et al.~\cite{Zhou2010a}}    &\multirow{3}*{MCMC-Random Testing} &Bessel  &93.02\%    &\\\cline{4-5}
&   &  &Ellint   &28.28\%   &\\\cline{4-5}
&   &  &Laguerre    &90.67\%   &\\

\cline{2-5}

  &\multirow{3}*{Chen et al.~\cite{Chen2008}}    &\multirow{3}*{FSCS/ECP-FSCS*} &Cubic   &0.42\% -- 3.48\%    &\\\cline{4-5}
  &   &   &Quadratic   &0.20\% -- 4.95\%   &\\\cline{4-5}
  &   &    &Tcas     &-1.99\% -- 4.72\%   &\\

\cline{2-5}

&\multirow{8}*{Walkinshaw and Fraser~\cite{Walkinshaw2017}}    &\multirow{8}*{FSCS} &BesselJ   &0.02\%    &\\\cline{4-5}
&   &   &Binomial   &0.59\%   &\\\cline{4-5}
&   &   &DaysBetween   &-0.94\%   &\\\cline{4-5}
&   &   &DerivativeSin   &-2.21\%   &\\\cline{4-5}
&   &   &Erf   &-0.62\%   &\\\cline{4-5}
&   &   &Gamma   &-0.63\%   &\\\cline{4-5}
&   &   &PeriodToWeeks   &8.84\%   &\\\cline{4-5}
&   &   &Romberg Integrator   &-0.19\%   &\\

\hline

\multirow{35}*{Object-Oriented Programs}
&\multirow{8}*{Ciupa et al.~\cite{Ciupa2008}} &\multirow{8}*{FSCS}	&Action\_sequence  &91.49\% &-38.29\%\\\cline{4-6}
&   &   &Array  &48.95\%    &-527.46\%\\\cline{4-6}
&   &   &Arrayedlist  &93.77\%    &-7.71\%\\\cline{4-6}
&   &   &Boundedstack  &20.91\%    &-1017.90\%\\\cline{4-6}
&   &   &Fixedtree  &28.01\%    &6.52\%\\\cline{4-6}
&   &   &Hashtable  &78.74\%    &-149.67\%\\\cline{4-6}
&   &   &Linkedlist  &61.38\%    &-442.39\%\\\cline{4-6}
&   &   &String  &82.09\%    &41.07\%\\
\cline{2-6}

&\multirow{6}*{Lin et al.~\cite{Lin2009}} &\multirow{6}*{FSCS}	&Math.geometry &87.36\% &86.31\%\\\cline{4-6}
&   &   &Math.util  &85.48\%    &99.05\%\\\cline{4-6}
&   &   &Lang  &20.96\%    &64.87\%\\\cline{4-6}
&   &   &Lang.text  &91.05\%    &89.78\%\\\cline{4-6}
&   &   &Collections.list  &86.85\%    &87.05\%\\\cline{4-6}
&   &   &Siena  &92.13\%    &82.80\%\\
\cline{2-6}

&\multirow{17}*{Chen et al.~\cite{Chen2017}}    &\multirow{17}*{FSCS + Forgetting}
            &CCoinBox	&2.46\% -- 63.83\%	&-725.00\% -- -144.44\%\\\cline{4-6}
	&	&	&Calendar	&0.00\% -- 76.41\%	&-210.53\% -- -13.46\%\\\cline{4-6}
	&	&	&Stack	&0.00\% -- 56.02\%	&-230.86\% -- -63.41\%\\\cline{4-6}
	&	&	&Queue	&0.00\% -- 55.21\%	&-118.67\% -- -5.81\%\\\cline{4-6}
	&	&	&WindShieldWiper	&5.11\% -- 67.36\%	&-457.89\% -- -70.97\%\\\cline{4-6}
	&	&	&SATM	&4.99\% -- 62.28\%	&-325.81\% -- -94.12\%\\\cline{4-6}
	&	&	&BinarySearchTree	&10.94\% -- 77.93\%	&-428.24\% -- -93.53\%\\\cline{4-6}
	&	&	&RabbitAndFox	&0.56\% -- 76.30\%	&-354.41\% -- -83.93\%\\\cline{4-6}
	&	&	&WaveletLibrary	&2.88\% -- 72.09\%	&-153.24\% -- -3.53\%\\\cline{4-6}
	&	&	&BackTrack	&3.09\% -- 76.87\%	&-128.57\% -- -1.05\%\\\cline{4-6}
	&	&	&NSort	&3.61\% -- 60.20\%	&-593.75\% -- -126.53\%\\\cline{4-6}
	&	&	&SchoolManagement	&3.14\% -- 79.69\%	&-510.53\% -- -109.64\%\\\cline{4-6}
	&	&	&EnterpriseManagement	&0.00\% -- 75.50\%	&-233.64\% -- -80.30\%\\\cline{4-6}
	&	&	&ID3Manage	&14.66\% -- 70.58\%	&-690.00\% -- -92.68\%\\\cline{4-6}
	&	&	&IceChat	&13.00\% -- 122.96\%	&-463.93\% -- -26.47\%\\\cline{4-6}
	&	&	&CSPspEmu	&8.88\% -- 92.79\%	&-169.94\% -- -25.36\%\\\cline{4-6}
	&	&	&Poco-1.4.4: Foundation	&6.72\% -- 86.12\%	&-144.67\% -- -20.80\%\\

\cline{2-6}

&\multirow{4}*{Jaygarl et al.~\cite{Jaygarl2009}}    &\multirow{4}*{FSCS}
        &Apache Ant &58.16\%	&65.87\%\\\cline{4-6}
&   &   &ASM    &85.99\%	&92.14\%\\\cline{4-6}
&   &   &ISSTA Containers   &41.26\%	&43.82\%\\\cline{4-6}
&   &   &Java Collections   &89.07\%	&98.76\%\\

\hline

\multirow{20}*{Configurable Systems} &\multirow{6}*{Huang et al.~\cite{Huang2014}}   &\multirow{6}*{FSCS}
        &Count  &11.16\% -- 21.41\% &\multirow{20}*{NR}\\\cline{4-5}
&   &   &Series   &9.93\% -- 21.28\%   &\\\cline{4-5}
&   &   &Tokens   &9.69\% -- 20.14\%  &\\\cline{4-5}
&   &   &Ntree   &11.19\% -- 21.73\%   &\\\cline{4-5}
&   &   &Nametbl   &9.03 -- 20.91\%  &\\\cline{4-5}
&   &   &Flex   &4.36\% -- 22.80\%   &\\\cline{4-5}
\cline{2-5}

&\multirow{14}*{Barus et al.~\cite{Barus2016}}   &
        &Cal    &31.91\% -- 87.05\% &\\\cline{4-5}
&   &   &Comm   &56.28\% -- 88.25\%   &\\\cline{4-5}
&   &   &Grep   &-114.76\% -- 72.60\%   &\\\cline{4-5}
&   &   &Look   &-133.58\% -- 58.40\%   &\\\cline{4-5}
&   &   &Printtokens   &37.61\% -- 64.59\%   &\\\cline{4-5}
&   &   &Printtokens2   &-17.69\% -- 65.36\%   &\\\cline{4-5}
&   &FSCS   &Replace   &-309.64\% -- 69.68\%  &\\\cline{4-5}
&   &FSCS + Mirroring   &Schedule   &-403.38\% -- 87.84\%   &\\\cline{4-5}
&   &   &Schedule2   &-2365.22\% -- 80.29\%   &\\\cline{4-5}
&   &   &Sort   &-52.01\% -- 84.21\%  &\\\cline{4-5}
&   &   &Spline   &-207.42\% -- 79.68\%   &\\\cline{4-5}
&   &   &TCAS   &-129.07\% -- 65.74\%  &\\\cline{4-5}
&   &   &Totinfo   &-264.70\% -- 62.64\%   &\\\cline{4-5}
&   &   &Uniq   &-30.14\% -- 86.31\%   &\\
\hline


\Xhline{1.0pt}
\end{tabular}
\end{table*}

\begin{table*}[!t]
\scriptsize
\centering
\begin{tabular}{l|l|l|l|r|r} \Xhline{1.0pt}
\multirow{2}*{\textbf{Application Domain}} &\multirow{2}*{\textbf{Original study}} &\multirow{2}*{\textbf{Method}} &\multirow{2}*{\textbf{Program Name}} &\textbf{\%Effectiveness} &\textbf{\%Efficiency} \\
&&&&\textbf{Improvement}&\textbf{Improvement}\\\Xhline{0.6pt}
\multirow{2}*{Web Services and Applications}   &\multirow{2}*{Selay et al.~\cite{Selay2014}} &FSCS + Forgetting	&\multirow{2}*{RWWA1-7}  &\multirow{2}*{-2.22\% -- 16.16\%}  &\multirow{2}*{-1.82\% -- 3.63\%}\\
& &FSCS + Mirroring   &   &   &\\\hline

Embedded Systems    &Iqbal et al.~\cite{Iqbal2012} &FSCS	&IC  &3.00\% &\multirow{15}*{NR}\\\cline{1-5}

Simulation and Modelling &Sun et al.~\cite{Sun2012}  &FSCS &UAV    &5.00\% -- 43.00\%  &\\\cline{1-5}

&\multirow{13}*{Shahbazi and Miller~\cite{Shahbazi2016}}  &   &Validation &4.50\% -- 94.60\%  &\\\cline{4-5}
&   &   	&PostCode	&22.10\% -- 137.00\% 	&\\\cline{4-5}	
&   &   	&Numeric	&43.20\% -- 458.00\% 	&\\\cline{4-5}	
&   &   	&DateFormat	&43.40\% -- 462.10\% 	&\\\cline{4-5}	
&   &   	&MIMEType	&-19.30\% -- 12.00\% 	&\\\cline{4-5}	
&   &FSCS   	&ResourceURL	&-17.20\% -- 15.60\% 	&\\\cline{4-5}	
Other Domains (String Programs)   &   &GA   	&URI	&-22.50\% -- -3.30\% 	&\\\cline{4-5}	
   &   &Multi-Objective GA   	&URN	&-47.60\% -- 15.60\% 	&\\\cline{4-5}	
&   &   	&TimeChecker	&-27.40\% -- -0.60\% 	&\\\cline{4-5}	
&   &   	&Clocale	&38.30\% -- 321.00\% 	&\\\cline{4-5}	
&   &   	&Isbn	&-28.20\% -- 65.60\% 	&\\\cline{4-5}	
&   &   	&BIC	&4.40\% -- 103.80\% 	&\\\cline{4-5}	
&   &   	&IBAN	&23.80\% -- 90.40\% 	&\\
\hline

Other Domains (SDN OpenFlow Switches)    &Koo and Park~\cite{Koo2016}  &FSCS   &Open vSwitch &66.67\%  &62.55\%\\

\Xhline{1.0pt}
\multicolumn{6}{l}{*ECP-FSCS is \textit{Edge-Center-Partitioning based FSCS}, which combines FSCS with static partitioning~\cite{Chen2008}.}\\
\multicolumn{6}{l}{$^\dag$``NR" indicates that some details were not reported in the original paper.}
\end{tabular}
\end{table*}

\subsubsection{Summary of Main Results for Application Domains}

Table~\ref{TAB:applicationResults} summarizes the main results for some of the original studies involving application of ART to different domains (``NR" indicates that some details were not reported in the original paper). 
Some of the studies did not explicitly provide results of the comparison between RT and ART, and so have been omitted from the table.
Information about the programs tested in each application domain is available in Table~\ref{TAB:A1}, in the appendix.

In Table~\ref{TAB:applicationResults}, the columns ``\%Effectiveness Improvement" and ``\%Efficiency Improvement" show the percentage improvements of testing effectiveness and efficiency of ART over RT, respectively. 
Given an evaluation metric $M$ (for example, to measure testing effectiveness or efficiency), if $M_a$ represents the value for ART, and $M_r$ the value for RT, then the percentage improvement (of ART over RT) can be calculated as:
\begin{itemize}
    \item $100 * (M_r - M_a)/M_r$, if lower values mean improvements; and
    \item $100 * (M_a - M_r)/M_r$, if greater values mean improvement.
\end{itemize}

Because of the additional computations involved in ART, it is intuitive that it should take longer than RT to generate the same number of test cases.
The ``Efficiency" data here, therefore, refers to the time taken to achieve the stopping criterion ---
for example, the time taken to detect the first failure (the \emph{F-time}~\cite{Chen2006a}), which includes the combined generation and execution time of all test cases executed before causing the first failure).
The ``Efficiency" can be considered the \textit{cost-effectiveness} metric.

Based on Table~\ref{TAB:applicationResults}, we have the following observations:
\begin{itemize}
    \item Across all application domains, ART usually achieves better testing effectiveness than RT, especially for numeric and object-oriented programs.
    \item For a particular application domain, due to the different characteristics of the various programs, ART may perform differently with different programs.
    \item Many of the original studies surveyed (including the numeric programs and configurable systems) only provided the information for ``\%Effectiveness Improvement",  not for ``\%Efficiency Improvement".
    \item Some studies found ART to be more cost-effective than RT (e.g., \cite{Lin2009,Koo2016,Jaygarl2009}), with some observing ART to require less time than RT to identify the first failure (e.g., \cite{Ciupa2008,Chen2017}). 
\end{itemize}

\subsection{Other Testing Applications}

A number of the surveyed studies considered ART as a strategy to support another testing method, with a goal of enhancing the effectiveness and applicability of that target method.
Fig.~\ref{FIG:applicationDistribution} shows that the most popular target testing application is regression testing~\cite{Yoo2012} (14\%), followed by combinatorial testing~\cite{Nie2011} (4\%), software reliability testing~\cite{Mus1996} (3\%), fuzzing~\cite{Purdom1972} (1\%), integration testing~\cite{Haley1984} (1\%), and active testing~\cite{Sen2008} (1\%). 

\subsubsection{Regression Testing}

\textit{Adaptive random sequences}~\cite{Chen2010} have been extensively applied to regression testing, including for both prioritization and selection~\cite{Yoo2012}.
Jiang et al.~\cite{Jiang2009} first used adaptive random sequences (obtained with FSCS~\cite{Chen2001}) to prioritize regression test suites, proposing a family of approaches that use code coverage to measure the distance between test cases. 
Other studies~\cite{Zhou2012a,Zhou2012,Sinaga2017} have also used code coverage to guide test case prioritization, but with different distance measures or different ART approaches.
Jiang \& Chan~\cite{Jiang2013,Jiang2015} proposed a family of novel input-based randomized local beam search techniques to prioritize test cases.
Chen et al.~\cite{Chen2016a} proposed a method using two clustering algorithms to construct adaptive random sequences for object-oriented software.
Zhang et al.~\cite{Zhang2014}, based on work on string test case prioritization~\cite{Ledru2012}, introduced a method to construct adaptive random sequences using string distance metrics.
They later used a different distance metric, based on CPM~\cite{Ostrand1988}, to propose another method for prioritizing test cases~\cite{Zhang2016}.
Huang et al.~\cite{Huang2014a} applied adaptive random prioritization to interaction test suites for combinatorial testing~\cite{Nie2011}.
Zhou~\cite{Zhou2010b} used the same code coverage distance information as Zhou et al.~\cite{Zhou2010a} to support regression test case selection.
Chen et al.~\cite{Chen2016} extracted tokens reflecting fault-relevant characteristics of the SUT (such as statement characteristics, type and modifier characteristics, and operator characteristics), and used these tokens to represent test cases as vectors for prioritizing programs for C compilers.

Previous investigations have shown that although adaptive random sequences generally incur higher computational costs than random sequences, they are usually significantly more effective in terms of fault detection. 
Furthermore, adaptive random sequences are also sometimes comparable to test sequences obtained by traditional regression testing techniques~\cite{Rothermel2001}, in terms of both testing effectiveness and efficiency. 

\subsubsection{Combinatorial Testing}

Huang et al.~\cite{Huang2012} used two popular ART techniques (FSCS and RRT) to construct an effective test suite (a \textit{covering array}~\cite{Cohen2003}) for combinatorial testing~\cite{Nie2011}.
Nie et al.~\cite{Nie2015} investigated covering arrays constructed by RT, ART, and combinatorial testing, in terms of their ability to identify interaction-triggered failures.

Overall, ART requires far fewer combinatorial test cases to construct covering arrays than RT, and can also detect more interaction-triggered failures for the same number of test cases~\cite{Huang2012}. 
ART also performs comparably to traditional combinatorial testing, especially for identifying interaction failures caused by a large number of influencing parameters~\cite{Nie2015}. 

\subsubsection{Reliability Testing}

Liu and Zhu~\cite{Liu2008} used mutation analysis~\cite{Jia2011} to evaluate the reliability of ART's fault-detection ability by analyzing the variation in fault detection, concluding that it is more reliable than RT.  
Cotroneo et al.~\cite{Cotroneo2016} evaluated the reliability improvement of two ART techniques, FSCS~\cite{Chen2001} and evolutionary ART~\cite{Tappenden2009}:
Based on the same operational profile, for the same test budget, they found that ART had comparable delivered reliability~\cite{Frankl1998} to traditional operational testing. 
For a given reliability level, however, for the same operational profile, ART typically requires significantly fewer test inputs, compared to traditional operational testing techniques. 

\subsubsection{Active, Fuzzing, and Integration Testing}

Based on FSCS-ART~\cite{Chen2001}, Yue et al.~\cite{Yue2015} proposed two input-driven active testing approaches for multi-threaded programs, with experimental evaluations indicating that the proposed methods are more cost-effective than traditional active testing.
Similarly, Sim et al.~\cite{Sim2011} applied FSCS-ART \cite{Chen2001} to fuzzing the Out-Of-Memory (OOM) Killer on an embedded Linux distribution, with results showing that their ART approach for fuzzing requires significantly fewer test cases than RT to identify an OOM Killer failure.
Shin et al.~\cite{Shin2010} proposed an algorithm based on normalized ART for integration and regression tests of units integrated with a front-end software module.
The related simulation studies showed that the proposed ART method could be useful for the integration tests.

\vspace{0.2cm}\noindent \fbox{\parbox[c]{0.97\linewidth} {
\textit{\textbf{Summary of answers to RQ3:}}
\begin{itemize}
    \item[1)] \textit{ART has been applied in many different application domains, including:
    numeric programs, object-oriented programs, embedded systems, configurable systems, and simulations and models.
    According to the surveyed studies, ART generally has better testing effectiveness than RT for most application domains (with respect to various evaluation metrics, including the number of test case executions necessary to identify the first failure).} 

    \item[2)] \textit{ART has also been used to augment or enhance other testing techniques, such as regression testing, combinatorial testing, and reliability testing.
    Similarly, the ART enhancement is generally better than the RT version.} 


    \item[3)] \textit{For each application domain, while ART is generally more effective than RT with respect to different evaluation metrics, it may still be less cost-effective overall.
    Nevertheless,  some ART approaches do provide better cost-effectiveness than RT.}    
\end{itemize}
}}

%
%

\section{Answer to RQ4: How have empirical evaluations in ART studies been performed?\label{SEC:Evaluations}}

\subsection{Distribution of Empirical Evaluations}

\begin{figure}[!t]
    \centering
    \graphicspath{{Graphs/summary/}}
    \includegraphics[width=0.3\textwidth]{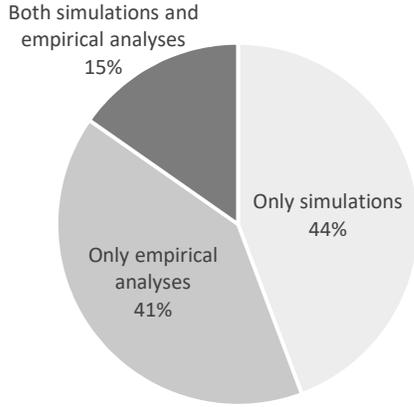}
    \caption{Distribution of empirical evaluations.}
    \label{FIG:experimentDistribution}
\end{figure}

Of the 140 primary studies examined, 131 (94\%) involved empirical evaluations.
Fig.~\ref{FIG:experimentDistribution} shows the distribution of the empirical evaluations in these 131 studies, with 58 papers (44\%) having only simulations, 53 (41\%) only experiments with real programs, and 20 (15\%) containing both simulations and experiments.
It can be observed that the number of studies containing only simulations is comparable to the number with only experiments.


\subsection{Simulations}

Simulations that attempt to construct failures in the numeric input domain to resemble real testing situations have frequently been used to evaluate ART techniques.
As discussed by Chen et al.~\cite{Chen2007c}, three factors\footnote{Simulations examining modeling or distance-calculation errors~\cite{Qi2017} have also appeared, but are very rare.}
are typically considered in the design of simulations:
the dimensionality ($d$) of the input domain;
the failure rate ($\theta$); and
the failure pattern.
In spite of their popularity, simulations have limitations as evaluation tools, including that:
(1) they are mostly only used to simulate numeric input domains;
(2) the assumed failure patterns may not be realistic; and
(3) simulations that do account for test execution time often have very similar execution times, which means that
the simulations are effectively comparing {\em generation}, rather than execution, time, which may decrease their applicability in practice. 

In general,  primary studies involving simulations assume the input domain $\mathcal{D}$ to be $[0,1.0)^d$ 
---
a unit hypercube (each dimension of $\mathcal{D}$ ranging from $0.0$ to $1.0$).
Excluding those few studies using simulations for a special testing environment (such as for combinatorial testing~\cite{Nie2011}), 73 of the 78 papers involving simulations used numeric input domains. 
In this section, we review these 73 studies according to the three main simulation design factors ($d$, $\theta$, and the failure pattern).

\subsubsection{Dimensionality Distribution}

\begin{figure}[!b]
    \centering
    \graphicspath{{Graphs/summary/}}
    \includegraphics[width=0.485\textwidth]{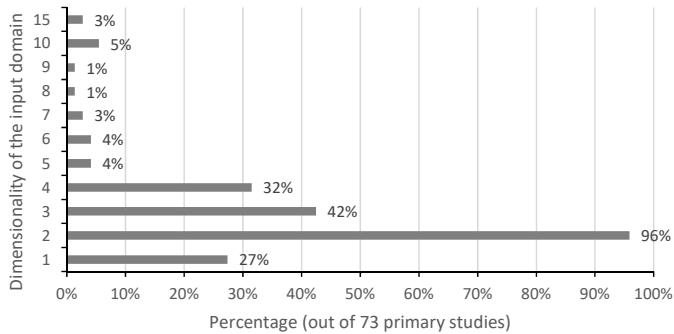}
    \caption{Input domain dimensionality distribution.}
    \label{FIG:dimensionalityDistribution}
\end{figure}

Fig.~\ref{FIG:dimensionalityDistribution} shows the input domain dimensionality ($d$) distribution across the 73 primary studies.
It can be observed that $d$ ranges from 1 to 15, with $d=2$ being the most popular (96\%), followed by $d=3$ (42\%), $d=4$ (32\%), and $d=1$ (27\%).
Only a maximum of four papers (5\%) involved each $d$ greater than $4$.
In other words, most simulations have been conducted in low dimensional  ($d\leq4$) numeric input domains.

\begin{figure}[!t]
    \centering
    \graphicspath{{Graphs/summary/}}
    \includegraphics[width=0.485\textwidth]{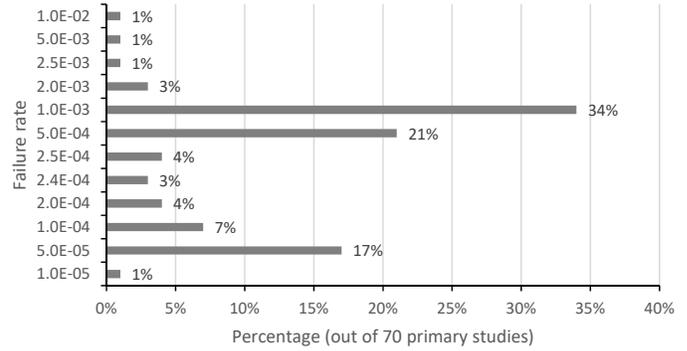}
    \caption{Distribution of failure rates.}
    \label{FIG:failureRateDistribution}
\end{figure}

\subsubsection{Failure Rate Distribution}



Of the 73 primary studies involving simulations, 70 simulated failures in the input domain.
Among these 70, the maximum failure rate ($\theta_H$) used was $1.0$, and the minimum ($\theta_L$) was $1.0\cdot 10^{-5}$.
Fig.~\ref{FIG:failureRateDistribution} shows the distribution of lowest simulation failure rates  ($\theta_L$) across these 70 papers: 
As shown in Fig.~\ref{FIG:failureRateDistribution},
more than half of the simulations involved a minimum failure rate of either $1.0 \cdot 10^{-3}$ (24 papers: 34\%) or $5.0\cdot 10^{-4}$ (15 papers: 21\%).
The next two most commonly used $\theta_L$ values are $5.0\cdot 10^{-5}$ (17\%), and $1.0\cdot 10^{-4}$ (7\%).
In total, only one paper had $\theta_L=1.0\cdot 10^{-5}$, indicating that the failure rates used in simulations have not been very low.
This lack of simulation data for very low failure rates contrasts with Anand et al.'s report that ``lower failure rates are actually favorable scenarios for ART with respect to F-measures''~\cite{Anand2013}.
Thus, it would be interesting and worthwhile to conduct more simulations involving lower failure rates to better evaluate ART performance.

\begin{figure}[!b]
    \centering
    \graphicspath{{Graphs/summary/}}
    \includegraphics[width=0.49\textwidth]{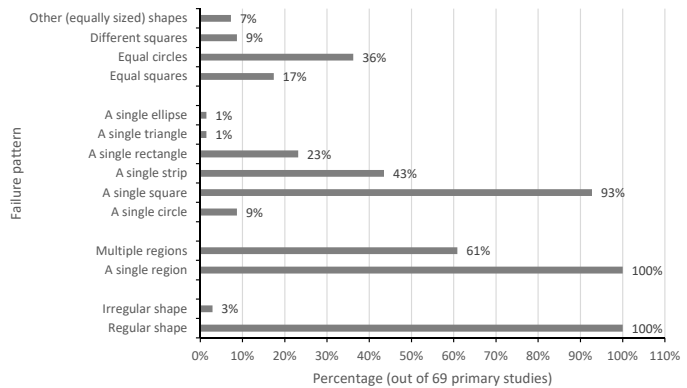}
    \caption{Distribution of failure patterns.}
    \label{FIG:failurePatternDistribution1}
\end{figure}

\subsubsection{Failure Pattern Distribution}

Of the 73 primary studies involving simulations, 69 involved specific failure pattern designs, which, according to a simple classification, can be categorized as either \textit{regular} or \textit{irregular} shapes~\cite{Chen2006b,Chen2006c}.
As shown in Fig.~\ref{FIG:failurePatternDistribution1}, all 69 papers (100\%) included regular shapes that, in a two-dimensional input domain
($d=2$)\footnote{For ease of description, the failure patterns in Fig.~\ref{FIG:failurePatternDistribution1} focus on two-dimensions, but the categories are similar for higher $d$:
For example, when $d=3$, a square becomes a cube, and a circle becomes a sphere.},
could be described as:
\textit{square}, \textit{rectangle},
\textit{strip}\footnote{The rectangle type is a special case of the strip type, with the main difference being that each side of the rectangle type is parallel to the corresponding dimension of $\mathcal{D}$, but strips are not necessarily so, and may not be parallelograms~\cite{Chan1996}.},
\textit{circle}, \textit{ellipse}, or \textit{triangle}.
Only two papers (3\%) used irregular shapes~\cite{Chen2006b,Chen2006c}, constructed by combining between two and five regular shapes.

\begin{figure}[!b]
    \centering
    \graphicspath{{Graphs/summary/}}
    \includegraphics[width=0.33\textwidth]{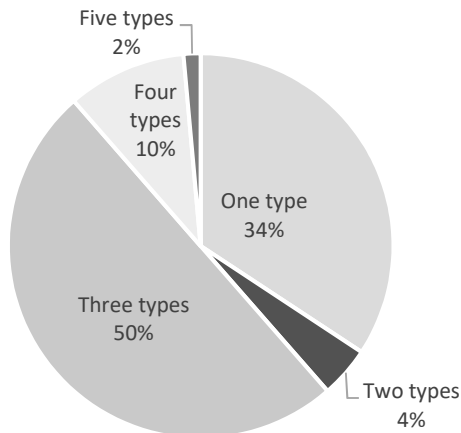}
    \caption{Distribution of failure pattern types.}
    \label{FIG:failurePatternDistribution2}
\end{figure}

Fig.~\ref{FIG:failurePatternDistribution1} shows that all 69 papers (100\%) had simulations involving a single region, and 42 papers (61\%) also investigated multiple regions.
The most popular shape used in single-region simulations was the square (93\%), followed by strip (43\%), rectangle (23\%), and circle (9\%);
both triangle and ellipse were used in only one primary study (1\%) each. 
In simulations with multiple failure regions, most studies used \textit{equal circles} (36\% of the 69 papers), \textit{equal squares} (17\%), \textit{different squares}\footnote{Previous studies~\cite{Chen2007c} designed a predominant region by assigning $q\%$ of the failure region to one square (e.g., $q$ may be equal to 30, 50, or 80), with the other squares sharing the remaining percentage of the failure region in a random manner.} (9\%),
or other (equally-sized) \textit{shapes}\footnote{The unknown shapes refer to situations where the shape details were not provided.} (7\%).

As shown in Fig.~\ref{FIG:failurePatternDistribution1}, there are many types of failure patterns.
Fig.~\ref{FIG:failurePatternDistribution2} presents how many failure pattern types were used in the 69 studies involving failure patterns.
Half of the papers examined three types, followed by one (34\%), four (10\%), and two types (4\%).
Only one paper (2\%) looked at five types of failure pattern in the simulations~\cite{Chen2007c}.
A conclusion from this analysis is that many studies have not designed the simulations comprehensively enough to accurately evaluate ART.

\subsection{Experiments with Real Programs}

In this section, we summarize some details about the ART experiments involving real programs.

\subsubsection{Subject Programs}

We collected the details of each subject program used in the ART experiments, including its name, implementation language, size, description, and references to the primary studies that reported results for that program.
This information is summarized in the appendix, in Table~\ref{TAB:A1} ---
``NR" again indicates that some details were not reported in the original paper.
For ease of illustration, we ordered the programs according to the number of references (the last column), listing the most studied ART subject programs at the top of the table.

\begin{figure}[!t]
    \centering
    \graphicspath{{Graphs/summary/}}
    \includegraphics[width=0.3\textwidth]{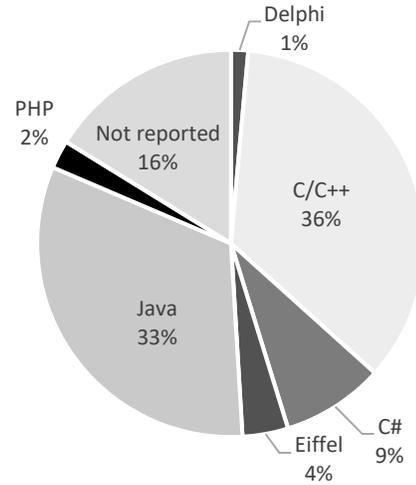}
    \caption{Programming language distribution of subject programs.}
    \label{FIG:programLanguage}
\end{figure}

In total, as can be seen from Table~\ref{TAB:A1}, 211 subject programs were found, ranging in size from 8 to 4,727,209 lines of code.
Fig.~\ref{FIG:programLanguage} shows the distribution of programming languages used to implement the programs.
It can be observed that most programs were written in C/C++ (36\%) and Java (33\%), followed by C\# (9\%).


\begin{figure}[!b]
    \centering
    \graphicspath{{Graphs/summary/}}
    \includegraphics[width=0.95\linewidth]{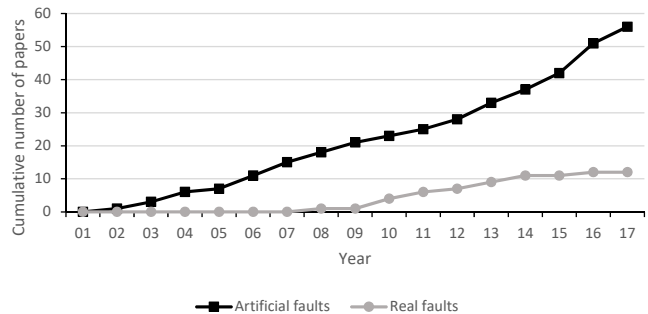}
    \caption{Real versus artificial faults.}
    \label{FIG:faultDistribution}
\end{figure}

\subsubsection{Types of Faults}

The testing effectiveness of ART is generally evaluated according to its ability to identify failures caused by faults in the subject programs.
Because actual defects are not always available in real programs, artificially faulty programs, in the form of \textit{mutants}, are often used.
The mutants can be created manually, or with an automatic mutation tool~\cite{Jia2011}.
Similar to previous surveys~\cite{Segura2016}, we investigated the relationship between artificial and real faults in the empirical evaluations of ART by calculating the cumulative number of primary studies using each, as shown in Fig.~\ref{FIG:faultDistribution}.
It can be seen that the first study involving artificial faults was reported in 2002, while the first with real faults was in 2008.
Furthermore, although both studies with artificial and real faults are increasing, the rate of increase for artificial faults is much higher than that for real faults.
By the end of 2017, more than 55 primary studies had used artificial faults, compared with only 12 identifying real faults, indicating that relatively few studies have used ART to detect real bugs.
Nevertheless, the number of ART studies detecting real faults has been increasing.

%
%
%
%
%
%
%
%

\subsubsection{Additional Information used to Support ART}

ART uses information from previously executed tests to guide generation of subsequent test cases to achieve an even-spreading over the input domain. 
Each test case contains some intrinsic information:
A test case $(0.5,0.5)$, for example, is a point in a 2-dimensional input domain, with intrinsic information represented by its location;
another test case, ``xyz", is a list of characters whose intrinsic information is represented by a string.
In addition to a test case's intrinsic information, some further information may also be available that can be extracted and applied to support ART execution.
In this section, we review some additional information obtained from the ART studies.

Most studies made some use of white-box information (including branch coverage, statement coverage, and mutation score) to guide test case generation.
Several studies~\cite{Zhou2012,Zhou2010b,Jiang2009} have used branch coverage information, but have adopted different representations.
For a given SUT with a list of $\beta$ branches, denoted  $\mathcal{BR}=\{ br_1,br_2,\cdots,br_\beta\}$,
a test case ${tc}$ could cover a set of these branches, denoted  ${BR}({tc})$, where ${BR}({tc}) \subseteq \mathcal{BR}$.
Some previous studies \cite{Zhou2012,Zhou2010b} have used a binary vector $(x_1,x_2,\cdots,x_\beta)$ where each element $x_i~(1\leq i \leq \beta)$ represents whether or not the branch $br_i$ is covered by ${tc}$:
if $br_i$ is covered, then $x_i=1$, otherwise $x_i=0$.
This information can also be represented by a set of branches~\cite{Jiang2009}, i.e., ${BR}({tc})$.
Zhou et al.~\cite{Zhou2012a} also used a test vector based on branch information $(y_1,y_2,\cdots,y_\beta)$, however, each element $y_i~(1\leq i \leq \beta)$ in their vector represents the number of times that $br_i$ is covered by ${tc}$.
Jiang et al.~\cite{Jiang2009} used sets of statements or methods for each test case.
Both Hou et al.~\cite{Hou2013} and Sinaga et al.~\cite{Sinaga2017} used the program path to represent each test case by constructing a \textit{Control Flow Graph} (CFG)~\cite{Allen1970} for the program.

Tappenden and Miller~\cite{Tappenden2014} also used a binary vector for individual test cases to represent the existence (or lack) of certain cookies within a global cookie collection.
Patrick and Jia~\cite{Patrick2015,Patrick2017} used mutation scores to construct a probability distribution for test case selections.
Some previous studies~\cite{Hemmati2010,Hemmati2011,Hemmati2013,Arcuri2010} have described test cases using UML state machine test paths, considering each test path as either a set or sequence.
Iqbal et al.~\cite{Iqbal2012}, using the same UML state machine as in other studies~\cite{Hemmati2010,Hemmati2011,Hemmati2013,Arcuri2010}, used a test data matrix to represent test cases.
Matinnejad et al.~\cite{Matinnejad2015} represented test cases using a sequence of signals that could be described as a function over time;
and Liu et al.~\cite{Liu2010a} represented test cases with an event sequence.
Indhumathi and Sarala~\cite{Indhumathi2014} used .NET Solution Manifest files to generate test case scenarios, each one producing at least one test case. 
Nikravan et al.~\cite{Nikravan2015} applied the path constraints of input parameters to support ART.
Nie et al.~\cite{Nie2009} enhanced ART testing effectiveness through the use of I/O relations.
When testing C compilers, where each test case was a C program, Chen et al.~\cite{Chen2016} counted the occurrence of certain tokens in each program, constructing a numeric vector to represent each test case.
Hui and Huang~\cite{Hui2016} applied metamorphic relations to support ART test case generation, and Yuan et al.~\cite{Yuan2011} have incorporated program invariant information into ART.

\subsection{Evaluation Metrics}

Various metrics have been used to evaluate the testing effectiveness and efficiency of ART approaches.
In this section, we review those metrics used in the primary studies.

\subsubsection{Effectiveness Metrics}

The effectiveness metrics, which are used to evaluate the effectiveness of ART techniques, can be classified into three categories:
fault-detection; test-case-distribution; and structure-coverage.

1) \textit{Fault-detection metrics}:
These metrics assess the fault-detection ability of ART, and include the \textit{F-measure}~\cite{Chen2004}, \textit{E-measure}~\cite{Chen2006}, and \textit{P-measure}~\cite{Chen2006}.
The F-measure is the expected \textit{F-count}~\cite{Chen2008b} (the number of test cases required to detect a failure in a specific test run);
the E-measure refers to the expected number of failures to be identified by a set of test cases; and
the P-measure is the probability of a test set identifying at least one program failure.
Liu et al.~\cite{Liu2012} proposed a variant of the F-measure, the \textit{$F^m$-measure}, which they defined as the expected number of test cases required to identify the first $m$ failures. 
These metrics may have different application environments:
the E-measure and P-measure, for example, are appropriate for the evaluation of automated testing systems~\cite{Shahbazi2013}; while the F-measure is more realistic for situations where testing stops once a failure is detected.

In addition to these three metrics, another widely-used one is the \textit{fault detection ratio}, which is defined as the ratio of faults detected by a test set to the total number of faults present~\cite{Rothermel2001}.
It should be noted that in the context of artificial faults (mutants), the fault detection ratio can be interpreted as the \textit{mutation score}~\cite{Jia2011}.

2) \textit{Test-case-distribution metrics}:
These metrics are used to evaluate the distribution of a test set, i.e., how evenly spread the test cases are.
For ease of description in the following, assume a test set $T=\{tc_1,tc_2,\cdots,tc_n\}$, of size $n$, from input domain $\mathcal{D}$.

\begin{itemize}
    \item \textit{Discrepancy}~\cite{Chen2007}:
    The definition of discrepancy was given in Eq. (\ref{Eq:5.1}).


    \item \textit{Dispersion}~\cite{Chen2007}:
    The dispersion of $T$ is calculated as the maximum distance among all pairs of nearest neighbor distances.
    Its definition is: 
        \begin{equation}\tag{7.1}
         	Dispersion(T) = \max_{1 \leq i \leq n}{\min_{1 \leq j \neq i \leq n}}{dist}(tc_i,tc_j).
        \end{equation}

    \item  \textit{Diversity}~\cite{Bueno2007,Bueno2014}:
    The diversity is similar to the dispersion, but uses the sum (not maximum) of all nearest neighbor distances.
    Its definition is: 
    \begin{equation}\tag{7.2}
         Diversity(T) = \sum_{1 \leq i \leq n}{\min_{1 \leq j \neq i \leq n}}{dist}(tc_i,tc_j).
    \end{equation}

    \item \textit{Divergence}~\cite{Lin2009}:
    Similar to diversity,  \textit{divergence}~\cite{Lin2009} is defined as:
    \begin{equation}\tag{7.3}
         Divergence(T) = \sum_{1 \leq i \leq n}{\sum_{1 \leq j \leq n}}{dist}(tc_i,tc_j).
    \end{equation}

    \item \textit{Spatial distribution}~\cite{Mayer2006a,Mayer2006e,Chen2007a,Chen2007e,Chen2007}:
    This refers to the position of each test case over the entire input domain $\mathcal{D}$, and can only be used for numeric input domains.
    The most intuitive version of spatial distribution depicts the locations of test cases:
    Mayer and Schneckenburger~\cite{Mayer2006a}, for example, recorded the locations of the $i$-th test case in a 2-dimensional input domain using 10 million test sets, generating a picture of pixels.
    However, each picture only shows up to the $i$-th test case, and their method cannot depict spatial distributions for input domains with more than three dimensions. 
   Some methods have tried to project the test case positions onto a single dimension:
   Chen et al.~\cite{Chen2007a},  for instance,  projected test cases from $T$ onto one dimension (the $x$-axis), dividing it into 100 equally-sized bins.
    The number of test cases within each bin was then counted, and the spatial distribution of $T$ was thus described with a histogram.

    Other methods have described the spatial distribution of $T$ by dividing $\mathcal{D}$ into a number of equally-sized, disjoint subdomains, from $\mathcal{D}$'s edge to its centre:
    Chen et al.~\cite{Chen2007}, for example, partitioned $\mathcal{D}$ into two subdomains, the \textit{edge} and \textit{center}  regions, defining a new measure of spatial distribution as:
        \begin{equation}\tag{7.4}
        	Edge:Center=\frac{|T_{Edge}|}{|T_{Center}|},
        \end{equation}
        where $T_{Edge}$ and $T_{Center}$ are the sets of test cases from $T$ located in the edge and center regions, respectively.
         Chen et al.~\cite{Chen2007e} also partitioned $\mathcal{D}$ into 128 subdomains, and analyzed the frequency distribution of test cases in each subdomain.
        Similarly, Mayer and Schneckenburger~\cite{Mayer2006e} divided $\mathcal{D}$ into 100 subdomains, and formalized the relative distance of a test case $tc \in T$ to the center of $\mathcal{D}$:
        \begin{equation}\tag{7.5}
        	dist_{max}(c, tc)=\max_{i=1,2,\cdots,d}{|c_i-tc_i|},
        \end{equation}
        where $c$ is the center of $\mathcal{D}$, and $d$ is its dimensionality.
\end{itemize}

3) \textit{Structure-coverage metrics}~\cite{Zhu1997}:
These metrics, which make use of structural elements in the SUT, have been widely used in the evaluation of many testing strategies.
Among them, two popular categories are \textit{control-flow coverage}~\cite{Huang1975} and \textit{data-flow coverage}~\cite{Laski1983}.
Control-flow coverage focuses on some control constructs of the SUT, such as \textit{block}, \textit{branch}, or \textit{decision}~\cite{Huang1975}.
Data-flow coverage, in contrast, checks patterns of data manipulation\footnote{Patterns of data manipulation refer to the definition of some data (\textit{def}, where values are assigned to the data), and its usage (\textit{use}, where the values are used by an operation).
Additionally, \textit{use} can be categorized into \textit{c-use} (where data are used as an output or in a computational expression), and \textit{p-use} (where data appears in a predicate within the program)~\cite{Laski1983}.}  within the SUT, such as \textit{p-uses}, \textit{c-uses}, and \textit{all-du-paths}~\cite{Laski1983}. 
These metrics have also been used to evaluate (fixed-size) test sets generated by ART. 

\begin{figure}[!b]
    \centering
    \graphicspath{{Graphs/summary/}}
    \includegraphics[width=0.49\textwidth]{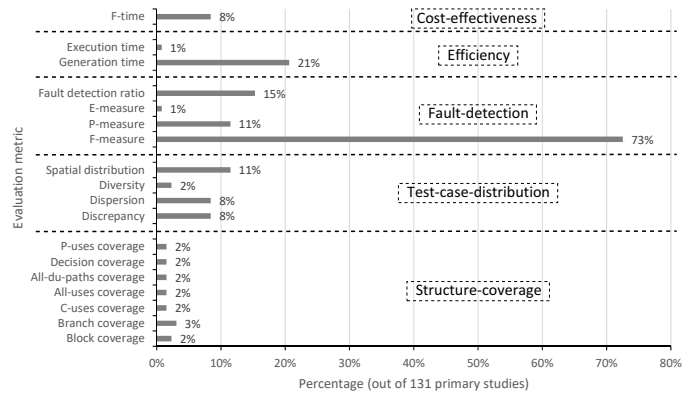}
    \caption{Application of evaluation metrics.}
    \label{FIG:metricDistribution}
\end{figure}

\begin{figure*}[!t]
\centering
\graphicspath{{Graphs/summary/}}
    \subfigure[Simulations.]
    {
        \includegraphics[width=0.485\textwidth]{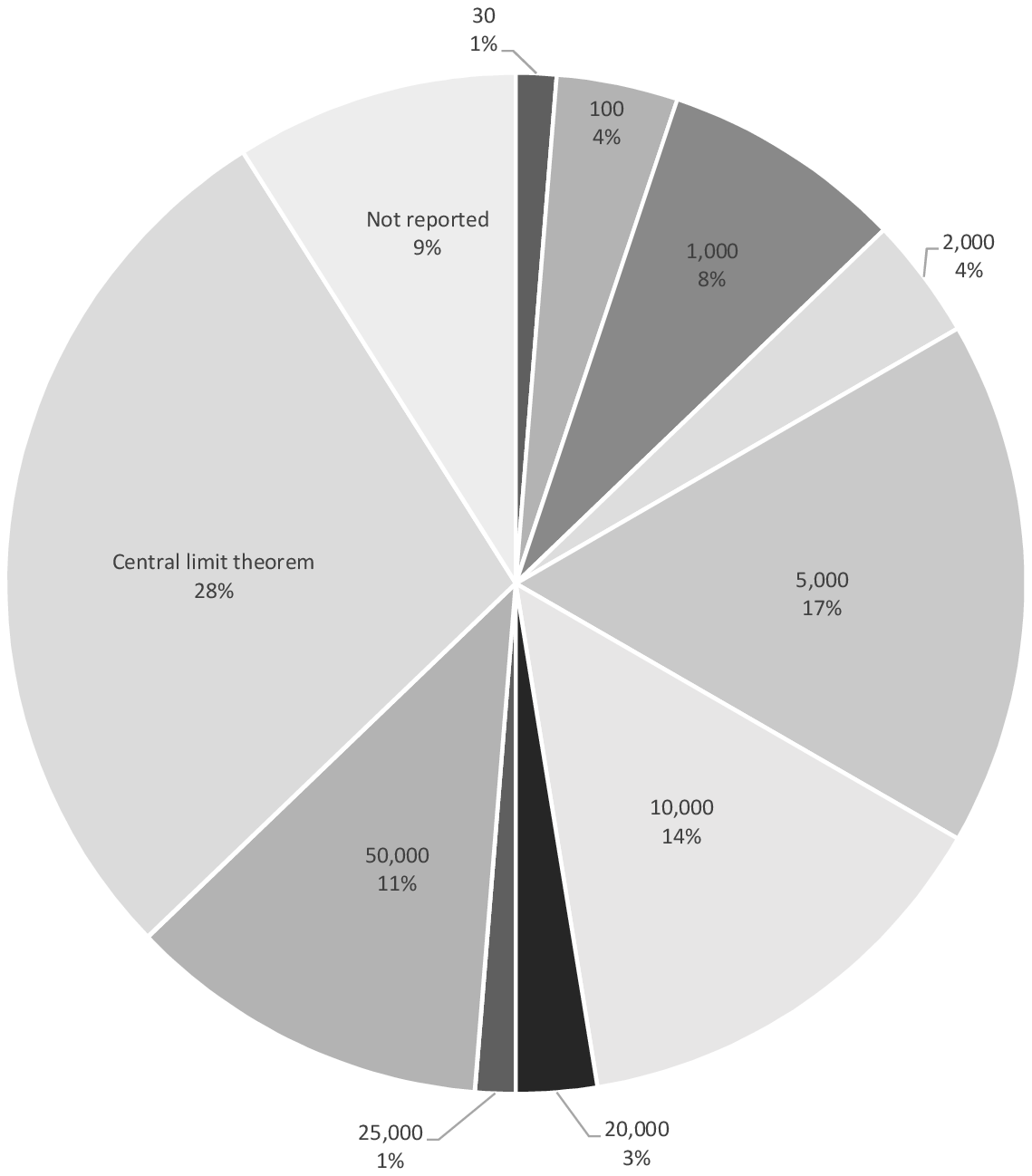}
        \label{FIG:repetitionDistribution1}
    }
    \subfigure[Experiments with real programs.]
    {
        \includegraphics[width=0.485\textwidth]{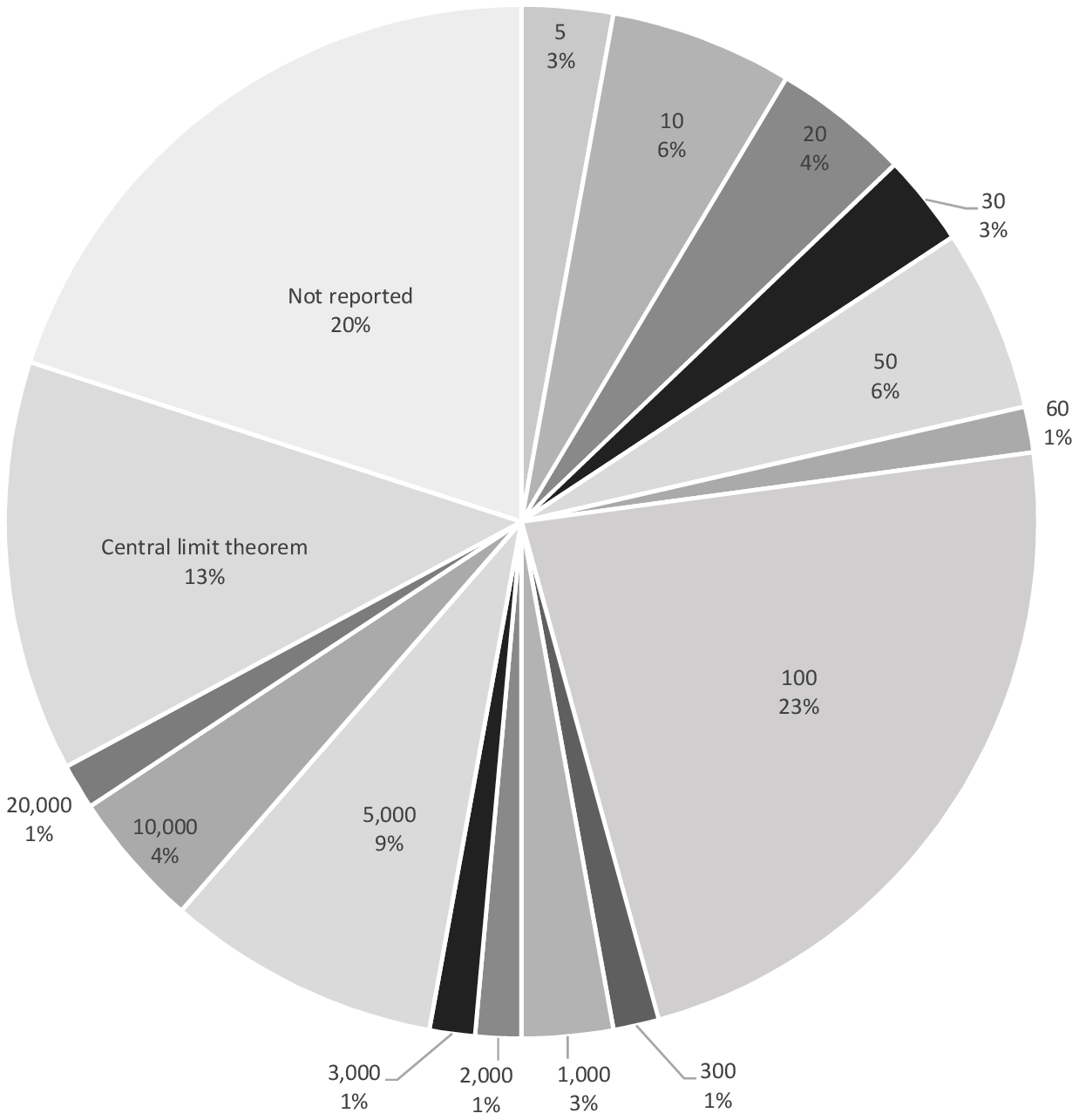}
        \label{FIG:repetitionDistribution2}
    }
    \caption{Distribution of the number of algorithm runs reported in the empirical studies.}
    \label{FIG:repetitionDistribution}
\end{figure*}

\subsubsection{Efficiency Metrics}

There have generally been two metrics used to evaluate the testing efficiency of ART:
the \textit{generation time}, and the \textit{execution time}.
The generation time reflects the computational cost of generating $n$ test cases;
while the execution time refers to the time taken to execute the SUT with $n$ test cases.
On the one hand, because RT has fewer computations involved in the test case generation, it is intuitive that it should have a much lower generation time than ART:
Given the same amount of time, RT typically generates more test cases than ART.
On the other hand, the variation in execution time depends mainly on the SUT.

\subsubsection{Cost-Effectiveness Metric}

The \textit{F-time}~\cite{Chen2006a} is defined as the running time taken to find the first failure.
Suppose the testing process requires $n$ test cases to identify the first failure (i.e., the F-count is equal to $n$), then the F-time comprises the generation time for these $n$ test cases, and the execution time for running them on the SUT.
The F-time, therefore, not only shows the testing efficiency of ART, but also reflects its effectiveness:
it is a cost-effectiveness metric~\cite{Anand2013}.

\subsubsection{Application of Evaluation Metrics}

Fig.~\ref{FIG:metricDistribution} presents the frequency of each applied metric from the 131 primary studies involving empirical evaluations.

Among the effectiveness metrics, the fault-detection metrics were the most used, followed by test-case-distribution metrics.
Very few studies used the structure-coverage metrics.
The majority of papers (73\%) used the F-measure to evaluate ART fault-detection effectiveness, followed by the fault detection ratio (15\%), and the P-measure (11\%).
Only one paper (1\%) used the E-measure, which reflects one of its main criticisms:
that higher E-measures do not necessarily imply more distinct failures or faults~\cite{Chen2004a}.

Regarding the efficiency metrics, 21\%  of the 131 papers used the generation time, whereas only 1\% used the execution time.
Finally, about 8\% of the studies adopted the F-time as the cost-effectiveness metric.

\subsection{Number of Algorithm Runs}

To accommodate the randomness in test cases generated by both RT and ART, empirical evaluations require that the techniques be run in an independent manner, a certain number of times (called the \textit{number of algorithm runs},  $\mathcal{S}$)~\cite{Arcuri2014}.
In this section, we analyze the number of algorithm runs used in each study\footnote{If a paper used different numbers of algorithm runs in different empirical studies, the minimum number was selected for $\mathcal{S}$.}.
Because some studies involving experiments may have had practical constraints (such as limited testing resources), to present the results in an unbiased way, we investigated the numbers of algorithm runs for both simulations and for experiments with real programs.

Fig.~\ref{FIG:repetitionDistribution} presents the paper classification based on the number of algorithm runs reported, with Fig.~\ref{FIG:repetitionDistribution1} showing the distribution of the 78 studies with simulations, and Fig.~\ref{FIG:repetitionDistribution2} showing the 74 studies involving experiments.
According to the \textit{Central limit theorem}~\cite{Tijms2004}, to estimate the mean of a set of evaluation values (such as F-measures), with an accuracy range of $\pm r$ and a confidence level of $(1 - \alpha) \times 100\%$, the size of $\mathcal{S}$ should be at least:
\begin{equation}\tag{7.6}
	\mathcal{S} = \bigg(\frac{100 \cdot z \cdot \sigma}{r \cdot \mu}\bigg)^2,
\end{equation}
where $z$ is the normal variate of the desired confidence level, $\mu$ is the population mean, and $\sigma$ is the population standard deviation. 

As shown in Fig.~\ref{FIG:repetitionDistribution1}, other than 5\% of papers with $\mathcal{S} \leq 100$, and 9\% ``Not reported'', all other simulation papers used either a value of $\mathcal{S}$ determined by the central limit theorem (28\%), or had at least $1,000$ algorithm runs (58\%).
Most studies determined $\mathcal{S}$ based on the central limit theorem, followed by $\mathcal{S}=5,000$ (17\%), $10,000$ (14\%), and $50,000$ (11\%).
On the other hand, as shown in Fig.~\ref{FIG:repetitionDistribution2},
about 46\% of papers involving experiments had $100$ or less algorithm runs ($\mathcal{S} \leq 100$), followed by 19\% with $1,000$ or more ($\mathcal{S} \geq 1,000$).
Only 13\% of experiment papers used a value of $\mathcal{S}$ calculated using the central limit theorem.
According to Arcuri and Briand's practical guidelines~\cite{Arcuri2014}, algorithms involving randomness should be run at least one thousand times ($\mathcal{S}=1,000$) for each artifact (exceptions being for heavy time-consuming SUTs, such as embedded systems~\cite{Hemmati2010,Hemmati2011,Hemmati2013,Arcuri2010,Iqbal2012}).
Therefore, while overall the number of algorithm runs for ART simulations was sufficient, it appears that the number of runs in some studies involving experiments was not.

\subsection{Statistical Significance}

One of the initial motivations behind developing ART was to enhance the testing effectiveness of RT.
It is natural, therefore, to compare each new ART technique with RT, in terms of testing effectiveness.
As already discussed, because test cases generated by both RT and ART contain randomness, it is necessary to determine the  \textit{statistical significance} of any comparison~\cite{Arcuri2014}.
Statistical tests can, amongst other things, determine whether or not there is sufficent empirical evidence to support, with a high level of confidence, that there is a difference between the performance of two algorithms $\mathcal{A}$ and $\mathcal{B}$.
Furthermore, when $\mathcal{A}$ does outperform $\mathcal{B}$, it is also important to quantify the magnitude of the improvement.
In this section, we report on the application of statistical tests in the empirical evaluations of the ART studies.

Of the 78 primary studies involving simulations, only four papers (5\%) used statistical tests.
However, of the 73 papers with experiments, 26 (36\%) examined the statistical significance when comparing two techniques, with the most used statistical tests being:
the $t$\textit{-test};
the \textit{Mann-Whitney U-test};
\textit{ANalysis Of VAriance (ANOVA) test}; and
\textit{Z-test}~\cite{Arcuri2014}.
The \textit{effect size} was the statistic most often used to measure the magnitude of improvements~\cite{Arcuri2014}, with
two papers using it for simulations~\cite{Tappenden2009,Shahbazi2013}, and six including it for experimental data~\cite{Shahbazi2016,Matinnejad2015,Patrick2017,Patrick2015,Arcuri2010,Shahbazi2013}.
The two main approaches used to calculate the effect sizes are from the work of Cohen~\cite{Cohen1992}, and Vargha and Delaney~\cite{Vargha2000}.

In summary, it appears that relatively few ART empirical studies have used sufficient and appropriate statistical testing.

\vspace{0.2cm}\noindent \fbox{\parbox[c]{0.97\linewidth} {
\textit{\textbf{Summary of answers to RQ4:}}
\begin{itemize}
    \item[1)] \textit{Studies involving simulations generally focused on the dimensionality of the input domain, the failure rate, and the failure pattern.
    However, many of these simulations appear not to have been comprehensively designed, and thus may not be accurate evaluations of ART.}
    \item[2)] \textit{The results of the experiments with real programs were influenced by both the subject programs themselves, and the types of faults.
    It was also noted that, so far, papers reporting the detection of real faults still represent only a small proportion of all empirical studies.}
    \item[3)] \textit{The F-measure (number of test case executions required to detect the first failure) was the most popular metric for evaluating ART testing effectiveness;
    test case generation time was the most widely-used for testing efficiency; and
    F-time (execution time required to detect the first failure) was the most commonly used for cost-effectiveness.}
    \item[4)] \textit{Overall, the number of algorithm runs was sufficient in simulations, but inadequate for the experiments with real programs.
    Furthermore, only a small proportion of empirical studies used statistical tests.}
\end{itemize}
}}

\section{Answer to RQ5: What misconceptions surrounding ART exist?\label{SEC:Misconceptions}}



During the development of ART, a number of misconceptions and misunderstandings have arisen, leading to confusion or incorrect conclusions.
Some misconceptions have been discussed previously~\cite{Anand2013}, indicating that they have existed for multiple potential ART users, especially those just beginning to apply it.
Two main misconceptions are discussed in this section.

\subsection{Misconception 1: ART is Equivalent to FSCS}

Anand et al.~\cite{Anand2013} noted that, because FSCS was the first published ART algorithm~\cite{Chen2001}, many studies have presented FSCS as {\em being} ART, or being equivalent or exchangeable.
As discussed in Section~\ref{SEC:Techniques}, FSCS is an ART implementation belonging to the STFCS category, and there are many other STFCS implementations.
There are also other ART implementation categories.
ART refers to a family of testing approaches in which randomly-generated test cases are evenly spread over the input domain~\cite{Chen2010}. 
FSCS is only one of many ART algorithms, and hence ART and FSCS are not equivalent.

\subsection{Misconception 2: ART Should Always Replace RT}

Although RT requires very little information when testing, ART does make use of additional information (such as locations of previously executed test cases) to guide the test case generation.
It may, therefore, seem reasonable that ART should always be better than RT, and thus always replace it.
From the perspective of testing effectiveness, however, Chen et al.~\cite{Chen2007c} found that ART's effectiveness is influenced by many factors, including the failure rate, failure pattern, and dimensionality of the input domain.
They identified several favorable conditions for ART, including a small failure rate, a low dimensionality, a compact failure region, and a small number of failure regions. 
Furthermore, because different approaches to achieve an even spread of test cases have resulted in different ART implementations, each implementation also has its own relative advantages and disadvantages (resulting in favorable and unfavorable conditions for its application).
In other words, there are situations where ART can have similar, or even worse, testing effectiveness compared to RT.
In terms of testing efficiency, compared with RT, in spite of several overhead-reduction algorithms~\cite{Chan2005,Chen2004b,Chan2006a,Shahbazi2013}, ART still incurs more computational overheads.
Consequently, even though ART may have better testing effectiveness than RT, there needs to be a balance between effectiveness and efficiency when choosing either RT or ART:
If the ART test case generation time is considerably less than the test setup and execution time, then it would be appropriate to replace RT with ART;
otherwise, RT may be more appropriate~\cite{Anand2013}. 
Nevertheless, it should be feasible to use ART rather than RT as a baseline when evaluating the state-of-the-art techniques for test case generation, especially from the perspective of testing effectiveness. 

\vspace{0.2cm}\noindent \fbox{\parbox[c]{0.97\linewidth} {
\textit{\textbf{Summary of answers to RQ5:}}
\begin{itemize}
    \item \emph{Two main misconceptions exist in much of the literature:
    that ART is equivalent to FSCS; and
    that RT should always be replaced by ART.}
\end{itemize}
}}

\section{Answer to RQ6: What are the remaining challenges and other future ART work?\label{SEC:Challenges}}

A number of open ART research challenges remain, requiring further investigation and additional (future) work.

\subsection{Challenge 1: Guidelines for Simulation Design}

Although simulations may have limitations compared with real-life programs (because they may not easily and accurately simulate real-world environments, especially complex ones), they are indispensable in the field of ART research.
For any given SUT, the fault details ---
including the size, number, location, and shape of failure regions ---
are fixed (but unknown) before testing begins.
Intuitively, therefore, it is reasonable that studies attempt to simulate faults by controlling and adjusting the factors that create different failure patterns (resulting in different faults).
Although some such simulated faults may seldom occur in real-world programs, they may nonetheless be representative of potential real-world situations, especially for numeric programs.
Furthermore, for a number of reasons, it can be challenging to obtain real-world faulty programs:
their existence or availability, for example, may be limited
Simulations, therefore, can be used to compensate for this lack of appropriate real-world programs.

As discussed in Section~\ref{SEC:Evaluations}, many studies (58 papers) have used simulations to evaluate ART, and different papers may have different simulation designs.
However, these studies only simulated numeric and configurable programs. 
Furthermore, there is a lack of reliable guidelines regarding simulation design, especially from the perspective of those factors that influence ART effectiveness (such as the failure rate and failure pattern details).
The existence of such guidelines could help testers when choosing simulations for experimental evaluations.

\subsection{Challenge 2: Extensive Investigations Comparing Different ART Approaches}

As discussed in Section~\ref{SEC:Evaluations}, although many ART studies are based on simulations and experiments with real programs, all studies have used simulations with failure rates greater than $10^{-6}$, and very few \cite{Mayer2006f,Arcuri2011} have used experiments with failure rates less than $10^{-6}$~\cite{Anand2013}. 
Similarly, few studies have investigated the favorable and unfavorable conditions for each ART approach~\cite{Chen2007c}.
Furthermore, only a very limited number of studies have used statistical testing with a sufficient number of algorithm runs to evaluate ART. 
It is therefore necessary to more fully and extensively investigate and compare the different ART approaches.
This investigation and comparison needs to address not only the strengths and weaknesses of each ART approach, but should also seek to confirm those theoretical results not yet empirically supported (including, as discussed in Section~\ref{Subsection:ReliabilityEstimation}, the potential for ART to support  software reliability estimation).

\subsection{Challenge 3: ART Applications}

As discussed in Section~\ref{SEC:Applications}, ART has been used to test many different applications.
Although ART could theoretically be applied to test any software applications where RT can be used, there remain some applications that have only been tested by RT, such as SQL database systems~\cite{Bati2007}.
It will therefore be interesting and significant to apply ART to these domains.
Furthermore, to date, only those ART approaches using the concept of \textit{similarity}, such as STFCS and SBS, have been used in different applications ---
other approaches, such as PBS and QRS, have mainly been confined to numeric input domains. 
It will therefore also be important to apply more different ART approaches to different applications.

A goal of ART is to achieve an even-spreading of test cases over different input domains (including nonnumeric input domains).
Unlike numeric input domains, nonnumeric domains cannot be considered Cartesian spaces, making visualization of test input locations and failure pattern shapes infeasible.
A key requirement for the application of ART in nonnumeric input domains, therefore, is the availability of a suitable \textit{dissimilarity} or \textit{distance} metric to enable comparison of the nonnumeric inputs.
Consider, for example, a program that checks whether or not an input string of the form ``\textit{YYYY-MM-DD}'' is a valid date:
Given three potential string input tests ---
$tc_1=\textrm{``\textit{2019-01-31}"}$,
$tc_2=\textrm{``\textit{2019-01-3X}"}$, and
$tc_3=\textrm{``\textit{1998-12-24}"}$ ---
some string dissimilarity metrics (e.g., \emph{Hamming distance}, \emph{Levenshtein distance}, and \emph{Manhattan distance}) may indicate that $tc_3$ is farther away from $tc_1$ than from $tc_2$~\cite{Shahbazi2016}.
However, while both $tc_1$ and $tc_3$ are valid inputs, $tc_2$ is invalid, and is thus likely to trigger different behavior and output.
If different test inputs trigger different functionalities and computations, they are also likely to have different failure behavior (including detecting or not detecting failures), which means that they are dissimilar to each other.
This suggests that it would be desirable to incorporate the \emph{semantics} of nonnumeric inputs into their dissimilarity metrics.
If a dissimilarity metric exists that accurately captures the semantic differences between test cases (based on functionality and computation), then ART should be considered.

\subsection{Challenge 4: Cost-effective ART Approaches}

ART cost-effectiveness is critical for real-life applications, and a number of approaches to reduce the computational overheads while  attempting to maintain testing effectiveness have been proposed~\cite{Chan2005,Chen2003,Chen2004b,Kuo2009,Huang2015,Shahbazi2013,Chan2006a,Barus2016}.
However, some approaches are only applicable to numeric inputs~\cite{Chen2003,Chen2004b,Kuo2009,Huang2015,Shahbazi2013}, using the location information of disjoint subdomains to enable their division. 
The main obstacles to applying these cost-reduction techniques to nonnumeric domains include:
(1) how to partition a nonnumeric input domain into disjoint subdomains; and
(2) how to represent the ``\textit{locations}" of these subdomains.

ART based on the concept of mirroring (MART)~\cite{Chen2003,Chen2004b,Kuo2009,Huang2015} first partitions the numeric input domain into equally-sized, disjoint subdomains, designating one as the \textit{source subdomain} and others as \textit{mirror subdomains}.
A mapping relation is used to translate test cases between the source domain and each mirror domain.
For example, consider a two-dimensional input domain $\mathcal{D}$, divided into four equally-sized subdomains, $\mathcal{D}_1$, $\mathcal{D}_2$, $\mathcal{D}_3$, and $\mathcal{D}_4$.
Without loss of generality, assuming that $\mathcal{D}_1$ is the source subdomain, and the others are mirror subdomains, then once a new test case is generated in $\mathcal{D}_1$ using ART (e.g., FSCS or RRT), a mapping relation between $\mathcal{D}_1$ and $\mathcal{D}_i~(i=2,3,4)$ maps the test case to three other test cases in $\mathcal{D}_2$, $\mathcal{D}_3$, and $\mathcal{D}_4$.
Although, intuitively speaking, subdomain locations can only be visualized/identified in a numeric input domain, not in a nonnumeric one, 
if partitioning and subdomain location assignment can be applied to nonnumeric input domains, then MART can be used.

While other overhead-reduction approaches~\cite{Chan2006a,Barus2016} may be applied to both numeric and nonnumeric input domains, they may also involve discarding some information, which may decrease their testing effectiveness.
It is therefore necessary to investigate more cost-effective ART approaches for different applications.

\subsection{Challenge 5: Framework for Selection of ART Approaches}

A framework for the selection of an ART approach could help guide testers to apply ART in practice, especially when facing a choice among multiple approaches.
Anand et al.~\cite{Anand2013} discussed two simple application frameworks, but only at a very high level, and a lot of technical details remain to be determined.
The framework design will also need to address the favorable and unfavorable conditions for each ART approach, as identified in the various studies.

\subsection{Challenge 6: ART Tools}

Although many approaches have been proposed for ART, there are very few tools~\cite{Jaygarl2009,Ciupa2008,Chen2017,Lin2009,Shahbazi2013}, some of which are:
AutoTest, which supports ART for object-oriented  (ARTOO) programs written in Eiffel~\cite{Ciupa2008};
ARTGen, which supports divergence-oriented ART for Java programs~\cite{Lin2009};
\textit{Practical Extensions of Random Testing} (PERT), which supports testing for various input types~\cite{Jaygarl2009}; and
OMISS-ART, which supports FSCS for C++ and C\# programs.
However, these tools are not publicly available.
The only publicly available tool~\cite{RBCVT} was developed to support FSCS, RRT, evolutionary ART, RBCVT, and RBCVT-Fast~\cite{Shahbazi2013}, but this can only be used for purely numeric input domains.
Currently, testers wanting to use ART have to implement the corresponding algorithm themselves.
There is, therefore, a desire and need to develop and make available more ART tools to support both research and actual testing.

\vspace{0.2cm}\noindent \fbox{\parbox[c]{0.97\linewidth} {
\textit{\textbf{Summary of answers to RQ6:}}
\begin{itemize}
    \item \emph{Six current challenges have been identified for ART that will require further investigation.
    These are the current lack of:
    (i) guidelines for the design of ART simulations;
    (ii) extensive investigations comparing different ART approaches;
    (iii) ART applications;
    (iv) cost-effective ART approaches;
    (v) a framework for the selection of ART approaches; and
    (vi) ART tools.}
\end{itemize}
}}

\section{Conclusion\label{SEC:Conclusions}}

In this article, we have presented a survey covering 140 ART papers published between 2001 and 2017.
In addition to tracing the evolution and distribution of ART topics, we have classified the various ART approaches into different categories, analyzing their strengths and weaknesses.
We also investigated the ART application domains, noting that it has been applied in multiple domains, and has been integrated with various other testing techniques.
Furthermore, we have identified that different types of failure patterns have been used in the various reported simulations, and that there has been an increasing number of real faults detected and reported. 
Finally, we discussed some misconceptions about ART, and listed some current and future ART challenges requiring further investigation.
We believe that this article represents a comprehensive reference for ART, and may also guide its future development.

\ifCLASSOPTIONcompsoc
  \section*{Acknowledgments}
\else
  \section*{Acknowledgment}
\fi
We would like to thank the anonymous reviewers for their many constructive comments.
We would also like to thank T. Y. Chen for his many helpful suggestions for this article.
This work is supported by the National Natural Science Foundation of China, under grant nos. 61502205 and 61872167; the China Postdoctoral Science Foundation, under grant no. 2019T120396; and the Senior Personnel Scientific Research Foundation of Jiangsu University, under grant no.~14JDG039. This work is also supported by the Young Backbone Teacher Cultivation Project of Jiangsu University, and the Postgraduate Research \& Practice Innovation Program of Jiangsu Province, under grant no.~KYCX19\_1614. Rubing Huang is the corresponding author.
We dedicate this article to the memory of our friend and colleague, Dr. Diana Fei-Ching Kuo (29/9/1973---6/10/2017), one of the pioneers of Adaptive Random Testing.
\bibliographystyle{IEEEtran}
\bibliography{ref}

\begin{IEEEbiography}[{\includegraphics[width=1in,height=1.25in,clip,keepaspectratio]{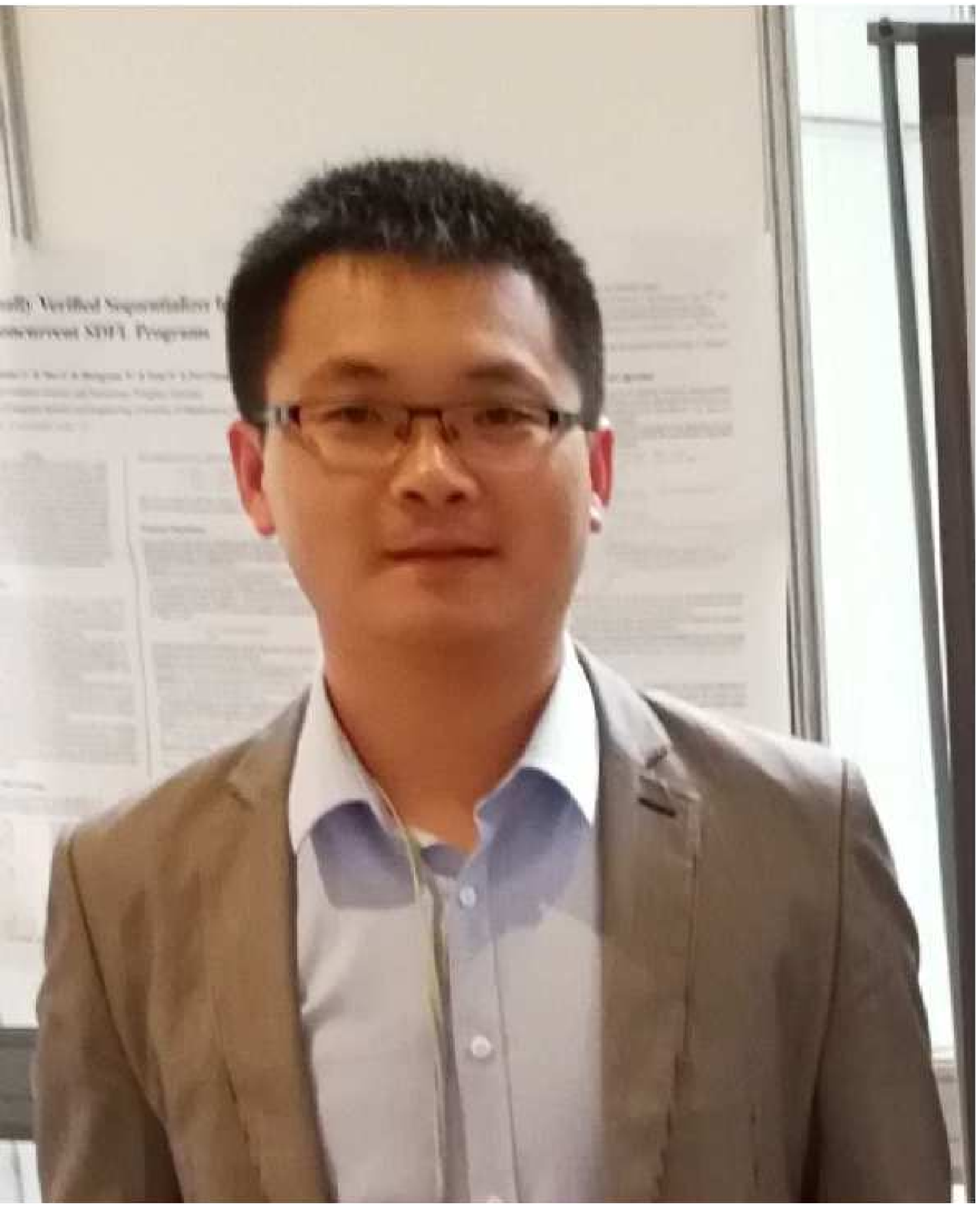}}]{Rubing Huang}
received the Ph.D. degree in computer science and technology from the Huazhong University of Science and Technology, Wuhan, China, in 2013. From 2016 to 2018, he was a
visiting scholar at Swinburne University of Technology and at Monash University, Australia.
He is an associate professor in the Department of Software Engineering, School of Computer Science and Communication Engineering, Jiangsu University, Zhenjiang, China. His current research interests include software testing (including adaptive random testing, random testing, combinatorial testing, and regression testing), debugging, and maintenance. He has more than 50 publications in journals and proceedings, including in IEEE Transactions on Software Engineering, IEEE Transactions on Reliability, Journal of Systems and Software, Information and Software Technology, IET Software, International Journal of Software Engineering and Knowledge Engineering, ICSE, and COMPSAC. He is a senior member of the IEEE and the China Computer Federation, and a member of the ACM. More about him and his work is available online at https://huangrubing.github.io/.
\end{IEEEbiography}
\begin{IEEEbiography}[{\includegraphics[width=1in,height=1.25in,clip,keepaspectratio]{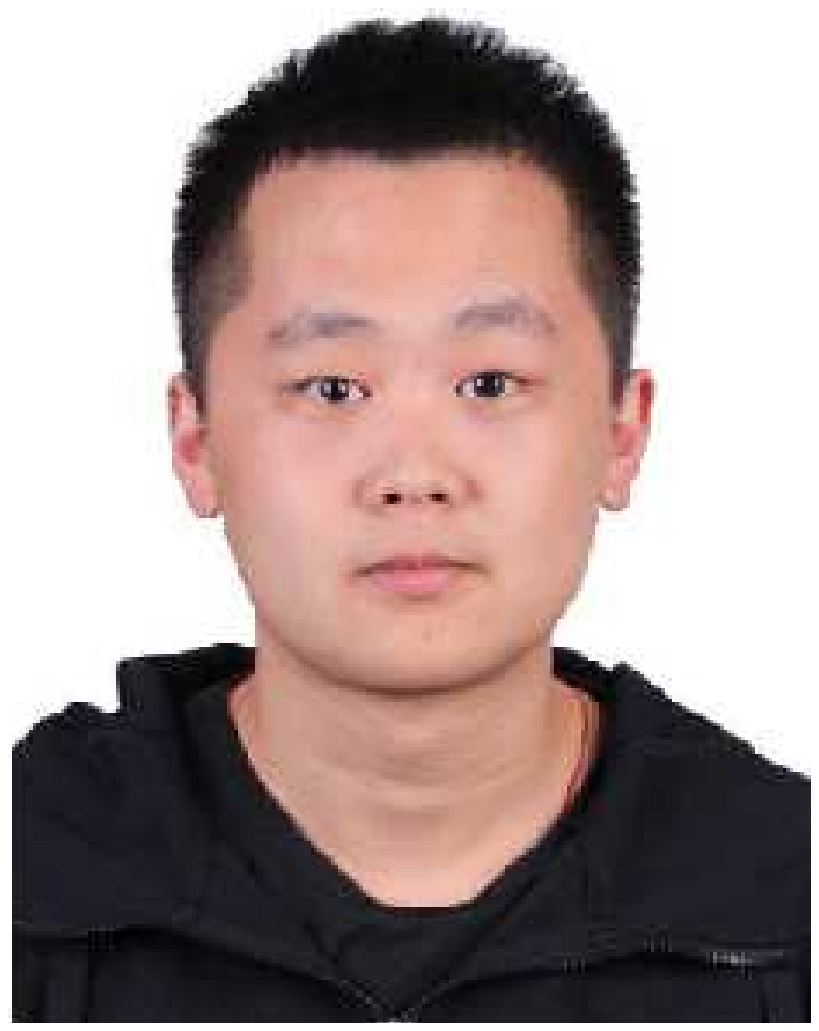}}]{Weifeng Sun}
received the B.Eng. degree in computer science and technology in 2018 from Jiangsu University, Zhenjiang, China, where he is currently working toward the M.Eng. degree with the School of Computer Science and Communication Engineering.
His current research interests include software testing and software debugging.
\end{IEEEbiography}

\begin{IEEEbiography}[{\includegraphics[width=1in,height=1.25in,clip,keepaspectratio]{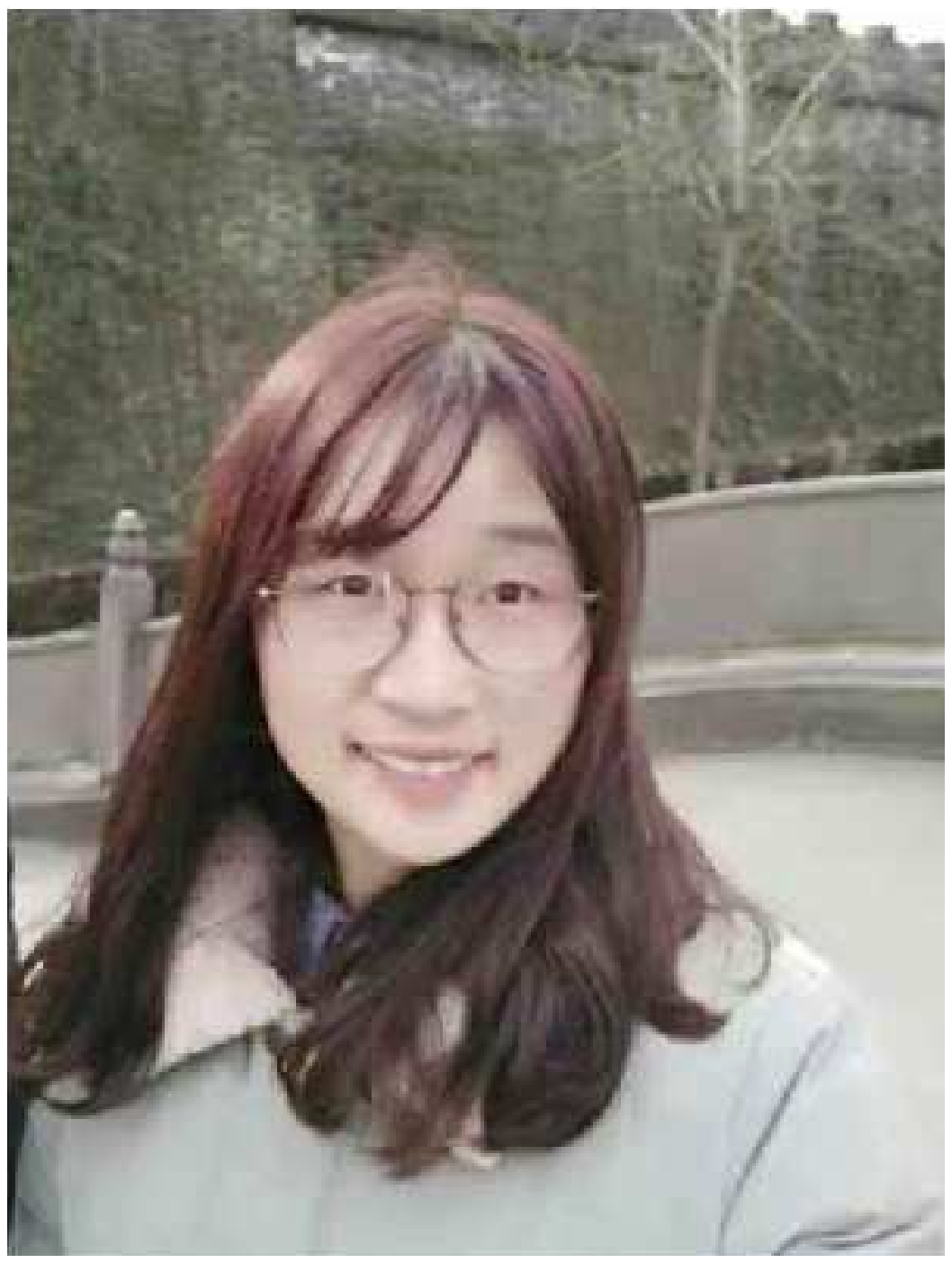}}]{Yinyin Xu} received the B.Eng. degree in Computer Science and Technology from Jinling Institute of Technology, Nanjing, China, in 2018. She is currently working toward the M.Eng. degree with the School of Computer Science and Communication Engineering, Jiangsu University, Zhenjiang, China.
Her current research interests include software maintenance.
\end{IEEEbiography}

\begin{IEEEbiography}[{\includegraphics[width=1in,height=1.25in,clip,keepaspectratio]{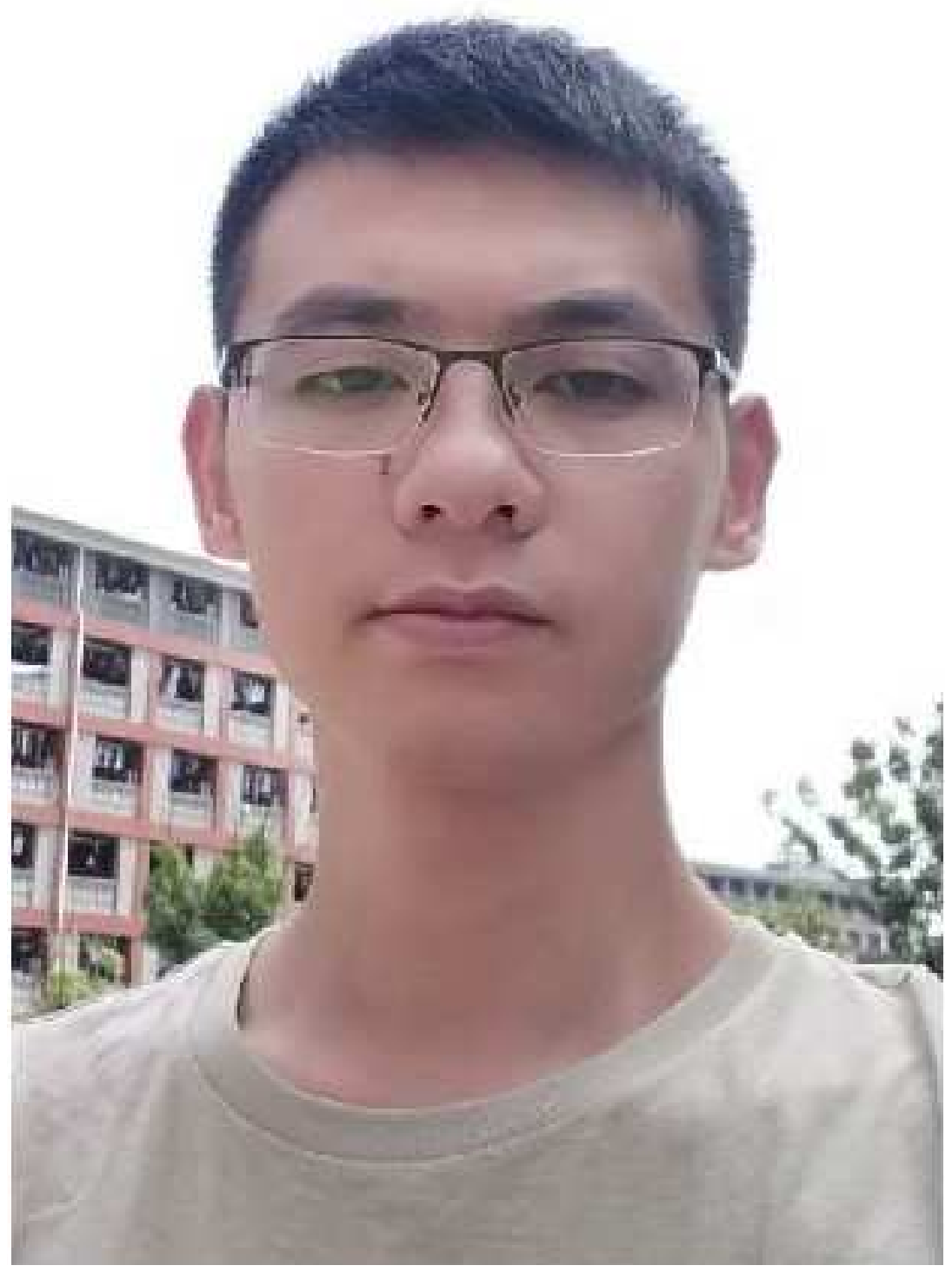}}]{Haibo Chen} received the B.Eng. degree in Computer Science and Technology from Changzhou Institute of Technology, Changzhou, China, in 2018. He is currently working toward the M.Eng. degree with the School of Computer Science and Communication Engineering, Jiangsu University, Zhenjiang, China.
His current research interests include software testing.
\end{IEEEbiography}

\begin{IEEEbiography}[{\includegraphics[width=1in,height=1.25in,clip,keepaspectratio]{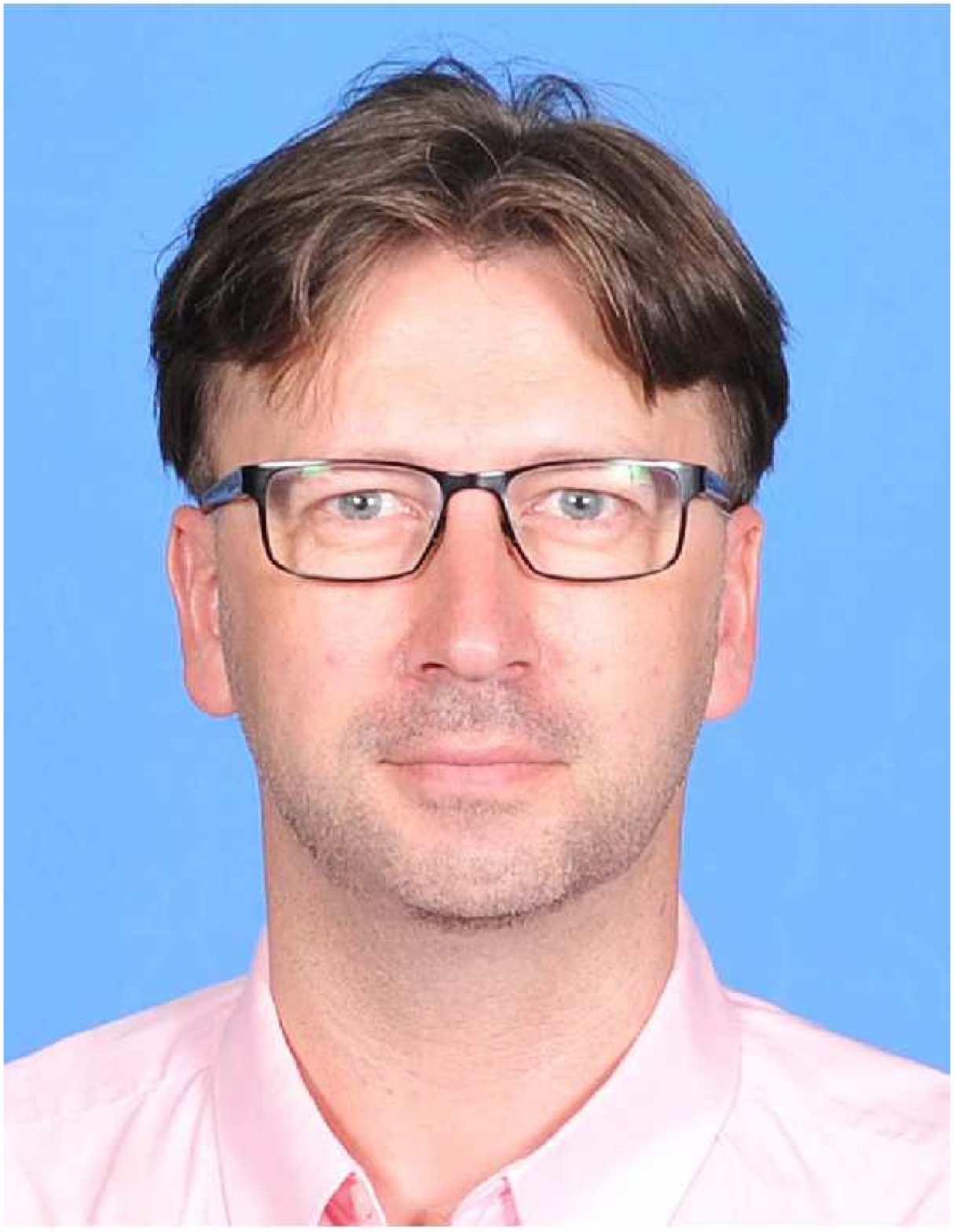}}]{Dave Towey} received the B.A. and M.A. degrees in computer science, linguistics, and languages from the University of Dublin, Trinity College, Ireland; the M.Ed. degree in education leadership from the University of Bristol, U.K.; and the Ph.D. degree in computer science from The University of Hong Kong, China.
He is an associate professor at University of Nottingham Ningbo China (UNNC), in Zhejiang, China, where he serves as the director of
teaching and learning, and deputy head of school, for the School of Computer Science. He is also the deputy director of the International Doctoral
Innovation Centre at UNNC. He is a member of the UNNC Artificial Intelligence and Optimization research group. His current research interests include software testing (especially adaptive random testing, for which he was amongst the earliest researchers who established the field, and metamorphic testing), computer security, and technology-enhanced education.
He co-founded the ICSE International Workshop on Metamorphic Testing in 2016.
He is a member of both the IEEE and the ACM.
\end{IEEEbiography}

\begin{IEEEbiography}[{\includegraphics[width=1in,height=1.25in,clip,keepaspectratio]{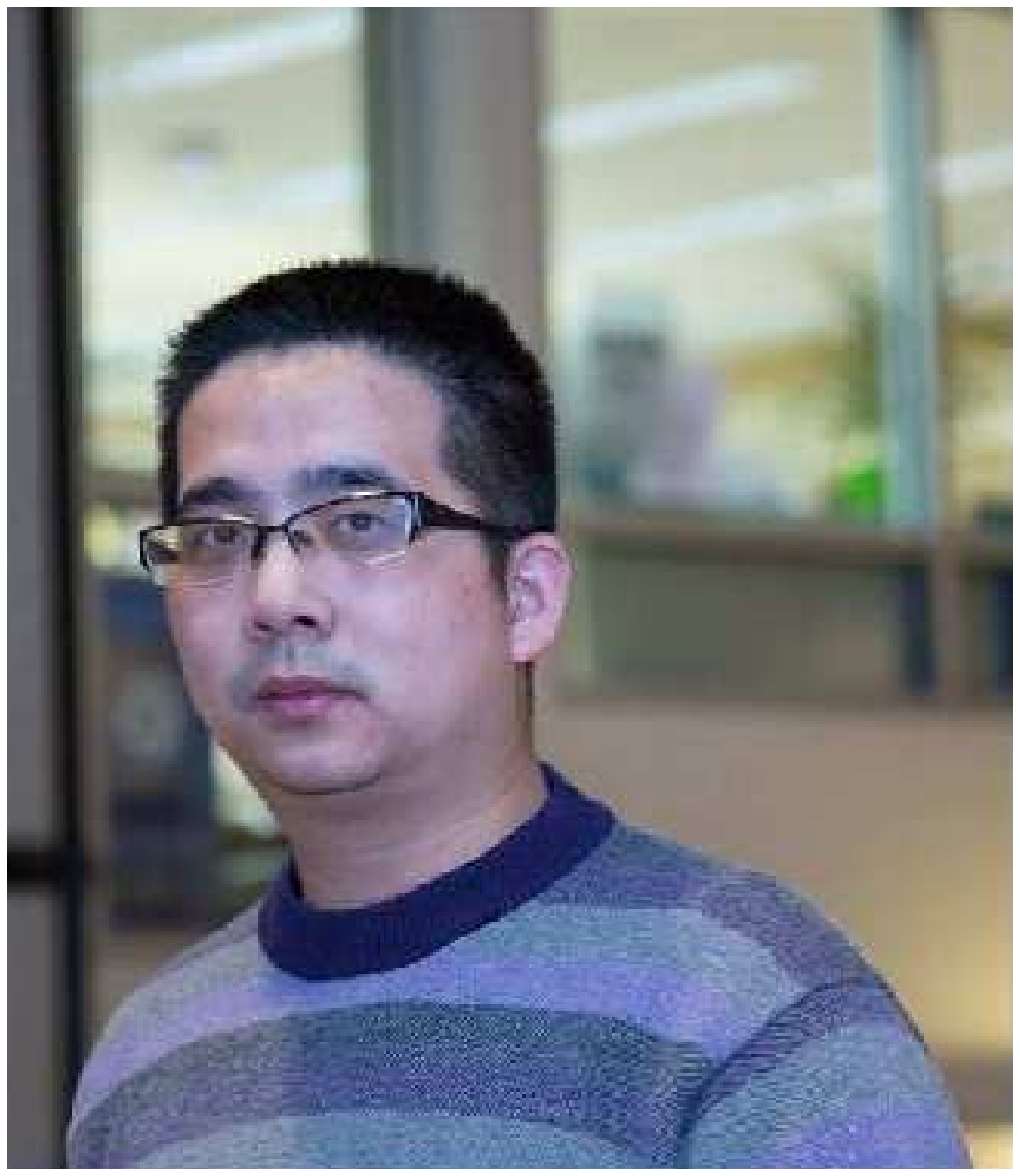}}]{Xin Xia} is a lecturer at the Faculty of Information Technology, Monash University, Australia. Prior to joining Monash University, he was a post-doctoral research fellow in the software practices lab at the University of British Columbia in Canada, and a research assistant professor at Zhejiang University in China. Xin received the Ph.D and bachelor degrees in computer science and software engineering from Zhejiang University in 2014 and 2009, respectively. To help developers and testers improve their productivity, his current research focuses on mining and analyzing rich data in software repositories to uncover interesting and actionable information. More information at: https://xin-xia.github.io/.
\end{IEEEbiography}


\appendices
\section{Subject Programs Used in ART}
\renewcommand{\thetable}{\thesection.\arabic{table}}
\addtocounter{table}{-6}

\begin{table*}[!t]
\scriptsize
\centering
\caption{Subject Programs in ART\label{TAB:A1}}
\begin{tabular*}{\textwidth}{rlllll} \Xhline{1.0pt}
\textbf{No.} &\textbf{Name} &\textbf{Language} &\textbf{Size} &\textbf{Description} &\textbf{Reference} \\\Xhline{0.6pt}
1.    &Bessj	&C/C++	&NR	&Bessel function of general integer order	&\cite{Chan2003}, \cite{Chen2004a}, \cite{Chan2002}, \cite{Chan2003a}, \cite{Chan2006b}, \\
&&&&&\cite{Chen2009}, \cite{Chan2004b}, \cite{Chen2005a}, \cite{Chen2007g}, \cite{Liu2016}, \\
&&&&&\cite{Chen2008a}, \cite{Mao2017}, \cite{Huang2015}, \cite{Liu2011}, \cite{Chen2007c} \\
&&&&&\cite{Chen2008}, \cite{Chan2005}, \cite{Liu2012}\\	

2.    &Plgndr	&C/C++	&36	&Compute the associated Legendre polynomial $P_l^m(x)$	&\cite{Chen2004a}, \cite{Chan2002}, \cite{Chan2003a}, \cite{Chan2006b}, \cite{Chen2009},\\ &&&&&\cite{Chan2004b}, \cite{Chen2005a}, \cite{Chen2007g}, \cite{Liu2016}, \cite{Chen2008a},\\
&&&&&\cite{Mao2017}, \cite{Huang2015}, \cite{Liu2011}, \cite{Chan2005},\\
&&&&&\cite{Chen2007c}, \cite{Patrick2015}, \cite{Patrick2017}, \cite{Liu2012}	\\

3.    &Gammq	&C/C++	&89--106	&Normalized incomplete gamma function	&\cite{Chan2003}, \cite{Chen2004a}, \cite{Chan2002}, \cite{Chan2003a}, \cite{Chan2006b},\\
&&&&&\cite{Chan2004b}, \cite{Zhou2010a}, \cite{Zhou2013}, \cite{Chen2005a}, \cite{Chen2007g},\\
&&&&&\cite{Mao2017}, \cite{Huang2015}, \cite{Liu2011}, \cite{Chen2007c}, \cite{Chan2005},\\
&&&&&\cite{Patrick2015}, \cite{Patrick2017}	\\
4.    &Bessj0	&C/C++	&28	&Bessel function of the first kind	&\cite{Chan2003}, \cite{Chen2004a}, \cite{Chan2002}, \cite{Chan2003a}, \cite{Chan2006b},\\
&&&&&\cite{Chan2004b}, \cite{Chen2005a}, \cite{Chen2007g}, \cite{Liu2016}, \cite{Mao2017},\\
&&&&&\cite{Huang2015}, \cite{Liu2011}, \cite{Chen2007c}, \cite{Chan2005}, \cite{Patrick2015},\\
&&&&&\cite{Patrick2017}, \cite{Liu2012}	\\
5.    &Erfcc	&C/C++	&14	&Complementary error function erfc($x$)	&\cite{Chan2003}, \cite{Chen2004a}, \cite{Chan2002}, \cite{Chan2003a}, \cite{Chan2006b},\\
&&&&&\cite{Chan2004b}, \cite{Chen2005a}, \cite{Chen2007g}, \cite{Mao2017}, \cite{Huang2015},\\
&&&&&\cite{Liu2011}, \cite{Chen2007c}, \cite{Chan2005}, \cite{Patrick2015}, \cite{Patrick2017}	\\
6.    &Cel	&C/C++	&NR	&Cel function	&\cite{Chan2003}, \cite{Chen2004a}, \cite{Chan2002}, \cite{Chan2003a}, \cite{Chan2006b},\\
&&&&&\cite{Chan2004a}, \cite{Chan2004b}, \cite{Chen2005a}, \cite{Chen2007g}, \cite{Mao2017},\\
&&&&&\cite{Huang2015}, \cite{Liu2011}, \cite{Chen2007c}, \cite{Chan2005}	\\
7.    &Sncndn	&C/C++	&NR	&Returns the Jacobian elliptic functions	&\cite{Chan2003}, \cite{Chen2004a}, \cite{Chan2002}, \cite{Chan2003a}, \cite{Chan2006b},\\
&&&&&\cite{Chan2004b}, \cite{Chen2005a}, \cite{Chen2007g}, \cite{Mao2017}, \cite{Huang2015},\\
&&&&&\cite{Liu2011}, \cite{Chen2007c}, \cite{Chan2005}, \cite{Liu2012}	\\
8.    &Airy	&C/C++	&43	&Evaluate the Airy functions $A_i(z)$ and $B_i(z)$ and their derivatives	&\cite{Chen2004a}, \cite{Chan2006b}, \cite{Chen2009}, \cite{Chen2005a}, \cite{Chen2007g},\\
&&&&&\cite{Chen2008a}, \cite{Mao2017}, \cite{Huang2015}, \cite{Liu2011}, \cite{Chen2007c},\\
&&&&&\cite{Chen2008}, \cite{Patrick2015}, \cite{Patrick2017}	\\
9.    &El2	&C/C++	&NR	&El2 function	&\cite{Chen2004a}, \cite{Chan2006b}, \cite{Chen2009}, \cite{Chen2005a}, \cite{Chen2007g},\\
&&&&&\cite{Chen2008a}, \cite{Mao2017}, \cite{Huang2015}, \cite{Liu2011}, \cite{Chen2007c}	\\
10.    &Probks	&C/C++	&22	&Probks function	&\cite{Chen2004a}, \cite{Chan2006b}, \cite{Chen2005a}, \cite{Chen2007g}, \cite{Mao2017},\\
&&&&&\cite{Huang2015}, \cite{Liu2011}, \cite{Chen2007c}, \cite{Patrick2015}, \cite{Patrick2017}	\\
11.    &Golden	&C/C++	&NR	&Golden section search function	&\cite{Chan2003}, \cite{Chen2004a}, \cite{Chan2006b}, \cite{Chen2005a}, \cite{Chen2007g},\\
&&&&&\cite{Mao2017}, \cite{Huang2015}, \cite{Liu2011}, \cite{Chen2007c}	\\
12.    &Tanh	&C/C++	&NR	&Tanh function	&\cite{Chen2004a}, \cite{Chan2006b}, \cite{Chen2005a}, \cite{Chen2007g}, \cite{Mao2017},\\
&&&&&\cite{Huang2015}, \cite{Liu2011}, \cite{Chen2007c}	\\
13.    &TCAS	&C/C++	&133--206	&A traffic collision avoidance system (Siemens suite)	&\cite{Jiang2009}, \cite{Bueno2014}, \cite{Bueno2007}, \cite{Chen2008}, \cite{Barus2016},\\
&&&&&\cite{Patrick2015}, \cite{Patrick2017}, \cite{Hui2016}	\\
14.    &Gammq	&Java	&89	&Normalized incomplete gamma function	&\cite{Schneckenburger2007}, \cite{Mayer2006d}, \cite{Shahbazi2013}, \cite{Mayer2006g}, \cite{Mayer2006f},\\
&&&&& \cite{Arcuri2011}	\\
15.    &Replace	&C/C++	&508--516	&Regular expression matching and substitutions (Siemens suite)	&\cite{Jiang2009}, \cite{Barus2016}, \cite{Zhou2012a}, \cite{Zhou2012}, \cite{Zhang2016}	\\
16.    &Tot\_info	&C/C++	&272--346	&Information measure (Siemens suite)	&\cite{Jiang2009}, \cite{Barus2016}, \cite{Zhou2012}, \cite{Zhang2016}, \cite{Zhou2010b}	\\
17.    &Grep	&C/C++	&3,161--10,068	&Regular expression processor	&\cite{Jiang2009}, \cite{Barus2016}, \cite{Jiang2013}, \cite{Jiang2015}, \cite{Cotroneo2016}	\\
18.    &Printtokens2	&C/C++	&350--402	&Lexical analyzer (Siemens suite)	&\cite{Jiang2009}, \cite{Barus2016}, \cite{Zhou2012}, \cite{Zhang2016}	\\
19.    &Schedule2	&C/C++	&261--297	&Priority scheduler (Siemens suite)	&\cite{Jiang2009}, \cite{Barus2016}, \cite{Zhou2012}, \cite{Zhang2016}	\\
20.    &Flex	&C/C++	&8,426--10,124	&Fast lexical analyzer	&\cite{Jiang2009}, \cite{Huang2014}, \cite{Jiang2013}, \cite{Jiang2015}	\\
21.    &Space	&C/C++	&5,905--9,564	&An interpreter for an array definition language 	&\cite{Zhou2012a}, \cite{Zhou2012}, \cite{Sinaga2017}, \cite{Zhou2010b}	\\
22.    &Bessj	&Java	&131	&Bessel function of general integer order	&\cite{Schneckenburger2007}, \cite{Mayer2006d}, \cite{Arcuri2011}	\\
23.    &Exprint	&Java	&86	&NR	&\cite{Schneckenburger2007}, \cite{Shahbazi2013}, \cite{Arcuri2011}	\\
24.    &Schedule	&C/C++	&291--299	&Priority scheduler (Siemens suite)	&\cite{Jiang2009}, \cite{Barus2016}, \cite{Zhang2016}	\\
25.    &Printtokens	&C/C++	&341--483	&Lexical analyzer (Siemens suite)	&\cite{Jiang2009}, \cite{Barus2016}, \cite{Zhang2016}	\\
26.    &Sed	&C/C++	&4,756--9,289	&Stream editor that perform text transformations on an input stream	&\cite{Jiang2009}, \cite{Jiang2013}, \cite{Jiang2015}	\\
27.    &Gzip	&C/C++	&4,081--5,159	&File compression and decompression	&\cite{Jiang2009}, \cite{Jiang2013}, \cite{Jiang2015}	\\
28.    &Look	&C/C++	&135	&Find words in the system dictionary or lines in a sorted files	&\cite{Chen2013}, \cite{Barus2016}, \cite{Zhang2016}	\\
29.    &Spline	&C/C++	&289	&Interpolate smooth curve based on given data	&\cite{Chen2013}, \cite{Barus2016}, \cite{Zhang2016}	\\
30.    &Cal	&C/C++	&163	&Print a calendar for a specified year or month	&\cite{Chen2013}, \cite{Barus2016}, \cite{Zhang2016}	\\
31.    &Bessel	&C/C++	&NR	&Calculate the regular cylindrical Bessel function of order $n$	&\cite{Zhou2010a}, \cite{Zhou2013}	\\
32.    &Laguerre	&C/C++	&NR	&Calculate generalized Laguerre polynomials	&\cite{Zhou2010a}, \cite{Zhou2013}	\\
33.    &Ellint	&C/C++	&NR	&Calculate incomplete elliptic integral to the accuracy	&\cite{Zhou2010a}, \cite{Zhou2013}	\\
34.    &Triangle	&Java	&26	&Classification of isosceles and equilateral triangles	&\cite{Shahbazi2013}, \cite{Arcuri2011}	\\
35.    &Triangle2	&Java	&41	&Classification of isosceles and equilateral triangles	&\cite{Shahbazi2013}, \cite{Arcuri2011}	\\
36.    &Median	&Java	&20	&Calculate the median value	&\cite{Shahbazi2013}, \cite{Arcuri2011}	\\
37.    &Remainder	&Java	&48	&NR	&\cite{Shahbazi2013}, \cite{Arcuri2011}	\\
38.    &Variance	&Java	&22	&Calculate the variance value	&\cite{Shahbazi2013}, \cite{Arcuri2011}	\\
39.    &BubbleSort	&Java	&14	&Bubble sorting algorithm	&\cite{Shahbazi2013}, \cite{Arcuri2011}	\\
40.    &Encoder	&Java	&65	&NR	&\cite{Shahbazi2013}, \cite{Arcuri2011}	\\
41.    &Fisher	&Java	&71	&NR	&\cite{Shahbazi2013}, \cite{Arcuri2011}	\\
42.    &Select  &C/C++  &NR   &Find the $q$-th smallest element from an array of $p$ real numbers  &\cite{Liu2016}, \cite{Liu2012} \\					
43.    &Expint	&C/C++	&NR	&Compute the real exponential integral of a floating point variable	&\cite{Bueno2014}, \cite{Bueno2007}	\\
44.    &Tritype	&C/C++	&NR	&Classify the type of triangle and calculates its area	&\cite{Bueno2014}, \cite{Bueno2007}	\\
45.    &Alt-sep-test	&C/C++	&NR	&A function from program TCAS	&\cite{Bueno2014}, \cite{Bueno2007}	\\
46.    &IC    &Java  &NR &A seismic acquisition system that interacts with sensors and actuators &\cite{Iqbal2012}, \cite{Arcuri2010}  \\					
47.    &Sort	&C/C++	&842	&Sort and merge files	&\cite{Chen2013}, \cite{Barus2016}	\\
48.    &CCoinBox	&C/C++	&120	&Simulate a vending machine	&\cite{Chen2013}, \cite{Barus2016} \\	
49.    &WindShieldWiper	&C/C++	&233	&Simulate a windshield wiper	&\cite{Chen2017}, \cite{Chen2016a} \\	
50.    &SATM	&C/C++	&197	&Simulate an ATM	&\cite{Chen2017}, \cite{Chen2016a} \\	
51.    &RabbitAndFox	&C\#	&770    &Simulate a predator-prey model   &\cite{Chen2017}, \cite{Chen2016a}   \\			
52.    &WaveletLibrary	&C\#	&2406    &Do wavelet analysis    &\cite{Chen2017}, \cite{Chen2016a}   \\			
53.    &Multimedia system	&C/C++	&NR	&Manage sending and receiving of multimedia streams	&\cite{Hemmati2011}, \cite{Hemmati2013} \\

\Xhline{0.6pt}
\multicolumn{6}{r}{continued on next page}\\
\Xhline{1.0pt}
\end{tabular*}
\end{table*}

\begin{table*}[!t]
\scriptsize
\centering
\begin{tabular*}{\textwidth}{rlllll} \Xhline{1.0pt}
\textbf{No.}  &\textbf{Name} &\textbf{Language} &\textbf{Size} &\textbf{Description} &\textbf{Reference} \\\Xhline{0.6pt}
54.    &Safety control system	&C/C++	&NR	&A safety monitoring component	&\cite{Hemmati2011}, \cite{Hemmati2013} \\	
55.    &Comm	&C/C++	&144	&Select or reject lines common to two sorted files	&\cite{Chen2013}, \cite{Zhang2016}	\\
56.    &Uniq	&C/C++	&125	&Report or remove adjacent duplicate lines	&\cite{Chen2013}, \cite{Zhang2016}	\\
57.    &Triangle	&C/C++	&75	&Classification of isosceles and equilateral triangles	&\cite{Patrick2015}, \cite{Patrick2017}	\\
58.    &Action\_sequence	&Eiffel	&2,477   &System library of the class for sequence   &\cite{Ciupa2008}   \\			
59.    &Array	&Eiffel	&1,208   &System library of the class for array    &\cite{Ciupa2008}   \\			
60.    &Arrayed\_list	&Eiffel	&2,164   &System library of the class for arrayed list    &\cite{Ciupa2008}   \\			
61.    &Bounded\_stack	&Eiffel	&779   &System library of the class for bounded stack    &\cite{Ciupa2008}   \\			
62.    &Fixed\_tree	&Eiffel	&1,623   &System library of the class for fixed tree    &\cite{Ciupa2008}   \\			
63.    &Hash\_table	&Eiffel	&1,791   &System library of the class for hash table    &\cite{Ciupa2008}   \\			
64.    &Linked\_list	&Eiffel	&1,893   &System library of the class for linked list    &\cite{Ciupa2008}   \\			
65.    &String	&Eiffel	&2,980   &System library of the class for string    &\cite{Ciupa2008}   \\			
66.    &Biorhythms Problem	&Java	&NR	&A program from the ACM International Collegiate Programming Contest	&\cite{Hou2013}	\\
67.    &Packing Problem	&Java	&NR	&A program from the ACM International Collegiate Programming Contest	&\cite{Hou2013}	\\
68.    &Chocolate Problem	&Java	&NR	&A program from the ACM International Collegiate Programming Contest	&\cite{Hou2013}	\\
69.    &Multiply Problem	&Java	&NR	&A program from the ACM International Collegiate Programming Contest	&\cite{Hou2013}	\\
70.    &Josephus Problem	&Java	&NR	&A program from the ACM International Collegiate Programming Contest	&\cite{Hou2013}	\\
71.    &ConjugateGradient	&Java	&107	&Conjugate Gradient function	&\cite{Putra2013}	\\
72.    &DefaultFieldMatrixChangingVisitor	&Java	&18	&Create custom visitors without defining all methods	&\cite{Putra2013}	\\
73.    &EigenDecomposition	&Java	&344	&EigenDecomposition function	&\cite{Putra2013}	\\
74.    &Abs	&Java	&8	&Abs function	&\cite{Putra2013}	\\
75.    &Gaussian	&Java	&81	&Gaussian function	&\cite{Putra2013}	\\
76.    &HarmonicOscillator	&Java	&58	&Harmonic Oscillator function	&\cite{Putra2013}	\\
77.    &Sigmoid	&Java	&58	&Sigmoid function	&\cite{Putra2013}	\\
78.    &Minus	&Java	&10	&Minus function	&\cite{Putra2013}	\\
79.    &TD1 &NR &NR &Calculate employee pay with a simple version &\cite{Chi2006} \\					
80.    &TD2 &NR &NR &Calculate employee pay with a complex version   &\cite{Chi2006} \\					
81.    &Sncndn	&Java	&NR	&Returns the Jacobian elliptic functions	&\cite{Mayer2006f}	\\
82.    &Coom	&C/C++	&144	&File comparator	&\cite{Barus2016}	\\
83.    &Quadratic   &C/C++  &NR &Calculate complex roots of the quadratic equation  &\cite{Chen2008}    \\					
84.    &Cubic   &C/C++  &NR &Calculate complex roots of the cubic equation  &\cite{Chen2008}    \\					
85.    &Gamma	&Java	&783	&Regularized version of Gammaq	&\cite{Walkinshaw2017}	\\
86.    &BesselJ	&Java	&1,211	&Calculate Bessel function	&\cite{Walkinshaw2017}  \\	
87.    &Binomial	&Java	&501	&Calculate binomial coefficient function	&\cite{Walkinshaw2017}  \\	
88.    &DerivativeStructure	&Java	&306	&Calculate asinh function	&\cite{Walkinshaw2017}  \\	
89.    &Erf	&Java	&763	&Calculate erf function	&\cite{Walkinshaw2017}  \\	
90.    &RombergIntegrator	&Java	&735	&Calculate Romberg integration function	&\cite{Walkinshaw2017}  \\	
91.    &PeriodToWeeks	&Java	&1,128	&Convert a period to to standard weeks 	&\cite{Walkinshaw2017}  \\	
92.    &DaysBetween	&Java	&1,251	&Date Duration Calculator	&\cite{Walkinshaw2017}  \\	
93.    &Math.geometry	&Java	&340	&3D calculation	&\cite{Lin2009}	\\
94.    &Math.util	&Java	&1,161	&Mathematic functions	&\cite{Lin2009}	\\
95.    &Lang	&Java	&4,276	&Basic utility	&\cite{Lin2009}	\\
96.    &Lang.text	&Java	&1,475	&Text processing	&\cite{Lin2009}	\\
97.    &Collections.list	&Java	&823	&A container structure	&\cite{Lin2009}	\\
98.    &Siena	&Java	&1,438	&A wide-area event notification system	&\cite{Lin2009}	\\
99.    &Calendar	&C/C++	&287	&Calendar operation	&\cite{Chen2017}	\\
100.    &Stack	&C\#	&420	&Stack operation	&\cite{Chen2017}	\\
101.    &Queue	&C\#	&201	&Queue operation	&\cite{Chen2017}	\\
102.    &BinarySearchTree	&C\#	&588	&Binary search tree algorithms	&\cite{Chen2017}	\\
103.    &BackTrack	&C\#	&1,051	&Backtracking algorithms	&\cite{Chen2017}	\\
104.    &NSort	&C\#	&1,118	&Sorting algorithms	&\cite{Chen2017}	\\
105.    &SchoolManagement	&C\#	&1,726	&Manage school activities	&\cite{Chen2017}	\\
106.    &EnterpriseManagement	&C\#	&1,357	&Manage enterprise business	&\cite{Chen2017}	\\
107.    &ID3Manage	&C\#	&4,538	&Read and writ of ID3 tags in MP3 files	&\cite{Chen2017}	\\
108.    &IceChat	&C\#	&71,000	&Implement an IRC (Internet Relay Chat) Client	&\cite{Chen2017}	\\
109.    &CSPspEmu	&C\#	&406,808	&A PSP (PlayStation Portable) emulator	&\cite{Chen2017}	\\
110.    &Poco-1.4.4: Foundation	&C/C++	&149,547	&A platform abstraction layer	&\cite{Chen2017}	\\
111.    &ISSTA Containers	&Java	&2,000	&Container classes	&\cite{Jaygarl2009}	\\
112.    &Java Collections	&Java	&22,000	&Java collection library	&\cite{Jaygarl2009}	\\
113.    &ASM	&Java	&40,000	&A Java bytecode manipulation and analysis framework	&\cite{Jaygarl2009}	\\
114.    &Apache Ant	&Java	&209,000	&A Java-based build tool	&\cite{Jaygarl2009}	\\
115.    &BugTracker.net	&C\#, ASP.NET 	&52,147	&A time-tested website content management system   &\cite{Tappenden2014}   \\		
116.    &E107	&PHP 	&175,265	&A web-based bug tracking and issue tracking application   &\cite{Tappenden2014}   \\		
117.    &GeekLog	&PHP	&118,706	&Manage dynamic web content   &\cite{Tappenden2014}   \\		
118.    &PhpBB2	&PHP	&43,831	&A  flat-forum bulletin board software solution   &\cite{Tappenden2014}   \\		
119.    &PhpBB3	&PHP	&203,465	&A  flat-forum bulletin board software solution   &\cite{Tappenden2014}   \\		
120.    &PhpMyAdmin	&PHP	&194,870	&Handle the administration of MySQL over the Web   &\cite{Tappenden2014}   \\		
121.    &RWWA1	&NR	&NR	&A real world web application from theTickets.com Pty Ltd	&\cite{Selay2014}	\\
122.    &RWWA2	&NR	&NR	&A real world web application from theTickets.com Pty Ltd	&\cite{Selay2014}	\\
123.    &RWWA3	&NR	&NR	&A real world web application from theTickets.com Pty Ltd	&\cite{Selay2014}	\\
124.    &RWWA4	&NR	&NR	&A real world web application from theTickets.com Pty Ltd	&\cite{Selay2014}	\\
125.    &RWWA5	&NR	&NR	&A real world web application from theTickets.com Pty Ltd	&\cite{Selay2014}	\\
126.    &RWWA6	&NR	&NR	&A real world web application from theTickets.com Pty Ltd	&\cite{Selay2014}	\\
127.    &RWWA7	&NR	&NR	&A real world web application from theTickets.com Pty Ltd	&\cite{Selay2014}	\\
128.    &Stock	&NR	&NR	&Searching stock information	&\cite{Chen2014}	\\
129.    &Weatherforecast	&NR	&NR	&Weather forecast service	&\cite{Chen2014}	\\
130.    &E-Banking	&NR	&NR	&Online banking service	&\cite{Chen2014}	\\
131.    &Bookfinding	&NR	&NR	&Searching book information	&\cite{Chen2014}	\\

\Xhline{0.6pt}
\multicolumn{6}{r}{continued on next page}\\
\Xhline{1.0pt}
\end{tabular*}
\end{table*}

\begin{table*}[!t]
\scriptsize
\centering
\begin{tabular*}{\textwidth}{rlllll} \Xhline{1.0pt}
\textbf{No.}    &\textbf{Name} &\textbf{Language} &\textbf{Size} &\textbf{Description} &\textbf{Reference} \\\Xhline{0.6pt}

132.	&Domainfinding	&NR	&NR	&Searching domain and	&\cite{Chen2014}	\\
133.	&Petinformation	&NR	&NR	&Searching pet information	&\cite{Chen2014}	\\
134.	&Traintime	&NR	&NR	&Searching train timetable	&\cite{Chen2014}	\\
135.	&Planetime	&NR	&NR	&Searching aircraft flight	&\cite{Chen2014}	\\
136.	&QQcheckonline	&NR	&NR	&Searching QQ online	&\cite{Chen2014}	\\
137.	&Queryresults	&NR	&NR	&Searching student achievement	&\cite{Chen2014}	\\
138.	&Producedorder	&NR	&NR	&Searching production order	&\cite{Chen2014}	\\
139.	&Calculator	&NR	&NR	&Arithmetic calculating service	&\cite{Chen2014}	\\
140.	&Maxdivisor	&NR	&NR	&Finding the greatest	&\cite{Chen2014}	\\
141.	&Mod	&NR	&NR	&Finding the remainder	&\cite{Chen2014}	\\
142.	&Reversestring	&NR	&NR	&Reversing the string	&\cite{Chen2014}	\\
143.	&Stringcopy	&NR	&NR	&Copying the string	&\cite{Chen2014}	\\
144.	&Stringlength	&NR	&NR	&Obtaining the length	&\cite{Chen2014}	\\
145.	&Login	&NR	&NR	&User login	&\cite{Chen2014}	\\
146.	&Vote	&NR	&NR	&Getting the vote	&\cite{Chen2014}	\\
147.	&Echoinformation	&NR	&NR	&Echoing personal information	&\cite{Chen2014}	\\
148.	&Checkeq	&C/C++	&90	&Report missing or unbalanced delimiters and .EQ/.EN pairs	&\cite{Chen2013}	\\
149.	&Col	&C/C++	&274	&Filter reverse paper motions for nroff output for display on a terminal	&\cite{Chen2013}	\\
150.	&Crypt	&C/C++	&121	&Encrypt and decrypt a file using a user supplied password	&\cite{Chen2013}	\\
151.	&Tr	&C/C++	&127	&Translate characters	&\cite{Chen2013}	\\
152.	&Count	&C/C++	&42	&Count lines, words, and characters	&\cite{Huang2014}	\\
153.	&Series	&C/C++	&288	&Generate an additive series of numbers	&\cite{Huang2014}	\\
154.	&Tokens	&C/C++	&192	&Sort/count alphanumeric tokens	&\cite{Huang2014}	\\
155.	&Ntree	&C/C++	&307	&Functions for managing a tree	&\cite{Huang2014}	\\
156.	&Nametbl	&C/C++	&329	&Functions for a symbol table	&\cite{Huang2014}	\\
157.	&SCC &Delphi	&NR &A supercharger clutch controller that controls a supercharger clutch   &\cite{Matinnejad2015}  \\				
158.	&ASS &Delphi	&NR &An auto start-Stop controller that controls the engine torque   &\cite{Matinnejad2015}  \\				
159.	&GCS &Delphi	&NR &A guidance control system that controls the position of a missile   &\cite{Matinnejad2015}  \\				
160.	&UAV	&NR	&NR	&Unmanned aerial vehicle (UAV) cruise control system	&\cite{Sun2012}	\\
161.	&Validation	&Java	&80	&Check whether the string is a valid email address   &\cite{Shahbazi2016}    \\		
162.	&PostCode	&Java	&293	&Check whether the string is a valid UK postcodes   &\cite{Shahbazi2016}    \\		
163.	&Numeric	&Java	&217	&Validate strings that represent integers   &\cite{Shahbazi2016}    \\		
164.	&DateFormat	&Java	&236	&Validate a date in the format 'dd/mm/yyyy'   &\cite{Shahbazi2016}    \\		
165.	&MIMEType	&Java	&145	&Validate MIME types   &\cite{Shahbazi2016}    \\		
166.	&ResourceURL	&Java	&339	&Validate Resource URLs   &\cite{Shahbazi2016}    \\		
167.	&URI	&Java	&267	&Validate different types of URIs   &\cite{Shahbazi2016}    \\		
168.	&URN	&Java	&327	&Check whether the string is a valid URNs   &\cite{Shahbazi2016}    \\		
169.	&TimeChecker	&Java	&267	&Validate 24 hour time format supplied as strings   &\cite{Shahbazi2016}    \\		
170.	&Clocale	&Java	&751	&Validate POSIX locale identifiers   &\cite{Shahbazi2016}    \\		
171.	&Isbn	&Java	&420	&Check whether the string is a valid international standard book number   &\cite{Shahbazi2016}    \\		
172.	&BIC	&Java	&200	&Check whether the string is a valid bank identifiers code   &\cite{Shahbazi2016}    \\		
173.	&IBAN	&Java	&288	&Check whether the string is a valid international bank account numbers   &\cite{Shahbazi2016}    \\		
174.	&Bluetooth	&Java	&NR	&A Bluetooth application for file sharing	&\cite{Liu2010a}	\\
175.	&Contact	&Java	&NR	&Management of user contact	&\cite{Liu2010a}	\\
176.	&SMS	&Java	&NR	&SMS client for sending and receiving SMS	&\cite{Liu2010a}	\\
177.	&Bluetalk	&Java	&NR	&An VOIP Bluetooth application	&\cite{Liu2010a}	\\
178.	&Dialer	&Java	&NR	&Make and answering calls	&\cite{Liu2010a}	\\
179.	&Browser	&Java	&NR	&Mobile Web Browser for surfing Internet	&\cite{Liu2010a}	\\
180.	&Open\_vSwitch	&NR	&NR	&A OpenFlow-Switch &\cite{Koo2016}	\\	
181.	&Crossword Sage  &Java   &NR 	&Unambiguous GUIs for testers to design the test cases   &\cite{Zhang2014}   \\				
182.	&LLVM	&C/C++  &4,727,209  &Open-source C compiler &\cite{Chen2016}    \\				
183.	&GCC	&C/C++  &3,343,377  &Open-source C compiler &\cite{Chen2016}    \\				
184.	&GCD &Java   &NR &Calculate the greatest common divisor   &\cite{Liu2008} \\					
185.	&LCM &Java   &6,199 &Calculate the least common multiplier   &\cite{Liu2008} \\					
186.	&Siena   &Java   &6,035  &A framework for constructing event notification services   &\cite{Cotroneo2016}    \\					
187.	&NanoXML   &Java   &7,646  &A simple XML parser for Java   &\cite{Cotroneo2016}    \\					
188.	&Make   &C/C++   &35,545  &Unix build utility   &\cite{Cotroneo2016}    \\					
189.	&SP	&C/C++	&NR	&A simple multi-threaded program	&\cite{Yue2015}	\\
190.	&SV	&C/C++	&NR	&A simple multi-threaded program	&\cite{Yue2015}	\\
191.	&SPF	&C/C++	&NR	&A modified version of SP	&\cite{Yue2015}	\\
192.	&SVF	&C/C++	&NR	&A modified version of SV	&\cite{Yue2015}	\\
193.	&Fft	&C/C++	&NR	&A simple multi-threaded program	&\cite{Yue2015}	\\
194.	&Lu\_cb	&C/C++	&NR	&A simple multi-threaded program	&\cite{Yue2015}	\\
195.	&Radix	&C/C++	&NR	&A simple multi-threaded program	&\cite{Yue2015}	\\
196.	&TS-7260\_ARM &NR &NR &An embedded Linux system with Out-Of-Memory (OOM) Killer &\cite{Sim2011} \\					
197.	&TextureAtlas	&C/C++	&405	&Store and manipulate multiple textures efficiently for graphics libraries 	&\cite{Patrick2017}	\\
198.	&Chunkybar	&C/C++	&451	&Implement multi-piece progress bars used in bittorrent clients 	&\cite{Patrick2017}	\\
199.	&PseudoLRU	&C/C++	&472	&An implementation of the Pseudo-LRU (Least Recently Used) cache algorithm	&\cite{Patrick2017}	\\
200.	&QPHashMap	&C/C++	&568	&A hashmap data structure that makes use of quadratic probing in order to manage collisions	&\cite{Patrick2017}	\\
201.	&P1	&C\#	&NR	&A small C\# program	&\cite{Indhumathi2014}	\\
202.	&P2	&C\#	&NR	&A small C\# program	&\cite{Indhumathi2014}	\\
203.	&P3	&C\#	&NR	&A small C\# program	&\cite{Indhumathi2014}	\\
204.	&P4	&C\#	&NR	&A small C\# program	&\cite{Indhumathi2014}	\\
205.	&P5	&C\#	&NR	&A small C\# program	&\cite{Indhumathi2014}	\\
206.	&Foo &C/C++  &8  &A simple program with two integer inputs   &\cite{Nikravan2015}    \\					
207.	&TriSquare   &NR   &168 &Check whether 3 positive real numbers could construct a triangle   &\cite{Hui2016} \\					
208.	&Sin &C/C++  &99 &A Sin mathematic function  &\cite{Hui2016} \\					
209.	&Bessjy  &C/C++  &332	&The Bessel function	&\cite{Hui2016} \\			
210.	&Mulriple	&NR	&48	&Calculate the absolute value, reciprocal or multiple of the input data	&\cite{Yuan2011}	\\
211.	&MyMath	&NR	&90	&Calculate the most value of the input data or the simple arithmetic	&\cite{Yuan2011}	\\

\Xhline{1.0pt}
\end{tabular*}
\end{table*}


\end{document}